\documentclass[11pt, a4paper]{article}
\usepackage[utf8]{inputenc}
\usepackage{jheppub}
\usepackage{dsfont}
\usepackage{shuffle}
\usepackage{array}
\usepackage[table]{xcolor}

\usepackage{tikz}
\usetikzlibrary{arrows,shapes}
\usetikzlibrary{trees}
\usetikzlibrary{patterns}
\usetikzlibrary{matrix,arrows} 	% For commutative diagram
\usetikzlibrary{positioning}				% For "above of=" commands
\usetikzlibrary{calc,through}				% For coordinates
\usetikzlibrary{decorations.pathreplacing}  % For curly braces
\usepackage{pgffor}							% For repeating patterns
\usepackage{diagbox}
\usetikzlibrary{decorations.pathmorphing}	% For Feynman Diagrams
\usetikzlibrary{decorations.markings}
\tikzset{
    vector/.style={decorate, decoration={snake}, draw},
	provector/.style={decorate, decoration={snake,amplitude=2.5pt}, draw},
	antivector/.style={decorate, decoration={snake,amplitude=-2.5pt}, draw},
        smallvector/.style={decorate, decoration={snake,amplitude=1.5pt,post length=0.5mm}, draw},
    fermion/.style={draw=black, postaction={decorate},
        decoration={markings,mark=at position .55 with {\arrow[draw=black]{>}}}},
    fermionbar/.style={draw=black, postaction={decorate},
        decoration={markings,mark=at position .55 with {\arrow[draw=black]{<}}}},
    fermionnoarrow/.style={draw=black},
    gluon/.style={decorate, draw=black,
        decoration={coil,amplitude=4pt, segment length=5pt}},
    scalar/.style={dashed,draw=black, postaction={decorate},
        decoration={markings,mark=at position .55 with {\arrow[draw=black]{>}}}},
    scalarbar/.style={dashed,draw=black, postaction={decorate},
        decoration={markings,mark=at position .55 with {\arrow[draw=black]{<}}}},
    scalarnoarrow/.style={dashed,draw=black},
    electron/.style={draw=black, postaction={decorate},
        decoration={markings,mark=at position .55 with {\arrow[draw=black]{>}}}},
    bigvector/.style={decorate, decoration={snake,amplitude=4pt}, draw},
    arrow/.style={draw=black, postaction={decorate},
        decoration={markings,mark=at position 1 with {\arrow[draw=black]{>}}}},
}

\tikzstyle{block} = [draw, rectangle, 
    minimum height=3em, minimum width=6earticlem]

\newcommand{\sq}[1]{\left[#1\right]}
\newcommand{\ang}[1]{\langle #1\rangle}

\newcommand{\mn}{{\mu\nu}}

\newcommand{\reef}[1]{(\ref{#1})}

\def\be{\begin{equation}}
\def\ee{\end{equation}}
\def\bea{\begin{eqnarray}}
\def\eea{\end{eqnarray}}
\def\ba{\begin{array}}
\def\ea{\end{array}}
\def\bd{\begin{displaymath}}
\def\ed{\end{displaymath}}

\def\Tr{{\rm Tr}}

\def\pa{\partial}                              % curly d
\def\>{\rangle} %right angle
\def\<{\langle} %left angle
\def\Dsl{D \hskip-.6em \raise1pt\hbox{$ / $ } }
\def\to{\rightarrow}
\def\pa{\partial}

\newcommand{\eps}{\epsilon}
\newcommand{\lra}{\leftrightarrow}

\title{
    Electromagnetic Duality and D3-Brane Scattering Amplitudes Beyond Leading Order}

\author[a]{Henriette Elvang,}
\author[b]{Marios Hadjiantonis,}
\author[a]{Callum R.~T.~Jones,}
\author[a]{and Shruti Paranjape}

\affiliation[a]{
    Leinweber Center for Theoretical Physics,\\ 
    Randall Laboratory of Physics, Department of Physics, University of Michigan, \\
    450 Church St, Ann Arbor, MI 48109, USA}
\affiliation[b]{
    NORDITA, KTH Royal Institute of Technology and Stockholm University, \\
    Roslagstullsbacken 23, SE-10691 Stockholm, Sweden}

\emailAdd{elvang@umich.edu}
\emailAdd{marios.hadjiantonis@su.se}
\emailAdd{jonescal@umich.edu}
\emailAdd{shrpar@umich.edu}

\abstract{We use on-shell methods to study the non-supersymmetric and supersym\-metric low-energy S-matrix on a probe D3-brane, including both the 1-loop contributions of massless states as well as the effects of higher-derivative operators. Our results include: 
(1) A derivation of the duality invariance of Born-Infeld electrodynamics  as the dimensional oxidation of the group of spatial rotations transverse to a probe M2-brane; this is done using a novel implementation of subtracted on-shell recursion. (2) The first explicit 
loop-level BCJ double-copy in 
a non-gravitational model, namely the calculation of the 4-point 
self-dual amplitude of non-supersymmetric Born-Infeld. (3) From previous results for $n$-point self-dual 1-loop BI amplitudes and the conjectured dimension-shifting relations in Yang-Mills, we obtain an explicit all-multiplicity, at all orders in $\epsilon$, expression for the 1-loop integrand of the MHV sector of $\mathcal{N}=4$ DBI. (4) For all $n>4$, the  explicitly integrated duality-violating 1-loop amplitudes (self-dual and next-to-self-dual in pure BI as well as MHV in $\mathcal{N}=4$ DBI) are shown to be removable at $\mathcal{O}(\epsilon^0)$ by adding finite local counterterms; we propose that this may be true more generally at 1-loop order. (5) We find that in non-supersymmetric Born-Infeld, 
not all finite local counterterms needed to restore electromagnetic duality can  be constructed using the double-copy with higher-derivative corrections, suggesting a fundamental tension between electromagnetic duality and color-kinematics duality at loop-level.
Finally we comment on oxidation of duality symmetries in supergravity and the parallels it has to the M2-brane to D3-brane oxidation demonstrated in this paper. 
}

\begin{document}

{\hspace*{\fill} \tt LCTP-20-13}\\
{\hspace*{\fill} \tt NORDITA 2020-059}

\maketitle
\flushbottom

%%%%%%%%%%%%%%%%%%%%%%%%%%%%%%
\section{Introduction}
\label{sec:Introduction}

Electromagnetic (EM) duality is a continuous, global symmetry of certain models containing abelian gauge fields in $d=4$.\footnote{This notion of EM duality is related to, but not exactly the same as, 
the Montonen-Olive S-duality of $\mathcal{N}=4$ super Yang-Mills 
theory \cite{Montonen:1977sn}; S-duality is necessarily a 
discrete symmetry since it acts on the quantized charge lattice. In 
the zero-coupling limit, $g_{\text{YM}}\rightarrow 0$, the charged 
states decouple and the symmetry is enhanced to the continuous $SO(2)$ 
duality symmetry of free Maxwell theory. The electromagnetic duality 
considered in this paper is of the continuous kind, without charged 
states, but with non-linear self-interactions of the gauge field which 
preserve the duality symmetry.} On-shell, the charge eigenstates of the symmetry coincide with the helicity eigenstates of a massless spin-1 field (and also potentially additional states), and as such give a concrete realization of \textit{chiral symmetry} for spin $>1/2$. The primary consequences of such a symmetry for the physically observable S-matrix elements are selection rules, or on-shell Ward identities, of the form
\begin{equation}
\label{EMamps}
\mathcal{A}_{n}\big( \underbrace{\gamma^+ \ldots \gamma^+}_{n_+} \underbrace{\gamma^- \ldots \gamma^-}_{n_-}\big) = 0 ~~\text{for}~~n_+\ne n_-\,,
\end{equation}
where we are using the common convention that helicity states are labelled with all particles outgoing.

Off-shell the meaning of electromagnetic duality is more subtle. In the standard manifestly Lorentz-covariant formulation, the off-shell action for an abelian gauge field is constructed using field operators $A^\mu(x)$ that are not in one-to-one correspondence with physical states. It is perhaps not surprising that as a consequence, there is no off-shell symmetry of the action for which the selection rule (\ref{EMamps}) is the conservation of the associated Noether charge. There is however, a corresponding symmetry of the classical equations of motion. In the simplest case of duality-invariant non-linear electrodynamics, the equations of motion consist of a Bianchi identity constraint and an Euler-Lagrange equation,
\begin{equation}
    \partial_\mu \tilde{F}^{\mu\nu} = 0, \hspace{5mm} \partial_\mu \tilde{G}^{\mu\nu}[F] = 0, \hspace{5mm} \text{where} \hspace{5mm} \tilde{G}^{\mu\nu}[F]\equiv 2\frac{\partial \mathcal{L}}{\partial F_{\mu\nu}}\,.
\end{equation}
Given any solution $F_{\mu\nu}$ we can generate a one-parameter family of solutions by a so-called duality rotation
\begin{equation}
\label{dualityrotation}
    F_{\mu\nu} \rightarrow \cos(\theta)F_{\mu\nu} + \sin(\theta)G_{\mu\nu}[F]\,,
\end{equation}
if and only if the model satisfies a non-linear, on-shell constraint known as the \textit{Gaillard-Zumino condition} \cite{Gaillard:1981rj}
\begin{equation}
    \label{GZ}
    F_{\mu\nu}\tilde{F}^{\mu\nu} 
    + G_{\mu\nu}[F]\tilde{G}^{\mu\nu}[F]=0\,.  
\end{equation}

The connection between duality invariance of the equations of motion and the conservation of a chiral charge \reef{EMamps} in the on-shell scattering amplitudes, is perhaps not widely known. At tree-level, the connection was demonstrated for classically duality invariant models of non-linear electrodynamics in \cite{Novotny:2018iph},  generalizing earlier demonstrations in the context of the non-supersymmetric Born-Infeld (BI) model \cite{Rosly:2002jt}. More generally, the conservation of a chiral charge is an empirical fact about tree-level scattering amplitudes in a wide variety of classically duality-invariant models, including models of extended supergravity \cite{Bern:2017rjw}.  

The goal of this paper is to improve our understanding of duality invariance, in the form of the selection rule (\ref{EMamps}), using modern on-shell methods. This includes a novel understanding of why such a symmetry may, in certain classes of models, be present. Using D-brane worldvolume EFTs as theoretical laboratories in which to make explicit calculations, we see from one perspective, using dimensional reduction/oxidation together with subtracted on-shell recursion, that such a symmetry emerges naturally. Contrarily, from the perspective of the BCJ double-copy construction \cite{Kawai:1985xq,Bern:2008qj,Cachazo:2014xea},
\be
\begin{split}
    \label{BIequalschiPTYM} 
    &\Big( \text{Born-Infeld with } \mathcal{N} \text{ SUSY} \Big) \\
    &= \Big( U(N) \text{ Yang-Mills with } \mathcal{N} \text{ SUSY } \Big) ~\otimes ~ \bigg(\frac{U(N)\times U(N)}{U(N)} \text{ Nonlinear Sigma Model}\bigg),
\end{split}
\ee
at both tree- and loop-level, the appearance of such symmetries is an unexplained miracle. Specifically, amplitudes in YM theory do not satisfy a non-abelian analogue of the selection rule \reef{EMamps}, so for example  any odd-point YM amplitude and any MHV amplitude with more than 4 particles have to give zero upon being double-copied with NLSM amplitudes. Such cancellations do indeed happen, as is necessary for the double-construction to work, but it is an emergent rather than a manifest property. 

In the non-supersymmetric case of \reef{BIequalschiPTYM}, we
 analyze the higher-derivative corrections to both Yang-Mills (YM) and the nonlinear sigma model ($\chi$PT) and find that the double-copy 
by itself is not sufficient to guarantee duality invariance. The result is a mysterious subset of higher-derivative corrections, which are neither completely duality-preserving nor completely general.

A further goal of this paper, building on earlier work by the same authors \cite{Elvang:2019twd}, is to improve our understanding of electromagnetic duality in the context of \textit{quantum} corrections. That the duality rotation (\ref{dualityrotation}) is only a symmetry of the classical equations of motion, and not the off-shell action, might lead us to suspect that the selection rule (\ref{EMamps}) is only valid at tree-level. This conclusion is complicated by the fact that there exist alternative, non-manifestly Lorentz-covariant, formulations of duality invariant models which do realize duality rotations off-shell \cite{Deser:1976iy,Schwarz:1993vs}. It is an interesting and important problem to determine if classically duality-invariant models can be quantized (at least perturbatively) in a way that preserves both the selection rule (\ref{EMamps}) \textit{and} Lorentz invariance, or if there is an anomaly which requires breaking the former to preserve the latter. We address this question by explicit calculation in the context of the worldvolume EFT of a probe D3-brane, described by the abelian $\mathcal{N}=4$ \textit{Dirac-Born-Infeld (DBI)} model. A central result is that we obtain an explicit, all-multiplicity expression for the 1-loop integrand of the MHV sector of $\mathcal{N}=4$ DBI in dimensional regularization at all orders in $\epsilon$. For $n>4$ the resulting manifestly Lorentz-invariant expressions, integrated in $d=4-2\epsilon$ up to $\mathcal{O}(\epsilon^0)$, are shown to be cancelled by $\mathcal{N}=4$ invariant finite local counterterms, and hence there is no duality anomaly in this sector. From previous results obtained by the present authors in the self-dual (all-plus) and next-to-self-dual (one-minus) sectors of non-supersymmetric Born-Infeld \cite{Elvang:2019twd}, together with the above general analysis of the higher-derivative double-copy, we demonstrate that the requisite finite counterterms cannot always be constructed as a KLT product. This may indicate that there exists no loop-level regularization scheme that respects both electromagnetic duality and color-kinematics duality, suggesting a deep conflict between these two physical principles.

Below we give a brief review of the physics of probe D-brane worldvolume EFTs, with emphasis on the manifestation of physical properties in the S-matrix. This is followed by a more detailed overview of the structure and key results of the present paper. 
 
\subsection{Review of D3-Brane Worldvolume EFTs}
 
The models we consider in this paper describe the low-energy dynamics of the massless excitations of a probe D3-brane with up to $\mathcal{N}=4$ linear supersymmetry, embedded in a $D$-dimensional bulk Minkowski spacetime. The effective field theory on the 4d worldvolume takes the form of a $U(1)$ gauge theory with nonlinear self-interactions, coupled to $N_f$ Weyl fermions and $N_s$ real scalars. The bosonic part of the leading-order (DBI) effective action for a probe D$p$-brane has the form
\begin{equation}
\label{Dpaction}
    S_{Dp}[X^I,F_{\mu\nu}] = T_{Dp} \int \text{d}^{p+1}x \sqrt{-\text{det}\left(\eta_{\mu\nu}+k^2\partial_\mu X_I \partial_\nu X^I + k F_{\mu\nu}\right)}, \hspace{5mm} k=2\pi \alpha',
\end{equation}
where $I=1,2,...,N_s$, with $N_s=D-p-1$ and $T_{Dp}\sim  1/\alpha'^2$. The complete $\kappa$-symmetric action including fermionic degrees-of-freedom for $p=3$ is given explicitly in \cite{Aganagic:1996pe}. In this paper we take an on-shell perspective, so rather than calculating scattering amplitudes directly from the action (\ref{Dpaction}), we begin with the physical properties we expect of the model, and use these to \textit{bootstrap} physical S-matrix elements. The massless degrees of freedom and their symmetries have a direct physical interpretation:
\begin{itemize}
    \item The world-volume \textit{photon} $\gamma^\pm$ describes the motion of the endpoints of open strings ending on the D3-brane \cite{Fradkin:1985qd}. The associated classical equations of motion are invariant under a duality rotation (\ref{dualityrotation}),  so as discussed above the photon helicity states carry a conserved chiral charge $Q[\gamma^\pm] = \pm 1$. If the external momenta are restricted to a 3d subspace, then a particular linear polarization is identified with the scalar modulus of a D2-brane in a T-dual frame. The associated enhanced low-energy theorem is sufficient to bootstrap the entire tree-level S-matrix of the non-supersymmetric Born-Infeld model \cite{Cheung:2018oki}.
    \item  The $N_f$ Weyl fermions $\psi^\pm_A$ are Goldstone fermions arising from (potentially partial) spontaneous supersymmetry breaking. All of the fermions are identical and transform together in the fundamental representation of $U(N_f)$. The broken supersymmetry is non-linearly realized in the effective action \cite{Aganagic:1996pe}, and consequently the on-shell scattering amplitudes satisfy an Adler-zero type low-energy theorem
    \begin{equation}
    \label{softfermion}
        \mathcal{A}_n^{\text{D}3}\left(\{|p\rangle,|p]\}_\psi^+,...\right) \xrightarrow{|p\rangle \sim 0} \mathcal{O}\left(|p\rangle\right).
    \end{equation}
    For the special case $N_f=1$, $N_s=0$, the spectrum consists of a single $\mathcal{N}=1$ vector multiplet and the scattering amplitudes satisfy the associated $\mathcal{N}=1$ on-shell supersymmetry Ward identities. Together with the low-energy theorem (\ref{softfermion}) this can be used to prove that the amplitudes of the bosonic truncation (non-supersymmetric Born-Infeld) satisfy a multi-chiral low-energy theorem \cite{Cheung:2018oki}, sufficient to bootstrap the tree-level S-matrix.
    \item The $N_s$ scalar fields $X_I$ are spacetime Goldstone bosons, or moduli, arising from the spontaneous breaking of the underlying $D$-dimensional Poincar\'e symmetry 
    \begin{equation}
        \text{ISO}(D-1,1)\rightarrow \text{ISO}(3,1)\times \text{SO}(D-4).
    \end{equation}
    Naturally, the scalar fields transform in the vector representation of $SO(D-4)$. The spontaneously broken spacetime symmetries are non-linearly realized in the effective action (\ref{Dpaction}) with the scattering amplitudes satisfying an \textit{enhanced} Adler-zero low-energy theorem \cite{Cheung:2014dqa}
        \begin{equation}
    \label{softscalar}
        \mathcal{A}_n^{\text{D}3}\left(\{p\}_X,...\right) \xrightarrow{p \sim 0} \mathcal{O}\left(p^2\right).
    \end{equation}
\end{itemize}

The problem of constructing a model that satisfies all of the above properties is vastly over-constrained, and it is remarkable that any solution exists. A complete formula for the tree-level S-matrix of the bosonic truncation (\ref{Dpaction}), valid in $d$-dimensions, was found in the CHY formalism \cite{Cachazo:2014xea}. Recently, a beautiful CHY formula based on \textit{polarized} scattering equations, appeared for the complete tree-level S-matrix of the $\mathcal{N}=4$ probe D3-brane ($N_f=4$ and $N_s=6$), as well as similar formulae for the 6d parent theories describing probe $\mathcal{N}=(1,1)$ D5- and $\mathcal{N}=(2,0)$ M5-branes \cite{Heydeman:2017yww}. Such higher-dimensional constructions have recently been used to obtain 1-loop integrands for the D3-brane theory in dimensional regularization \cite{Wen:2020qrj}.

The existence of various $U(1)$ charges requires that any non-zero amplitude contains an \textit{even} number of states of a given spin. It is therefore consistent to consider truncations to the purely scalar ($p$-brane) sector as well as the purely fermionic (Akulov-Volkov or Goldstino) sector. Using the above vanishing low-energy theorems, it was shown in \cite{Cheung:2014dqa,Cheung:2015ota,Cheung:2016drk,Luo:2015tat,Elvang:2018dco} that we can bootstrap the leading-order contribution to the tree-level S-matrix, including in the former case, a finite number of higher-derivative or \textit{Galileon} contributions \cite{deRham:2010eu}. 

The CHY construction \cite{Cachazo:2014xea} has revealed the surprising fact that the well-known field theory KLT formula \cite{Kawai:1985xq}, relating Yang-Mills and Einstein gravity, can also be used to construct D$p$-brane world-volume EFTs. The generalization of the double-copy relation (\ref{BIequalschiPTYM}) takes the schematic form
\be
\label{gendcBI}
    \Big( \text{BI} \oplus N_f \text{ fermions} \oplus N_s \text{ scalars}\Big) = \Big(\text{YM}\oplus N_f \text{ fermions} \oplus N_s \text{ scalars}\Big)\;\; \otimes \;\; \chi\text{PT}\,,
\ee
where $\oplus$ in the Yang-Mills theory on the right-hand-side indicates whatever couplings between the gluons and massless adjoint fermions and scalars (including Yukawa and scalar potentials) are required to satisfy color-kinematics duality \cite{Bern:2008qj}. In this construction, the duality symmetry of the Born-Infeld photon is completely obscured since there is no analogue in the non-abelian Yang-Mills theory. Conceptually the double-copy paints an interesting picture of the massless degrees-of-freedom on the D$p$-brane. The \textit{pions} of $\chi$PT are Goldstone modes of a spontaneously broken internal symmetry, which combine through the double-copy with the scalar fields of the Yang-Mills theory to form the spacetime Goldstone modes of the D$p$-brane. Similarly, the pions combine with the adjoint fermions of the Yang-Mills theory to form Goldstone fermions. The Born-Infeld photon itself has no known interpretation as a Goldstone mode, but through the double-copy we see that it is the combination of a Goldstone scalar pion and a non-abelian gauge boson. 
 
\subsection{Outline of this Paper}

In Section \ref{sec:EMduality} we prove a theorem stating that at leading order in the derivative expansion, the existence of a global $U(1)$ symmetry in the 3d reduction of a model of non-linear electrodynamics is a necessary and sufficient condition for the existence of a duality symmetry in 4d. As a corollary, the $U(1)$ symmetry of the probe M2-brane world-volume EFT corresponding to the rotation of a pair of transverse dimensions \textit{oxidizes} to the duality symmetry of the D3-brane, giving a novel inductive proof of the latter using subtracted on-shell recursion.

In Section \ref{sec:DCloop} we present an explicit test of the loop-level BCJ double-copy conjecture. We use known color-kinematics duality satisfying BCJ numerators from the self-dual 1-loop 4-point (i.e.~all-plus) sector of pure Yang-Mills together with generic BCJ numerators for $\chi$PT to construct a conjectured representation for the self-dual 1-loop 4-point integrand of the pure Born-Infeld amplitude. The result is found to agree precisely with recent results obtained using unitarity methods.

In Section \ref{sec:BIall1loop} we review the results of \cite{Elvang:2019twd}. Explicit, all-multiplicity expressions are given for the 1-loop integrand of the self-dual and next-to-self-dual sectors of pure Born-Infeld. At 4-point we present the explicit results for the local counterterms needed to restore electromagnetic duality symmetry in the sense of setting the 1-loop self-dual BI amplitude $\mathcal{A}^\text{1-loop}_4(++++)$ to zero.

In Section \ref{sec:SD_MHV_relation} we present a conjectured explicit form of the 1-loop, MHV, integrand of $\mathcal{N}=4$ Dirac-Born-Infeld. This result is obtained by combining the conjectured dimension-shifting relation between the integrands of self-dual and maximally supersymmetric Yang-Mills with the loop-level BCJ double-copy. At $n=4$ the (cut-constructible part of the) expression is shown to precisely match the result expected from unitarity cuts, and at $n=6$ the expression agrees with recent calculations using dimensional reduction of M5-brane tree-amplitudes. 

In Section \ref{sec:rational}, based on the observation that $n$-point amplitudes in the SD and NSD sectors can be set to zero by appropriate choice of finite local counterterms, we present an argument (though not proof) that indicates that there is no possible violation of electromagnetic duality at 1-loop in BI theory in any sector. In other words, that there exists a scheme in which the electromagnetic duality $U(1)$ symmetry is unbroken at 1-loop. 

In Section \ref{sec:DChd} we study the tree-level double-copy relation \reef{BIequalschiPTYM} with higher-derivative terms. We present a general analysis at 4-point level. In particular we find that, perhaps surprisingly, this does not allow us to construct, by BCJ double-copy, the finite local counter\-term needed to restore electromagnetic duality. This is a new curious conflict between electromagnetic duality at loop-level and the double-copy construction. The results of the higher-derivative double-copy analysis are also compared to the $\alpha'$-corrections from string theory, both for the superstring and bosonic string. 

In Section \ref{sec:disc} we discuss a few interesting open questions. We also discuss in quite some detail duality symmetries in the context of pure 4d supergravity theories. This includes their emergence in the double-copy, the effect of higher-derivative corrections, and the ideas of possible oxidation of symmetries from 3d to 4d. The parallels, and contrasts, with EM duality in Born-Infeld theories is emphasized. 

The five Appendices contain various technical details of the results presented in the main text. 

%%%%%%%%%%%%%%%%%%%%%%%%%%%%%%

%%%%%%%%%%%%%%%%%%%%%%%%%%%%%%
\section{Dimensional Oxidation: the D3-brane and M2-Brane}
\label{sec:EMduality}
%%%%%%%%%%%%%%%%%%%%%%%%%%%%%%
We present the worldvolume reduction of the D3-brane action to the M2-brane action and show how the latter has a manifest $U(1)$ symmetry. The oxidation of this symmetry to 4d is very non-trivial, but nonetheless we prove that it is true in the leading order theory using a novel recursion relation. Subsequently we examine higher-derivative corrections and prove that there exist duality-violating operators in 4d that are not ruled out by the 3d $U(1)$ symmetry.

\subsection{World-Volume Analysis}
\label{s:WV}

It is well-known that in $D$ dimensions, the compactification of a D$p$-brane on a $p-p'$ torus along its worldvolume directions is T-dual to a flat D$p'$-brane, again in $D$ dimensions. This property is inherited by the low-energy effective theories on the branes. The $p+1$-dimensional vector boson states of the D$p$-brane, polarized in the compact dimensions, are physically identified with the additional $p-p'$ scalar moduli of the D$p'$-brane. Since all dimensions transverse to the world-volume are identical, the scalar moduli of the D$p$-brane realize a linear $SO(D-p-1)$ symmetry, while after dimensional reduction this is enhanced to $SO(D-p'-1)$.

A special case of this phenomenon occurs in the dimensional reduction of the D3-brane. For simplicity we restrict the discussion to the truncation to the vector boson sector described by pure Born-Infeld electrodynamics in $\mathbb{R}^4$, 
\begin{equation}
\label{D3}
S_{\text{D}3}[F_{\mu\nu}] = T_{D3} \int \text{d}^4x \sqrt{-\det\left(\eta_{\mu\nu}+k F_{\mu\nu}\right)}\,.
\end{equation}
This action does not manifest any continuous global symmetries, but as described in the introduction, this model has a hidden $U(1)$ electromagnetic duality symmetry that leads to the helicity selection rule (\ref{EMamps}). Following the above discussion, dimensional reduction on a circle of vanishing radius gives the action of a D2-brane embedded in $\mathds{R}^4$, 
\begin{equation} 
\label{D2}
S_{\text{D}2}[F_{\mu\nu},X] = T_{D2} \int \text{d}^3x \sqrt{-\det\left(\eta_{\mu\nu}+k^2\partial_\mu X \partial_\nu X+k F_{\mu\nu}\right)}\,.
\end{equation} 
Since there is only a single scalar modulus, this action again does not manifest any continuous global symmetries. But, in fact this model has a hidden $U(1)$ symmetry that is revealed off-shell by changing variables via a Legendre transformation. To do so, the implicit Bianchi identity constraint is replaced by an explicit Lagrange multiplier term \cite{Polchinski:1998rq}: 
\begin{equation}
\label{D2alt}
S_{\text{D}2}[F_{\mu\nu},X,Y] =T_{D2} \int \text{d}^3x\left[ \sqrt{-\det\left(\eta_{\mu\nu}+k^2\partial_\mu X \partial_\nu X+k F_{\mu\nu}\right)}-\frac{k^2}{2} Y \epsilon^{\mu\nu\rho}\partial_\rho F_{\mu\nu}\right].
\end{equation}
By integrating by parts in the final term, the roles of the fields $F_{\mu\nu}$ and $Y$ as dynamical and auxiliary, respectively, are interchanged. Integrating out $F_{\mu\nu}$ generates a \textit{dual} representation of the D2-brane model parametrized in terms of a pair of scalar fields $X$ and $Y$ \cite{Polchinski:1998rq}. Remarkably, this new representation is precisely the (truncated) action of an M2-brane,
\begin{equation}
\label{M2}
S_{\text{M}2}[X,Y] =T_{M2} \int \text{d}^3x \sqrt{-\det\left(\eta_{\mu\nu}+k^2\partial_\mu X \partial_\nu X+k^2\partial_\mu Y \partial_\nu Y\right)}\,.
\end{equation}
The physics of the model (\ref{M2}) is that it describes the spontaneous breaking of spacetime symmetry $ISO(4,1)\rightarrow ISO(2,1)\times SO(2)$, with the scalar brane-moduli $X$ and $Y$ identified with the Goldstone modes of the broken transverse translation symmetries. The linearly realized $SO(2)$ symmetry is manifest in the action (\ref{M2}) as a rotation of the vector $(X,Y)$ and has the physical interpretation as an isometric rotation between the two non-compact dimensions transverse to the membrane. 

The physical significance of this observation is well-known \cite{Schmidhuber:1996fy,Townsend:1995af}. The non-truncated version of this argument is a crucial test of the identification of the 10d embedding of the probe D2-brane with the 11d embedding of the probe M2-brane \cite{Polchinski:1998rq}. In that case, the D3-brane model has a manifest $SO(6)$ symmetry acting as a rotation of the moduli associated with the 6 transverse dimensions. Dimensional reduction and T-dualization produces a D2-brane model with a manifest $SO(7)$ symmetry acting on the 7 moduli. Legendre transforming the 3d gauge boson produces a model with 8 moduli and a manifest $SO(8)$ symmetry, with the additional scalar identified with the spontaneous breaking of translation invariance in the 11th, M-theory dimension. 

In the truncated model, for the purposes of calculating scattering amplitudes it is more convenient to form a single complex scalar field $Z=(X+iY)/\sqrt{2}$ and rewrite the action in the form
\begin{equation}
\label{M2action}
S_{\text{M}2}[Z,\bar{Z}] =T_{M2} \int \text{d}^3x \sqrt{-\det\left(\eta_{\mu\nu}+
k^2 \partial_{(\mu} Z \partial_{\nu)}\bar{Z}
\right)}\;.
\end{equation}
The manifest $U(1)$ symmetry in \reef{M2action} implies that the scattering amplitudes of this model satisfy the selection rule
\begin{equation}
\label{M2amp}
\mathcal{A}^{\text{M}2}_{n}\big( \underbrace{Z \ldots Z}_{n_Z}\; \underbrace{\bar{Z} \ldots \bar{Z}}_{n_{\bar{Z}}}\big) = 0 ~~\text{for}~~n_Z\ne n_{\bar{Z}}.
\end{equation}
While the existence of the conserved $U(1)$ charge of the M2-brane action \reef{M2action} is completely obscured in the D2-brane action (\ref{D2}), obtained by standard dimensional reduction from the D3-brane action \reef{D3}, it is almost trivial at the level of the on-shell scattering amplitudes. Since we are scattering massless Kaluza-Klein modes, the 3d tree-level scattering amplitudes are insensitive to the radius of the compactified dimension and so are formally equivalent to the 4d tree-level scattering amplitudes with the external momenta restricted to an arbitrary 3d subspace. In the commonly used 4d spinor-helicity variables this restriction can be efficiently made by making the replacement
\begin{equation}
\label{4dto3d}
|p]_a\longrightarrow \langle p|_{\dot{a}}\,.
\end{equation}
Since the models only contain bosonic degrees of freedom, the resulting expressions containing only angle-spinors can always be rewritten in terms of Mandelstam invariants: in 3d, $s_{ij} = (p_i + p_j)^2 = -\<ij\>^2 $. All that remains is to relabel the on-shell states in a 3d language as
\begin{equation}
\label{idZgamma}
\gamma^+ \longleftrightarrow Z, \hspace{10mm} \gamma^- \longleftrightarrow \bar{Z}.
\end{equation}
Using this dictionary, we find that {\em the $U(1)$ electromagnetic duality symmetry of the M2-brane amplitudes follows as a necessary consequence of the selection rule (\ref{EMamps}) of electromagnetic duality of the D3-brane tree amplitudes.} 

At the level of the amplitude selection rules \reef{EMamps} and \reef{M2amp}, along with the dictionary \reef{4dto3d}-\reef{idZgamma}, it is clear that no special properties of Born-Infeld were used. Therefore it implies the following general result:\\
\vspace{1pt}\\
{\bf \em Theorem 1.} 
\textit{Given any 4d local model of non-linear electrodynamics $S_{(4d)}[F_{\mu\nu}]$, we can construct a 3d model of a complex scalar $S_{(3d)}[Z,\bar{Z}]$ by dimensional reduction followed by a Legendre transformation. If the 4d model has a $U(1)$ electromagnetic duality symmetry of the equations of motion, then the 3d model must have an off-shell $U(1)$ symmetry of the action. }\\
\vspace{1pt}\\
The above argument includes the possibility of higher-derivative terms; at the level of the amplitude selection rules, this is obvious. At the level of the action, the Legendre transformation  becomes the procedure of systematically integrating out the field strength. 

The reverse of the above statement would be that the off-shell U(1) symmetry in 3d is oxidized to a duality symmetry in 4d. This would be    a more surprising and interesting property. We are going to prove this converse statement for $F^n$-theories: \\
\vspace{1pt} \\
{\bf \em Theorem 2.} 
\textit{If the action $S_{(4d)}[F_{\mu\nu}]$ depends only on operators of the form $F^n$ and $S_{(3d)}[Z,\bar{Z}]$ has an off-shell $U(1)$ symmetry, then the 4d model must have a $U(1)$ electromagnetic duality symmetry.}\\
\vspace{1pt}\\
This converse result is at first sight quite surprising. When a model is dimensionally reduced we typically lose some information about the higher-dimensional physics. One might expect the existence of 4d operators that violate the duality symmetry, but vanish when reduced to 3d and are therefore not ruled out by the 3d symmetry. Indeed this does happen, as we show in Section \ref{s:hdops}, but this is delayed to higher orders in the derivative expansion and is absent at the leading orders in the effective field theory. In the following two subsections, we prove this oxidation statement. 

In the context of the D3-brane, this argument explains (at least from one point of view) the otherwise mysterious fact that the probe D3-brane preserves the continuous EM duality symmetry of free Maxwell theory. This symmetry is the dimensional oxidation of the linearly realized transverse isometries of the M2-brane. It is interesting to contrast this result with the well-known argument that $\mathcal{N}=4$ super Yang-Mills inherits the modular symmetry $SL(2,\mathds{Z})$ from the compactification to four-dimensions of the $(2,0)$ SCFT on the world-volume of an M5-brane stack \cite{Witten:2007ct}. That argument gives an M-theory based \textit{geometric} explanation of the discrete S-duality form of electromagnetic duality, but gives no indication that it should enhance to a continuous symmetry in the abelian limit. 

\subsection{3d \texorpdfstring{$\to$}{->} 4d Oxidation: Contact Terms}
\label{s:cts}

The models of non-linear electrodynamics we are considering have the form
\begin{equation}
    \label{NLED}
    S_{(4d)}[F_{\mu\nu}] = \int \text{d}^4x \;\mathcal{L}\left(X,Y\right), \hspace{10mm} X = F_{\mu\nu}F^{\mu\nu}, \hspace{10mm} Y = F_{\mu\nu}\tilde{F}^{\mu\nu}.
\end{equation}
In 4d this form of the Lagrangian is completely general at leading-order in the derivative expansion; using the Cayley-Hamilton theorem it is straightforward to show that general operators of the form $\text{Tr}[F^n]$ can be reduced to functions of the invariants $X$ and $Y$. Furthermore, we  assume that in the weak field limit, the Lagrangian admits a low-energy EFT expansion with the leading order term given by the standard Maxwell Lagrangian
\begin{equation}
    S_{(4d)}[F_{\mu\nu}] = \int \text{d}^4x \left[-\frac{1}{4}X + \frac{1}{\Lambda^4}\mathcal{L}^{(2)}(X,Y) +...+ \frac{1}{\Lambda^{4k-4}}\mathcal{L}^{(k)}(X,Y)+...\right],
\end{equation}
where $\mathcal{L}^{(k)}$ is a homogeneous polynomial of degree $k$. Simple dimensional analysis then gives that tree-level scattering amplitudes with $n$ external photons are homogeneous rational functions of spinor-helicity variables $\mathcal{A}_n\left(t|i\rangle,t|i]\right) = t^{2n}\mathcal{A}_n\left(|i\rangle,|i]\right)$. Alternatively, stripping off the overall factor of $\Lambda^{-2n+4}$ the remaining kinematic part of the amplitude must have mass dimension $n$. 

Consider the dimensional reduction of  $S_{(4d)}[F_{\mu\nu}]$ to $S_{(3d)}[Z,\bar{Z}]$, as in Theorem 2 of Section \ref{s:WV}. Suppose the 3d amplitudes conserve a global $U(1)$ charge \reef{M2amp}. We want to prove that the 4d amplitudes must then conserve the duality charge (\ref{EMamps}). The dictionary \reef{idZgamma} makes it clear that any $U(1)$-conserving amplitude in 3d lifts to a 4d duality-conserving amplitude. What remains to be shown is that there are no 4d duality-violating amplitudes that vanish upon restriction of the external momenta to a 3d subspace.

There can be no odd-point amplitudes in a theory with only $F^n$ interactions, so the lowest multiplicity we need to consider is 4-point. As the lowest-point amplitudes, they cannot have any kinematic singularities, hence they must be polynomial functions of spinor-helicity variables. The most general ansatze for these amplitudes consistent with locality, Bose symmetry, mass dimension and little group constraints are
	\begin{align} 
	\label{4ptdv}
		\mathcal{A}_4(1_\gamma^+\, 2_\gamma^+\, 3_\gamma^+\, 4_\gamma^+)&= \alpha_+ \left( \sq{12}^2\sq{34}^2+\sq{13}^2\sq{24}^2+\sq{14}^2\sq{23}^2\right),\nonumber\\
		\mathcal{A}_4(1_\gamma^-\,2_\gamma^-\,3_\gamma^-\,4_\gamma^-)&= \alpha_- \left( \langle 12\rangle^2\langle 34\rangle^2+\langle 13\rangle^2\langle 24\rangle^2+\langle 14\rangle^2\langle 23\rangle^2\right),\nonumber\\
		\mathcal{A}_4(1_\gamma^+\,2_\gamma^+\,3_\gamma^+\,4_\gamma^-) &= \mathcal{A}_4(1_\gamma^-\,2_\gamma^-\,3_\gamma^-\,4_\gamma^+) = 0\,.
	\end{align}
	Reducing the non-zero amplitudes to 3d gives
\begin{equation}
    \mathcal{A}_4(1_\gamma^\pm\,2_\gamma^\pm\,3_\gamma^\pm\,4_\gamma^\pm) \xrightarrow{3d} \alpha_\pm \left(s^2+t^2+u^2\right),
\end{equation}
which is non-vanishing unless $\alpha_\pm=0$. So we establish that for $n=4$, the existence of a $U(1)$ symmetry after dimensional reduction to 3d requires duality conservation in the 4d model \reef{NLED}.

At higher multiplicity, we examine the factorization properties of duality-violating 4d amplitudes. Consider an $n$-point amplitude where the number of positive helicity states $n_+$ does not equal the number of negative helicity states $n_-\ne n_+$. On a factorization pole, an amplitude splits into $\mathcal{A}_L$ and $\mathcal{A}_R$ such that
\be
  \label{nLnR}
  \begin{split}
		n^L_++n^R_+&=n_++1\,,\\
		n^L_-+n^R_-&=n_-+1\,.
	\end{split}
\ee
	It is clear that $\mathcal{A}_L$ and $\mathcal{A}_R$ cannot both be duality-preserving amplitudes since $n^L_++n^R_+\ne n^L_-+n^R_-$ by assumption. Thus, at least one of $\mathcal{A}_L$ and $\mathcal{A}_R$ must be duality-violating.
	
	Since the 4-point duality-violating amplitudes vanish, this then means that any 6-point duality-violating amplitude must be local, i.e.~it must arise from a local contact term of the form $F^6$. 	By dimensional analysis, little group scaling, and Bose symmetry, the only possible options are (focusing on the mostly-plus sectors)
	\be
	\begin{split}
	\label{6ptdv}
		\mathcal{A}_6(1_\gamma^+\,2_\gamma^+\,3_\gamma^+\,4_\gamma^+\,5_\gamma^+\,6_\gamma^+)&= \beta_+ \left( \sq{12}^2\sq{34}^2\sq{56}^2+ \text{perms}\right),\nonumber\\
		\mathcal{A}_6(1_\gamma^+\,2_\gamma^+\,3_\gamma^+\,4_\gamma^+\,5_\gamma^+\,6_\gamma^-)&= 0\,,\nonumber\\
		\mathcal{A}_6(1_\gamma^+\,2_\gamma^+\,3_\gamma^+\,4_\gamma^+\,5_\gamma^-\,6_\gamma^-)&= \beta_\text{MHV} \<56\>^2\left( \sq{12}^2\sq{34}^2+ \text{perms}\right).
	\end{split}
	\ee
In 3d kinematics, the non-zero matrix elements become sums of products of Mandelstam variables {\em squared} and as such there can be no cancellations; they must be non-vanishing. This means that there are no 4d local, or --- by the factorization argument above ---  non-local, duality-violating 6-point amplitudes that are not ruled out by the 3d $U(1)$ symmetry. 

Given that electromagnetic duality is preserved at 6-point order, the 8-point duality-violating amplitudes must be polynomial. The structure observed at 4-point and 6-point for the local contact terms continues at higher point,  so the 8-point local duality-violating amplitudes do not vanish in 3d and are therefore also ruled out by the 3d $U(1)$ symmetry, and so on. 

	 A derivation of Theorem 2 naturally lends itself to an inductive argument on the number of particles. In the next section we introduce a new on-shell recursion relation that allows us to prove at all multiplicities that the 3d $U(1)$ symmetry implies 4d electromagnetic duality in $F^n$-type theories.

\subsection{3d \texorpdfstring{$\to$}{->} 4d Oxidation: Subtracted Recursion Relations}	
\label{s:recrel4d3d}
	
	We construct an inductive proof of Theorem 2 in Section \ref{s:WV} based on a new version of subtracted recursion relations that accesses information about how 4d amplitudes behave when the external particles are restricted to a 3d subspace. Consider a set of $n$ on-shell external momenta subject to momentum conservation. Define the shift 
	\begin{align}
		\hat{p}_i^\mu= (1-z)p_i^\mu+ z q_i^\mu\,,
	\end{align}
	subject to the usual constraints that the shifted momenta $\hat{p}_i^\mu$ are on-shell, $\hat{p}_i^2 = 0$, and satisfy momentum conservation for any value of $z$, i.e.~for each $i$ we require
	\begin{align}
		\label{eq:momcons}
		p_i\cdot q_i= q_i^2=0~~~\text{ and }~~~ \sum_{i =1}^n q_i&=0\,.
	\end{align} 
	We choose the shift vectors $q_i$ to be normal to some unit space-like vector $N^\mu$,
	\begin{align}
	\label{eq:3dsub}
		q_i\cdot N =0\,. 
	\end{align}
	Then, at $z=1$, the shifted momenta are projected onto the 3d subspace normal to $N^\mu$: $\hat{p}_i^\mu N_\mu \big|_{z=1} = 0$. We implement this shift in terms of spinor-helicity brackets via the holomorphic shift
	\be
	\begin{array}{llllll}
	\label{eq:spinorshift}
		\langle\hat{i}|=&(1-z)\langle i| + z\, a_i [i|N\,, &~~& |\hat{i}]=|i]\,, &&\text{for }~~h_i\ge 0\,,\\
		\langle\hat{i}|=&\langle i|\,, &~~& |\hat{i}]=(1-z)|i] + z\, a_i N|j\rangle\,, &&\text{for }~~h_i<0\,,
	\end{array}
	\ee
	where $N=N^\mu\sigma_{\mu}$	and $a_i$ are constants constrained by momentum conservation,
	\begin{align}
	\label{eq:momconsai}
		\sum_{i:\,h_i \ge 0} a_i |i][i|N+\sum_{i:\,h_i<0} a_i N|i\rangle\langle i|\,=0.
	\end{align}
	The shift \reef{eq:spinorshift} ensures that each shifted momentum is on-shell and the whole set of momenta lie in a 3d subspace when $z=1$. The latter follows from ($h_i\ge 0$ case)
	\be
	   N\cdot \hat{p}_i\big|_{z=1} \propto N^\mu N^\nu \Tr(\bar\sigma_\mu |i][i| \sigma_\nu) =N^\mu N^\nu[i|\sigma_\nu\bar\sigma_\mu|i] = 
	   -N^2 [ii] = 0\,.
	\ee
	Naively, equations \eqref{eq:momcons} and \eqref{eq:3dsub} place $3n+4$ constraints on the $4n$ components of the $q_i^\mu$'s. However, the projection \reef{eq:3dsub} means that momentum conservation in the $N$-direction is automatic (i.e.~the $q_i$'s are 3d vectors), so the number of constraints is actually $3n+3$. This gives $n-3$ free parameters. Alternatively, this can be seen from the fact that \reef{eq:momconsai} places 3 constraints on the set of $n$  parameters $a_i$ in \eqref{eq:spinorshift}. Thus the system is under-constrained, and a shift can always be constructed for $n\ge 4$. In Appendix \ref{app:explicitshift} we present an explicit solution for the $a_i$.
	
	It was shown in \cite{Elvang:2018dco} that under a linear holomorphic shift, such as \reef{eq:spinorshift}, effective field theory amplitudes scale as the following power of $z$,
	\begin{align}
		4-n-\frac{n-2}{v-2}[g_v]-\sum_i s_i\,,
	\end{align}
	when $z$ is large. Here $v$ is the valence of the lowest-point interaction, $[g_v]$ is the mass-dimension of the coupling and $s_i$ are spins of the external states. For models of non-linear electrodynamics (\ref{NLED}), $v=4$, $[g_v]=-4$ and $s_i=1$, and so
	\begin{align}
		\mathcal{A}_n(z)\overset{z\to\infty}{\sim} z^0\,.
	\end{align}
	This means that in a standard recursion relation based on a contour integral of $\mathcal{A}_n(z)/z$, there is pole at infinity whose residue cannot be determined by factorization. This simply reflects the fact that the action (\ref{NLED}) contains an infinite number of independent, gauge-invariant, local operators of the form $F^n$ that only begin to contribute to on-shell scattering amplitudes at multiplicity $\geq n$. Since such contributions are completely independent of the lower-valence operators, any attempt to derive on-shell recursion relations must fail unless it incorporates additional physical information sufficient to pick out a unique model. In the present context, the additional constraint we impose is the vanishing of duality-violating amplitudes when external momenta are restricted to a 3d subspace. To do so, we employ {\em subtracted} recursion relations based on a contour integral of $\mathcal{A}_n(z)/z(1-z)$. The key is to avoid picking up a pole at $z=1$.
	
	 Consider an $n$-point duality-violating 4d amplitude $\mathcal{A}_n$. By assumption of the 3d $U(1)$ symmetry, any 4d duality-violating amplitude must vanish when its momenta are  restricted to a 3d subspace; hence the shifted amplitude satisfies
	\begin{align}
	\label{eq:vanish3d}
	\mathcal{A}_n(z=1)=0\,.
	\end{align}
	The unshifted amplitude can be retrieved via the contour integral
	\begin{align}
	\label{eq:unshifted}
		\mathcal{A}_n(0)&=\oint_{\mathcal{C}_0}\frac{\text{d}z}{2\pi i}\frac{\mathcal{A}_n(z)}{z(1-z)}\,,
	\end{align}
	where $\mathcal{C}_0$ is a small circular contour surrounding $z=0$. The extra factor of $(1-z)$ does not affect the residue at $z=0$, nor does it introduce a pole in the integrand, precisely due to the property \eqref{eq:vanish3d}.  

	Cauchy's theorem is then used to re-express the integral \reef{eq:unshifted} as a sum of residues on poles away from the origin. Locality ensures that the only poles in a tree amplitude correspond to factorization singularities. All poles occur at finite values of $z$ where an intermediate momentum $\hat{P}_I= (1-z)P_I+zQ_I$ goes on-shell,
    \be
	 \hat{P}_I^2=(1-z)^2P_I^2+z^2Q_I^2+2z(1-z)P_I\cdot Q_I= (P_I-Q_I)^2(z-z^I_+)(z-z^I_-)\,,
	\ee
	and the residue on these poles is the factorized amplitude,
	\begin{align}
		\underset{P_I^2=0}{\text{Res}}\mathcal{A}_n&=\mathcal{A}_L(z)\mathcal{A}_R(z)\,.
	\end{align}
	Thus the unshifted amplitude can be expressed as 
	\be
	\label{recAn}
		\mathcal{A}_n(0)=\sum_I \underset{z=z^I_\pm}{\text{Res}} \frac{\mathcal{A}_L(z)\mathcal{A}_R(z)}{z(1-z)P_I^2(z)}\,.
	\ee
	Let us now use \reef{recAn} to prove Theorem 2 from Section \ref{s:WV} by induction. By the factorization argument surrounding \reef{nLnR}, any duality-violating amplitude in 4d factorizes into sub-amplitudes $\mathcal{A}_L$ and $\mathcal{A}_R$ of which at least one must also be duality-violating. If we assume that all duality-violating amplitudes with fewer than $n$ external states are zero, then from the recursive formula \reef{recAn} it follows that $n$-point duality-violating amplitudes must likewise vanish. The recursion relations we derived are valid for $n>3$, and since there are no non-zero 3-point amplitudes for a self-interacting abelian gauge boson consistent with Lorentz invariance and Bose symmetry, this serves as the basis of the induction, and the result is proven.

\subsection{Higher Derivative Corrections} 
\label{s:hdops}
Theorem 2, proven in the previous subsection, demonstrates that the global $U(1)$ symmetry of a 3d complex scalar obtained by dimensional reduction must oxidize to a duality symmetry in 4d $F^n$-theories. This was restricted to leading-order in the derivative expansion. Physically this means that we can only expect the oxidation to hold in the deep infrared, where terms with the smallest number of derivatives dominate. A natural question concerns extending this theorem to operators of the form $\partial^{2k} F^n$ with $k\neq 0$.

	\begin{table}
		\begin{center}
			\begin{tabular}{|c|c|c|c|}
				\hline
				\backslashbox{Operator}{Helicity Sector} & SD & NSD & MHV\\
				\hline
				$F^4$ & 1 & 0 & 
				\cellcolor{gray}\textcolor{white}{1}\\ 
				\hline
				$\partial^2 F^4$ & 1 & 1 & \cellcolor{gray}\textcolor{white}{1}\\ 
				\hline
				$\partial^4 F^4$ & 1 & 0 & 
				\cellcolor{gray}\textcolor{white}{2}\\ 
				\hline
				$\partial^6 F^4$ & 1 & 1 & 
				\cellcolor{gray}\textcolor{white}{2}\\ 
				\hline
				$\partial^8 F^4$ & 2 & 1 & \cellcolor{gray}\textcolor{white}{3}\\ 
				\hline
				$\partial^{10} F^4$ & 1 & 1 & 
				\cellcolor{gray}\textcolor{white}{3}\\ 
				\hline
				$\partial^{12} F^4$ & 2 & 1 & 
				\cellcolor{gray}\textcolor{white}{4}\\ 
				\hline
				$F^5$ & 0 & 0 & 0\\
				\hline
				$\partial^2 F^5$ & 0 & 0 & 0\\ 
				\hline
				$\partial^4 F^5$ & 0 & 1 & 1\\ 
				\hline
				$\partial^6 F^5$ & 0 & 2 & 5\\
				\hline
				$\partial^8 F^5$ & 1 & \cellcolor{lightgray}13 (1) & \cellcolor{lightgray}11 (2)\\
				\hline
				$F^6$ & 1 & 0 & 1\\
				\hline
				$\partial^2 F^6$ & 1 & 2 & 2\\ 
				\hline
				$\partial^4 F^6$ & 3 & 4 & 12\\
				\hline
				$\partial^6 F^6$ & $-$ & \cellcolor{lightgray}15 (2) & \cellcolor{lightgray}30 (5)\\
				\hline
			\end{tabular}
		\caption{The table shows the number of linearly independent 4d matrix elements of the given operator in the SD (self-dual = all-plus), NSD (next-to-self-dual = one minus), and MHV sectors. For 4-point, the MHV amplitudes (dark gray) conserve duality, but for completeness we include the count.
		No linear combinations of possible matrix elements vanish in 3d, except for operators in helicity sectors corresponding to the (light gray) shaded cells of the $\partial^8 F^5$ and $\partial^6 F^6$ operators. In each shaded cell, the number in parenthesis is the number of linearly independent matrix elements that do vanish when restricted to 3d kinematics. The ``$-$" indicates that we have not studied the SD sector of the $\partial^6 F^6$ operator. \label{table1}}
		\end{center}
	\end{table}

The recursion relation approach developed in the previous subsection is no longer valid at higher-derivative order, so we proceed by analyzing the matrix elements of operators $\partial^{2k} F^n$ at low-multiplicity. 
For each operator $\partial^{2k} F^n$ and each helicity assignment, we construct all possible linearly independent matrix elements allowed by mass-dimension, little group scaling, and Bose symmetry; the count is shown in Table 
\ref{table1}. We then test if any linear combination of the (independent in 4d) matrix elements vanishes in 3d kinematics. This tests if they escape the constraints of the 3d $U(1)$ symmetry. Table \ref{table1} shows that at multiplicity 4, there are no matrix elements that vanish in 3d, up to and including 16 derivative order. However, at 5-point and 6-point we do find such duality-violating amplitudes that vanish in 3d. The light-gray-shaded cells are those for which such duality-violating matrix elements exist. For example in the MHV sector of $\partial^8 F^5$, there are a total of 11 independent matrix elements and a 2-parameter family of these vanish in 3d. 

We now discuss a more systematic way to construct duality-violating, yet 3d $U(1)$-compatible, matrix elements. The simplest construction of a 4d matrix element that vanishes in 3d kinematics involves a Levi-Civita tensor. For example, for $n\ge5$ consider the scalar matrix element 
\begin{align}
\label{eq:WZW}
    \epsilon\left(1,2,3,4\right)=\epsilon_{\mu\nu\rho\sigma}p_1^\mu p_2^\nu p_3^\sigma p_4^\rho\,.
\end{align}
For $n=5$, this is the matrix element of the Wess-Zumino-Witten (WZW) term and by momentum conservation it is fully antisymmetric in all five momenta.

Suppose we construct a polynomial in the spinor-helicity brackets that has little-group scaling corresponding to five external photons with helicities $+++--$ and is {\em antisymmetric} in any exchanges of identical particles. The lowest dimension polynomial with these properties has mass-dimension 9 and there are in fact two such independent polynomials. 
Upon multiplication of these two polynomials with \reef{eq:WZW}, we obtain two spinor helicity polynomials of mass-dimension 13, with little-group scaling of photons in the MHV sector, and Bose symmetry in identical particles. They are therefore matrix elements of operators of the schematic form $\partial^8 F^5$ involving contractions with a single Levi-Civita tensor. The Levi-Civita ensures that the matrix elements vanish in 3d. These 2 matrix elements are the ones listed in the gray-shaded box for $\partial^8 F^5$ in the MHV sector and we explicitly match them to the ones found by the general analysis of linear combinations of 11 independent MHV matrix elements possible for any $\partial^8 F^5$ operator (see Table \ref{table1}). Similarly for the NSD (one minus) sector of $\partial^8 F^5$: the matrix element that vanishes in 3d is exactly the product of the WZW polynomial \reef{eq:WZW} and the unique spinor bracket polynomial of mass-dimension 9 with little group scaling of photons with helicity $++++-$, and full antisymmetry in identical states.

Since duality-violating amplitudes associated with operators of the form $\partial^m F^4$ are excluded for $m\le 12$, any 6-point duality-violating  amplitude must be polynomial in the spinor brackets and correspond to matrix elements of an operator of the form $\partial^{2k} F^6$. MHV matrix elements that vanish in 3d can then be constructed by multiplying $\epsilon\left(1,2,3,4\right)$ with a polynomial with MHV ($++++--$) little group scaling that is antisymmetric in $\{1, 2, 3, 4\}$ and symmetric in $\{5,6\}$. The lowest dimension of such a polynomial is 8 and it is unique. The result is a matrix element of an operator of the form $\partial^6 F^6$. To construct the NSD matrix element that vanishes in 3d, take 
\begin{align}
    \epsilon\left(1,2,3,4,5\right) \equiv 
    \epsilon\left(1,2,3,4\right)-\epsilon\left(1,2,3,5\right)+\epsilon\left(1,2,4,5\right)-\epsilon\left(1,3,4,5\right)+\epsilon\left(2,3,4,5\right)\,,
\end{align}
that is antisymmetric in labels $\{1,\cdots,5\}$ and multiply it with the unique dimension-8 NSD ($+++++-$) polynomial antisymmetric in identical helicity states. This construction is included in the $\partial^6 F^6$ matrix elements that we find to vanish in 3d in Table~\ref{table1}, although not all matrix elements reported there can be obtained using this method.

Another construction at 5-point is to multiply 5-point photon matrix elements (with proper Bose symmetry) with the quintic Galileon term,
\begin{align}
    \epsilon\left(1,2,3,4\right)^2\,.
\end{align}
By the results in Table \ref{table1}, the lowest dimension 5-point matrix element arises from $\partial^4 F^5$. When multiplied by the matrix element of the 8-derivative quintic Galileon term, we get a matrix element of $\partial^{12} F^5$ that is guaranteed to vanish in 3d. 

The point here is the existence of 4d duality-violating matrix elements that, since they vanish in 3d, are not excluded by the 3d $U(1)$ symmetry. We note that at low multiplicities, the duality violation is delayed until quite high order; the lowest order example given here is 12-derivative at 6-point (compared with the 6-derivative leading BI term $F^6$). 

In the context of string theory, the Born-Infeld action (\ref{D3}) describes the leading-order in $\alpha'$-contribution to the dynamics of the D3-brane. Matching to the world-sheet calculation of the open-string S-matrix, there is an infinite set of sub-leading corrections (the first ones calculated originally in \cite{Andreev:1988cb}). By dimensional analysis, the sub-leading $\alpha'$-contributions to a given amplitude correspond to derivative corrections to the effective action of the form $\partial^{2k}F^n$. 
A reasonable question to ask is whether, when dimensionally reduced to 3d, these sub-leading corrections are consistent with the M2-brane picture and preserve the hidden $U(1)$ symmetry. This is the question we have examined here in the generic context of higher derivative corrections. Whether the particular duality-violating operators we present here are in fact produced in string theory (or if they are compatible with supersymmetry) is a question beyond the scope of this paper. 

%%%%%%%%%%%%%%%%%%%%%%%%%%%%%%
\section{Loop Amplitudes from BCJ Double-Copy}
\label{sec:DCloop}
%%%%%%%%%%%%%%%%%%%%%%%%%%%%%%

The double-copy of Yang-Mills and chiral perturbation theory ($\chi$PT) was shown to result in Born-Infeld theory at tree-level \cite{Cachazo:2014xea}, using the CHY formalism \cite{Cachazo:2013hca,Cachazo:2013gna,Cachazo:2013iea}. BI tree amplitudes can be constructed either via the KLT relations or the BCJ double-copy.  The former is used in the context of higher-derivative corrections in Section \ref{sec:DChd}. In this section, we begin with the BCJ construction of the 4-point tree amplitude and then use loop-level BCJ to construct the integrand for the self-dual 1-loop 4-point amplitude.

%%%%%%%%%%%%%%%%%%%%%%%%%%%%%%
\subsection{Tree-level BCJ Double-Copy of BI}
\label{sec:BItreeBCJ}
%%%%%%%%%%%%%%%%%%%%%%%%%%%%%%
At 4-point, the BCJ form of a tree amplitude can be written as
\begin{equation}
\label{4ptBCJform}
  \mathcal{A}_4\left(1\,2\,3\,4\right) = \frac{c_{1342}\ n_{1342}}{t} + \frac{c_{1423}\ n_{1423}}{u},
\end{equation}
where we use the convention for the Mandelstam invariants 
\begin{equation}
    s=(p_1+p_2)^2, \hspace{5mm} t=(p_1+p_3)^2, \hspace{5mm} u=(p_1+p_4)^2,
\end{equation}
and the \textit{BCJ color tensors} are defined as
\begin{equation}
  c_{ijkl} = f_{a_i a_j b}f_{a_k a_l b}\,.
\end{equation}
The Jacobi identity $c_{1234} + c_{1423} + c_{1342} = 0$ has been used here to eliminate the term with color factor $c_{1234}$. 
The numerator factors $n_{ijkl}$ can be written in terms of color-ordered amplitudes as
\be
  \label{4ptnums}
  n_{1342} = -t\,\mathcal{A}_4[1\,3\,4\,2] 
  ~~~~~\text{and}~~~~~
  n_{1423} = u\,\mathcal{A}_4[1\,2\,3\,4]\,.
\ee
For our specific application, we use the ordered tree-level amplitudes for Yang-Mills and $\chi$PT
\begin{equation}
  \mathcal{A}_4^{\text{YM}}[1^+\,2^+\,3^-\,4^-] = - g_{\text{YM}}^2\frac{[12]^2\langle 34\rangle^2}{su}, \hspace{10mm} \mathcal{A}_4^{\chi\text{PT}}[1\,2\,3\,4] = \frac{1}{f^2_\pi}t.
\end{equation}
Using \reef{4ptnums}, the BCJ numerators take the form
\begin{align}
  &n_{1342}^{\text{YM}}=g_{\text{YM}}^2\frac{[12]^2\langle 34\rangle^2}{s}, \hspace{11mm} n_{1423}^{\text{YM}} =-g_{\text{YM}}^2\frac{[12]^2\langle 34\rangle^2}{s},\nonumber\\
  &n_{1342}^{\chi\text{PT}}=-\frac{1}{f^2_\pi}tu, \hspace{23.5mm} n_{1423}^{\chi\text{PT}} =\frac{1}{f^2_\pi}tu.
\end{align}
Together with $n_{1234}\equiv0$, these sets of BCJ numerators satisfy the \textit{kinematic Jacobi identity} $n_{1234}+n_{1342}+n_{1423}=0$. The BCJ double-copy, at tree-level and at 4-point takes the form
\begin{equation}
    \mathcal{A}_4^{A\otimes B}(1,2,3,4) = \frac{1}{\lambda^2} \left[\frac{n^A_{1234}\,n^B_{1234}}{s}+\frac{n^A_{1342}\,n^B_{1342}}{t}+\frac{n^A_{1423}\,n^B_{1423}}{u}\right],
\end{equation}
where $\lambda$ is some arbitrary constant with $[\lambda]=1$.\footnote{From the equivalence between the tree-level BCJ double-copy and the KLT formula \cite{Bern:2008qj}, together with the interpretation of the KLT kernel as the inverse of a matrix of bi-adjoint scalar amplitudes \cite{Cachazo:2013iea} we find that $\lambda$ has a physical interpretation as the $\phi^3$ coupling constant.}
The tree-amplitudes of BI are given as the double-copy
\begin{align}
  \mathcal{A}^{\text{BI}}_4 & \left(1^+\,2^+\,3^-\,4^-\right) \nonumber \\
  &= \frac{1}{\lambda^2} \left[\frac{n_{1342}^{\text{YM}}\,n_{1342}^{\chi\text{PT}}}{t}+\frac{n_{1423}^{\text{YM}}\,n_{1423}^{\chi\text{PT}}}{u}\right] \nonumber\\
                                                        &= \frac{1}{\lambda^2} \left[\frac{1}{t}\left(g_{\text{YM}}^2\frac{[12]^2\langle 34\rangle^2}{s}\right)\left(-\frac{1}{f_\pi^2}tu\right)+\frac{1}{u}\left(-g_{\text{YM}}^2\frac{[12]^2\langle 34\rangle^2}{s}\right)\left(\frac{1}{f_\pi^2}tu\right)\right] \nonumber\\
                                                        &= \frac{g_{\text{YM}}^2}{\lambda^2 f^2_{\pi}}[12]^2\langle 34\rangle^2.
\end{align}
Comparing this to the well-known result for the 4-point tree amplitude in Born-Infeld
\begin{equation}
  \mathcal{A}^{\text{BI}}_4\left(1^+\,2^+\,3^-\,4^-\right) = \frac{1}{\Lambda^4}[12]^2\langle 34 \rangle^2,
\end{equation}
we see that the BCJ double-copy indeed gives the correct result when the couplings are related as 
\begin{equation}
\label{gfLambda}
  \frac{g_{\text{YM}}^2}{\lambda^2 f_\pi^2} = \frac{1}{\Lambda^4}\sim \frac{1}{T_{D3}}\,.
\end{equation}
We use this in the following loop-calculations.

%%%%%%%%%%%%%%%%%%%%%%%%%%%%%%
\subsection{1-loop 4-point Self-Dual Amplitude as a Double-Copy}
\label{sec:DC4ptSD}
%%%%%%%%%%%%%%%%%%%%%%%%%%%%%%
For loop amplitudes, the BCJ construction is conjectured to be valid at the level of the integrand \cite{Bern:2010ue}. In this section we use it to construct the 1-loop 4-point BI integrand of the self-dual sector. 

The strategy is as follows: we construct the $\chi$PT 4-point 1-loop integrand directly using unitarity to ensure that it satisfies all cuts. In that expression we then replace the 1-loop color-factors by the color-kinematic duality-obeying numerator factors for the self-dual 4-point 1-loop YM amplitude obtained previously in \cite{Faller:2018vdz} in the FDH scheme \cite{Bern:2002zk}. This conjecturally gives the self-dual 1-loop BI integrand. We integrate the expression to show that the integrated result agrees exactly with \reef{genBIL} obtained by different techniques in \cite{Elvang:2019twd} and Appendix \ref{app:1loopcalc}.

\vspace{3mm}
\paragraph{YM Self-Dual 1-loop 4-point Integrand.}
The calculation is simplified significantly by the fact that there exists a representation of the self-dual 4-point 1-loop YM numerators where only the box-numerators are non-vanishing \cite{Faller:2018vdz}; specifically these box numerators take the form
\begin{equation} \label{ymnum}
  n^{(\text{box})}_{1234}(l) = 2g_{\text{YM}}^4\frac{[12][34]}{\langle 12 \rangle \langle 34 \rangle} \big(\mu^2\big)^2,
\end{equation}
where $g_{\text{YM}}$ is the Yang-Mills coupling and $\mu^2 = l^2_{-2\epsilon}$ is the square of the momentum in the $-2\eps$ extra dimensions, $l^\mu = l_{[4]}^\mu + l_{-2\epsilon}^\mu$. 
The external state momenta $p_i$, $i=1,2,3,4$, are assumed to be on-shell and strictly 4d. The self-dual 1-loop amplitude of YM then  takes the form
\be\label{BCJ4ptYM}
  \mathcal{A}_4^{\text{YM}}\left(1_g^+\,2_g^+\,3_g^+\,4_g^+\right) 
  = \!\! \int \!\!\frac{\text{d}^{4-2\epsilon}l}{(2\pi)^{4-2\epsilon}}
  \bigg[\frac{c^{(\text{box})}_{1234}n^{(\text{box})}_{1234}(l)}{l^2(l-p_2)^2(l-p_2-p_3)^2(l+p_1)^2} + (2 \lra 3) + (3\lra 4)\bigg].
\ee
The expression \reef{ymnum} is symmetric under permutation of the external states, so the box coefficients of all three independent box diagrams of the 1-loop 4-point amplitude are the same: 
$n^{(\text{box})}_{1324} = n^{(\text{box})}_{1243} = n^{(\text{box})}_{1234}$. The color factors\footnote{We work in conventions where 
$[t^a,t^b]=if^{abc}t^c$ and $\text{Tr}\left[t^a t^b\right] = \delta^{ab}$.} are 
\be
\label{cboxdef}
c^{(\text{box})}_{1234} = f_{a_1b_1b_4}\,f_{a_2b_2b_1}\,f_{a_3b_3b_2}\,f_{a_4b_4b_3}\,,
\ee
along with the permutations $(2 \lra 3)$ and $(3\lra 4)$ (which on the RHS acts on the $a$-subscripts only). We check the validity of the representation \reef{BCJ4ptYM} by computing the maximal cuts, e.g.~ 
\be
\raisebox{-1.35cm}{{\begin{tikzpicture}[scale=0.5, line width=1 pt]
    \draw [vector] (-1,1)--(0,0);
    \draw [vector] (-1,-3)--(0,-2);
    \draw [vector] (2,0)--(3,1);
    \draw [vector] (2,-2)--(3,-3);
    \draw [vector] (0,0)--(0,-2);
    \draw [vector] (0,-2)--(2,-2);
    \draw [vector] (2,-2)--(2,0);
    \draw [vector] (0,0)--(2,0);
    \draw [scalarnoarrow, color=red] (1,0.5)--(1,-0.5);
    \draw [scalarnoarrow, color=red] (-0.5,-1)--(0.5,-1);
    \draw [scalarnoarrow, color=red] (1,-1.5)--(1,-2.5);
    \draw [scalarnoarrow, color=red] (1.5,-1)--(2.5,-1);
    \node at (-1.5,1.2) {\small $1_g^+$};
    \node at (-1.5,-3.2) {\small$2_g^+$};
    \node at (3.5,1.2) {\small$4_g^+$};
    \node at (3.5,-3.2) {\small$3_g^+$};
    \draw [->] (-0.75,-0.5)--(-0.75,-1.5);
    \draw [->] (2.75,-1.5)--(2.75,-0.5);
    \draw [->] (0.5,-2.75)--(1.5,-2.75);
    \draw [<-] (0.5,0.75)--(1.5,0.75);
    \node at (-1.2,-1) {$l_1$};
    \node at (1.2,-3.3) {$l_2$};
    \node at (3.3,-1) {$l_3$};
    \node at (1,1.3) {$l_4$};
\end{tikzpicture}}}
~=~
\raisebox{-1.35cm}{{\begin{tikzpicture}[scale=0.5, line width=1 pt]
    \draw [vector] (-1,1)--(0,0);
    \draw [vector] (-1,-3)--(0,-2);
    \draw [vector] (2,0)--(3,1);
    \draw [vector] (2,-2)--(3,-3);
    \draw (0,0)--(0,-2);
    \draw (0,-2)--(2,-2);
    \draw (2,-2)--(2,0);
    \draw (0,0)--(2,0);
    \draw [scalarnoarrow, color=red] (1,0.5)--(1,-0.5);
    \draw [scalarnoarrow, color=red] (-0.5,-1)--(0.5,-1);
    \draw [scalarnoarrow, color=red] (1,-1.5)--(1,-2.5);
    \draw [scalarnoarrow, color=red] (1.5,-1)--(2.5,-1);
    \node at (-1.5,1.2) {\small $1_g^+$};
    \node at (-1.5,-3.2) {\small$2_g^+$};
    \node at (3.5,1.2) {\small$4_g^+$};
    \node at (3.5,-3.2) {\small$3_g^+$};
    \draw [->] (-0.75,-0.5)--(-0.75,-1.5);
    \draw [->] (2.75,-1.5)--(2.75,-0.5);
    \draw [->] (0.5,-2.75)--(1.5,-2.75);
    \draw [<-] (0.5,0.75)--(1.5,0.75);
    \node at (-1.2,-1) {$l_1$};
    \node at (1.2,-3.3) {$l_2$};
    \node at (3.3,-1) {$l_3$};
    \node at (1,1.3) {$l_4$};
\end{tikzpicture}}}
\ee
Here we are using the supersymmetric trick \reef{1loopSYSY} to replace the gluon in the loop with a complex scalar with 4d mass $\mu^2 =l_{-2\epsilon}^2$. On the cut, the integrand gives
\be
\begin{split} \label{cutpred}
  \text{Cut}_{1234}
  \left[\mathcal{I}_4^{\text{YM}}\left(1_g^+\,2_g^+\,3_g^+\,4_g^+\right)\right] 
  =&
2\,\mathcal{A}^{\text{tree}}_3\left(1_g^+\,(l_1)_\phi\,(-l_4)_{\overline{\phi}}\right)
\mathcal{A}^{\text{tree}}_3\left(2_g^+\,(l_2)_\phi\,(-l_1)_{\overline{\phi}}\right)
\\
&\times
\mathcal{A}^{\text{tree}}_3\left(3_g^+\,(l_3)_\phi\,(-l_2)_{\overline{\phi}}\right)
\mathcal{A}^{\text{tree}}_3\left(4_g^+\,(l_4)_\phi\,(-l_3)_{\overline{\phi}}\right) ,
\end{split}
\ee
where the factor of 2 accounts for the fact that the complex scalar can run in both directions. The scalar-scalar-gluon amplitude is 
\begin{equation}
\label{A3massivephi}
  \mathcal{A}_3^{\text{tree}}\left(1_g^+\,\ell_\phi\,\ell'_{\overline{\phi}}\right) = -i g_{\text{YM}}f_{a_1a_\ell a_{\ell'}}\frac{[1|\ell|q\rangle}{\langle 1 q\rangle},
\end{equation}
where $|q\rangle$ is an arbitrary spinor. The normalization in \reef{A3massivephi} is fixed by the 3-gluon amplitude via the SUSY Ward identities in the massless limit $\ell^2 ={\ell'}^2 \to 0$.

It is convenient to choose
\begin{equation}
  |q_1\rangle = |2\rangle,\;\; |q_2\rangle = |1\rangle, \;\; |q_3\rangle = |4\rangle, \;\; |q_4\rangle = |3\rangle,
\end{equation}
in which case the cut \reef{cutpred} gives
\begin{align}
\label{YMcut}
  &\text{Cut}_{1234}\left[\mathcal{I}_4^{\text{YM}}\left(1_g^+\,2_g^+\,3_g^+\,4_g^+\right)\right]\nonumber\\
  &= c_{1234}^{(\text{box})}2\,g_{\text{YM}}^4 \frac{[1|l_1 p_2 l_1|1\rangle[3 l_3 p_4 l_3|3\rangle}{\langle 12\rangle^2 \langle 34\rangle^2}\nonumber\\
  &= c^{(\text{box})}_{1234} \,2\,g_{\text{YM}}^4(\mu^2)^2 \frac{[12][34]}{\langle 12 \rangle \langle 34 \rangle},
\end{align}
where in the second step we use special 3-particle kinematics to show that 
$[1|l_1 p_2 l_1|1\rangle \times [3| l_3 p_4 l_3|3\rangle = -\langle 12 \rangle[12]\mu^2$ and similarly for the other angle-square brackets. Thus we see from \eqref{YMcut} that the product of tree-amplitudes \reef{cutpred} indeed reproduces the YM numerator factor \reef{ymnum} of \cite{Faller:2018vdz}.

\vspace{3mm}
\paragraph{$\chi$PT 1-loop 4-point Integrand.}
Next, we compute the $\chi$PT 1-loop 4-point integrand. We then replace its color factors $c^{(\text{box})}$ by the YM numerators to construct the 1-loop BI self-dual amplitude at 4-point order. The color-dressed $\chi$PT tree amplitudes can be written 
\begin{equation} \label{amp4}
  \mathcal{A}_4^{\text{tree}}\big(1_{a_1}\,2_{a_2}\,3_{a_3}\,4_{a_4}\big) =  
  \frac{2}{f_\pi^2}\Big[  f_{a_1 a_4 b}f_{a_2 a_3 b}\,(p_1\cdot p_3)
  +f_{a_1 a_3 b}f_{a_2 a_4 b}\,(p_1\cdot p_4) \Big].
\end{equation}
The $s$-channel cut of the 1-loop amplitude 
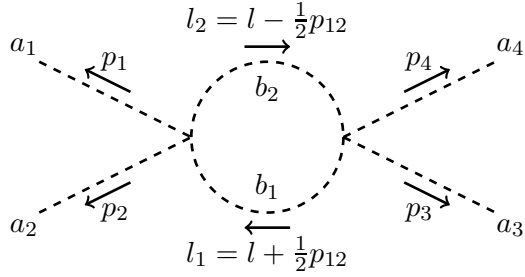
\begin{center}
{\begin{tikzpicture}[scale=1, line width=1 pt]
\draw [scalarnoarrow] (-2,1)--(0,0);
\draw [scalarnoarrow] (-2,-1)--(0,0);
\draw [scalarnoarrow] (0,0) arc (180:0:1);
\draw [scalarnoarrow] (2,0) arc (0:-180:1);
\draw [scalarnoarrow] (2,0)--(4,1);
\draw [scalarnoarrow] (2,0)--(4,-1);
\draw [->] (0.7,1.2)--(1.3,1.2);
\draw [<-] (0.7,-1.2)--(1.3,-1.2);
\draw [->] (-0.8,0.6)--(-1.4,0.9);
\draw [->] (-0.8,-0.6)--(-1.4,-0.9);
\draw [->] (2.8,0.6)--(3.4,0.9);
\draw [->] (2.8,-0.6)--(3.4,-0.9);
\node at (-1,1) {$p_1$};
\node at (-1,-1) {$p_2$};
\node at (3,1) {$p_4$};
\node at (3,-1) {$p_3$};
\node at (1,1.55) {$l_2 = l-\frac{1}{2}p_{12}$};
\node at (1,0.65) {$b_2$};
\node at (1,-1.55) {$l_1 =l+\frac{1}{2}p_{12}$};
\node at (1,-0.65) {$b_1$};
\node at (-2.2,1.2) {$a_1$};
\node at (-2.2,-1.2) {$a_2$};
\node at (4.2,1.2) {$a_4$};
\node at (4.2,-1.2) {$a_3$};
\end{tikzpicture}}
\end{center}
gives 
\begin{align}
  &\mathcal{A}_4^{\text{tree}}\Big(1_{a_1}\,2_{a_2}\,(-l_1)_{b_1}\,(l_2)_{b_2}\Big)
  \mathcal{A}_4^{\text{tree}}\left(3_{a_3}\,4_{a_4}\,(-l_2)_{b_2}\,(l_1)_{b_1}\right) 
  \nonumber\\[2mm]
  &\hspace{5mm}= \frac{4}{f_\pi^4}
  \Big[f_{a_1 b_2 b_3}f_{a_2 b_1 b_3}
     \big(-p_1\cdot l_1\big)
  +f_{a_1 b_1 b_3}f_{a_2 b_2 b_3} 
     \big(p_1\cdot l_2\big)\Big]
  \nonumber\\
  &\hspace{15mm}\times
  \Big[f_{b_2 a_4 b_4}f_{b_1 a_3 b_4}
      \big(-p_3\cdot l_2 \big)
  +f_{b_2 a_3 b_4}f_{b_1 a_4 b_4}
      \big(-p_4\cdot l_2\big)\Big] 
      \nonumber\\[2mm]
      &\hspace{5mm}=
        \frac{4}{f_\pi^4}c^{(\text{box})}_{1234}\Big[-(p_1\cdot l)(p_{12}\cdot l) 
  +\tfrac{1}{8}s^2\Big]
  +\frac{4}{f_\pi^4}c^{(\text{box})}_{1243}\Big[(p_1\cdot l)(p_{12}\cdot l) 
  + \tfrac{1}{8}s^2\Big]\,.
\end{align}      

To obtain the last line we expanded out the product, identified the box color-factors \reef{cboxdef} in terms of the structure constants, and then performed some simplifications. Terms linear in the loop-momentum $l$ were dropped since they integrate to zero. 

Including the propagators, this gives the $s$-channel contribution to the integrand. The $t$- and $u$-channels can be obtained by simple permutations of the external momentum labels, so the full integrand is

\be
  \label{chiPTinte12}
  \begin{split}
  \mathcal{I}_4^{\chi \text{PT}}\left(1_{a_1}\,2_{a_2}\,3_{a_3}\,4_{a_4};l,\mu^2\right)
 &= \Bigg(c^{(\text{box})}_{1234} \frac{2}{f_\pi^4} \frac{-(p_1\cdot l)(p_{12}\cdot l) +\frac{1}{8}s^2}{\left(l-\frac{1}{2}p_{12}\right)^2\left(l+\frac{1}{2}p_{12}\right)^2}
 \\& ~~~~~
 +c^{(\text{box})}_{1243} \frac{2}{f_\pi^4}\frac{(p_1\cdot l)(p_{12}\cdot l) + \frac{1}{8}s^2}{\left(l-\frac{1}{2}p_{12}\right)^2\left(l+\frac{1}{2}p_{12}\right)^2}\Bigg)
 +(2 \lra 3)+(2 \lra 4)
\,.
\end{split}
\ee
We could proceed to extract the box numerator factors for $\chi$PT, but it is a lot simpler to directly use the form \reef{chiPTinte12} in the double-copy. 

\vspace{3mm}
\paragraph{BI Self-Dual 1-loop 4-point Integrand and Amplitude.}
We obtain the self-dual loop-integrand for BI theory via the BCJ double-copy by replacing the box-color factors in \reef{chiPTinte12} by the YM numerators \reef{ymnum}. Since these are symmetric in the external labels, we have
\be
  c^{(\text{box})}_{ijkl} \to \frac{1}{\lambda^4} n^{(\text{box})}_{1234}(l) = \frac{2g_{\text{YM}}^4}{\lambda^4}\frac{[12][34]}{\langle 12 \rangle \langle 34 \rangle} \big(\mu^2\big)^2
  = \frac{2g_{\text{YM}}^4}{\lambda^4}\frac{[12]^2[34]^2}{s^2} (\mu^2)^2\,.
\ee 
This results in significant simplifications, in particular all dependence on the 4d part of the loop-momentum in the numerator cancels. We are left with 
\be
  \label{BIallplusBCJ1}
  \begin{split}
   \mathcal{I}_4^{\text{BI}}\left(1_{a_1}\,2_{a_2}\,3_{a_3}\,4_{a_4};l,\mu^2\right)
   &=
   \mathcal{I}_4^{\chi \text{PT}}\left(1_{a_1}\,2_{a_2}\,3_{a_3}\,4_{a_4};l,\mu^2\right)\Big|_{c^{(\text{box})} \to \lambda^{-4}n^\text{(box)\,YM}} 
   \\
   &= \frac{1}{\Lambda^8}[12]^2 [34]^2
    \frac{\big(\mu^2\big)^2}{\left(l-\frac{1}{2}p_{12}\right)^2\left(l+\frac{1}{2}p_{12}\right)^2}+(2 \lra 3)+(2 \lra 4)
\,,
 \end{split}
\ee 
where we used the identification \reef{gfLambda}. The loop integral can be evaluated using the dimension-shifting technique of \cite{Bern:1996ja}. The details of this are given in Appendix~\ref{app:dimregintegrals}. Using equation \eqref{eq:bubble_int} with $p = 2$, we find
\begin{equation}
    \int \frac{d^{4 - 2 \epsilon} l}{( 4 \pi )^{4 - 2\epsilon}} \frac{( \mu^2 )^2}{( l - \frac 1 2 p_{12} )^2 ( l + \frac 1 2 p_{12} )^2} = \frac{i}{( 4 \pi )^{2 - \epsilon}} \epsilon ( 1 - \epsilon ) \frac{\Gamma ( - 2 + \epsilon ) \Gamma^2 ( 3 - \epsilon )}{\Gamma ( 6 - 2 \epsilon )} s^{2 - \epsilon} = \frac{- i s^2}{960 \pi^2}\,,
\end{equation}
which finally gives
\begin{equation}
\label{SD4pt}
\boxed{  \mathcal{A}_4^{\text{1-loop BI}}\left(1_\gamma^+\,2_\gamma^+\,3_\gamma^+\,4_\gamma^+\right) = \frac{-i}{960\pi^2}\Big([12]^2[34]^2s^2+[13]^2[24]^2t^2+[14]^2[23]^2u^2\Big)
   + \mathcal{O}(\eps).}
\end{equation}
In this and subsequent formulae we suppress the $\Lambda$-dependent prefactor, this can be easily restored by dimensional analysis. The result \reef{SD4pt} agrees exactly with the result obtained by generalized unitarity in \cite{Elvang:2019twd} and offers an interesting different path to exploring BI at loop-level.

%%%%%%%%%%%%%%%%%%%%%%%%%%%%%%%
\section{1-loop SD and NSD Amplitudes from Unitarity}
\label{sec:BIall1loop}
%%%%%%%%%%%%%%%%%%%%%%%%%%%%%%%

We briefly review results from \cite{Elvang:2019twd} for the all-multiplicity self-dual and next-to-self-dual 1-loop amplitudes of pure Born-Infeld theory. 

\subsection{4-point and Counterterm}

The self-dual amplitude \reef{SD4pt} was derived in \cite{Elvang:2019twd} and can also be computed directly from Feynman rules (see Appendix \ref{app:1loopcalc}). The next-to-self-dual 4-point amplitude vanishes \cite{Elvang:2019twd}
 \begin{equation} \label{genBIL}
 \boxed{
 \mathcal{A}_4^\text{1-loop BI}\big(1_\gamma^+\, 2_\gamma^+\, 3_\gamma^+\, 4_\gamma^-\big) 
 =
 0 + \mathcal{O}(\epsilon)\,.
}
\end{equation}
The vanishing of 
$\mathcal{A}_4^{\text{1-loop BI}}\big(1_\gamma^+\,2_\gamma^+\,3_\gamma^+\,4_\gamma^-\big)$ is easy to understand: the amplitude has to be local, but little group scaling, Bose symmetry, and dimensional analysis show that there is no such possible local matrix element.

The non-vanishing of $\mathcal{A}_4^{\text{1-loop BI}}\big(1_\gamma^+\,2_\gamma^+\,3_\gamma^+\,4_\gamma^+\big)$, given in \reef{SD4pt}, indicates a potential violation of EM duality at loop-order. However, there is a local operator $\pa^4 F^4$ that generates this matrix element. We can construct this operator explicitly by using the spinorized form of the field strengths to make direct contact with the matrix elements in spinor helicity formalism. Define 
\be
  \label{spinorF}
  F_+ 
  = \frac 1 2 (F^{\mu\nu}  + i \tilde{F}^{\mu\nu})\sigma_{\mu\nu}, \hspace{10mm}  F_- 
  = \frac 1 2 (F^{\mu\nu}  - i \tilde{F}^{\mu\nu})\overline{\sigma}_{\mu\nu},
\ee
with 
$\sigma^{\mu\nu} = \tfrac{1}{4} (\sigma^{\mu} \bar{\sigma}^\nu-\sigma^{\nu} \bar{\sigma}^\mu)$, $\overline{\sigma}^{\mu\nu} = \tfrac{1}{4} (\bar{\sigma}^{\mu} \sigma^\nu-\bar{\sigma}^{\nu} \sigma^\mu)$
and $\tilde{F}^{\mu\nu} = -\tfrac{1}{2} \eps^{\mu\nu\rho\sigma} F_{\rho\sigma}$. 
Then the Feynman rule for a positive/negative helicity external photon with momentum $p$ is simply\footnote{The reference spinors of the polarizations drop out because $F_{\mu\nu}$ is gauge invariant.}

\be  
  \label{FpFm}
    F_+  ~~\lra~~ \sqrt{2} |p][p|
    ~~~~~~\text{and}~~~~~~
    F_-  ~~\lra~~ \sqrt{2} |p\>\<p| \,.
\ee
The $\pa^4 F^4$ operator that cancels the loop-contribution \reef{SD4pt} is 
 \begin{equation}
 \label{4ptCT}
   S_{\text{BI}} \rightarrow S_{\text{BI}} + \frac{1}{7680\pi^2\Lambda^8}\int \text{d}^4 x \left[(\partial_\mu F_{+\alpha\beta})(\partial^\mu F_+^{\alpha\beta})\right]^2 +\text{h.c.}
 \end{equation}
This then restores the duality symmetry at 4-point 1-loop order. 

\subsection{\texorpdfstring{$n$-point}{n-point}}

In a recent paper \cite{Elvang:2019twd}, we computed  1-loop amplitudes in the self-dual and next-to-self-dual sectors for any number of external states, using a combination of powerful modern methods.  In the  self-dual sector of pure BI, the 1-loop integrand is
\bea
\label{SDintegrand}
 &&\hspace{-5mm}\mathcal{I}_{2n}^{\text{BI}}\big(1_\gamma^+2_\gamma^+\cdots 2n_\gamma^+;l,\mu^2\big)
 \\[1mm]
\nonumber
&=& \left(\frac 1 2 \right)^{n - 1} [ 12 ]^2 [ 34 ]^2 \ldots [2n - 1, 2n ]^2 \frac{\left ( - \mu^2 \right )^{n}}{\prod_{i = 1}^{n} \left [ \left ( l - \sum_{j = 1}^{2i} p_j \right )^2 + \mu^2 \right ]} + \mathcal P ( 2,3,\ldots,2n ) \,,
\eea
where the $\mathcal P ( 2,3,\ldots,2n )$ stands for all permutations of momentum labels $2,3,\ldots,2n$.
Using the integral \reef{npinte}, it was shown in \cite{Elvang:2019twd} that \reef{SDintegrand} integrates to the 
local expression
\begin{equation}
    \label{SDallamp}
    \boxed{
    \begin{aligned}
    &\mathcal A_{2n}^{\text{BI}_4\; \text{1-loop}}\left(1_\gamma^+\, 2_\gamma^+\ldots 2n_\gamma^+\right) \\
    &\hspace{5mm}= \frac{i}{32 \pi^2} \left( - \frac 1 2 \right )^{n-1} \frac{1}{n ( n + 1 ) ( n + 2 ) ( n + 3 )} \\
    &\hspace{5mm}\times \Bigg [ [ 12 ]^2 [ 34 ]^2 \ldots [ 2n-1, 2n ]^2 \left ( \sum_{i < j}^n \sum_{k < l}^n a_{i j k l} \left ( \sum_{m = 2 i + 1}^{2 j} p_m \right )^2 \left ( \sum_{m = 2 k + 1}^{2 l} p_m \right )^2 \right ) \\
    &\hspace{10mm} + \mathcal P ( 2, 3, \ldots, 2n ) \Bigg] + \mathcal{O}(\epsilon)\,,
    \end{aligned}
    }
\end{equation}
with
\begin{equation}
a_{i j k l} = \left \{ \begin{array}{ll}
1 & \qquad \text{if all $i,j,k,l$ are different} \\
2 & \qquad \text{if exactly 2 of $i, j, k, l$ are identical} \\
4 & \qquad \text{if $i = k$ and $j = l$}
\end{array} \right .\,.
\end{equation}
It is straightforward to check that this result matches the results of the explicit calculations for the case of $n = 2$ presented above.

In the next-to-self-dual sector, the  1-loop amplitudes in pure BI theory are
\begin{equation}
    \label{NSDallamp_grey}
    \boxed{
    \begin{aligned}
        &\mathcal A_{2 n}^{\text{BI}_4\; \text{1-loop}}\left(1_\gamma^+\,2_\gamma^+\ldots (2n-1)_\gamma^+\,2n_\gamma^-\right)
        \\
        & = \frac{i}{32 \pi^2}\frac{( n - 2 )!}{( n + 2 )!}\left (- \frac 1 2 \right )^{n-1} \frac{[ 12 ]^2 \ldots[ 2n-3 \ 2n-2 ]^2  [ 2 n - 1|p_{2n-2}+p_{2n-3}| 2 n \rangle^2}{s_{2n,2n-2,2n-3}} \\
        &\times \left[\sum_{i < j}^{n-2} \sum_{k < l}^{n-2} a_{i j k l} \left(\sum_{m = 2i + 1}^{2j} p_m \right )^2 \left(\sum_{m = 2k + 1}^{2l} p_m \right )^2+4\sum_{i \leq j}^{n-2} \left(\sum_{m = 1}^{2i} p_m \right )^2 \left(\sum_{m = 1}^{2j} p_m \right )^2\right. \\
        &\left.+2\sum_{i =1}^{n-2}\sum_{k < l}^{n-2}a_{i (n-1) k l} \left(\sum_{m = 1}^{2i} p_m \right )^2 \left(\sum_{m = 2k + 1}^{2l} p_m \right )^2\right] + \mathcal P ( 1,2,\ldots,2n-1 ) +\mathcal{O}(\epsilon) 
        \\
        & + \text{local terms}
        .
    \end{aligned}
    }
\end{equation}
The local terms in this amplitude have been computed explicitly and can be found in equations (4.24)-(4.27) in \cite{Elvang:2019twd}.

As shown in \cite{Elvang:2019twd}, the 1-loop next-to-self-dual $+\dots+-$  amplitude \reef{NSDallamp_grey} has simple poles on which it factorizes into self-dual
$+\dots++$ amplitudes times a 4-point tree-level BI amplitude, e.g.
\bea
    &&
    \hspace{-5mm}\operatorname*{Res}_{p_f^2 = 0} \mathcal A_{2 n}^{\text{BI}_4\; \text{1-loop}}\left(1_\gamma^+\ldots (2n-1)_\gamma^+\,2n_\gamma^-\right) \\
    \nonumber
    &=& \mathcal A_{2n - 2}^{\text{BI}_4\; \text{1-loop}} \left ( 1_\gamma^+\ldots  ( 2 n - 3 )_\gamma^+\, \left ( p_f \right )_\gamma^+ \right ) \times \mathcal A_4^{\text{BI}_4} \left ( \left ( - p_f \right )_\gamma^-\, ( 2 n - 2 )_\gamma^+\, ( 2 n - 1 )_\gamma^+\, ( 2 n )_\gamma^- \right ).
\eea

There are no other poles of any kind in amplitudes in the self-dual and next-to-self-dual sectors. Therefore, if local counterterms are chosen to set all 1-loop self-dual amplitudes to zero, then the next-to-self-dual 1-loop  amplitudes are also local and can therefore be removed by local finite counterterms as well. This means that there is no violation of EM duality at the 1-loop level in the self-dual and next-to-self-dual sectors.

%%%%%%%%%%%%%%%%%%%%%%%%%%%%%%
\section{Supersymmetric (D)BI and MHV Amplitudes at 1-loop}
\label{sec:SD_MHV_relation}
%%%%%%%%%%%%%%%%%%%%%%%%%%%%%

We present supersymmetric extensions of BI theory and derive the $U(1)$ EM duality charges of the states in the supermultiplets. We then use the result from Section \ref{sec:BIall1loop} for the self-dual 1-loop integrand of pure BI theory to construct a conjectured expression for the MHV 1-loop integrand of $\mathcal{N}=4$ DBI using the  dimension-shifting  relation of \cite{Bern:1996ja}. This integrates to a local polynomial expression for the MHV 1-loop amplitudes in $\mathcal{N}=4$ DBI that agrees at $n=4,6$ with known results.   

%%%%%%%%%%%%%%%%%%%%%%%%%%%%%
\subsection{Supersymmetric Born-Infeld}
%%%%%%%%%%%%%%%%%%%%%%%%%%%%%
Consider Born-Infeld theory supersymmetrically coupled to $N_f$ Weyl fermions and $N_s$ complex scalars. We have 
\be
  \label{SUSYDBI}
  \begin{array}{ll}
      \text{$\mathcal{N}=1$ BI:}&~~~~ N_f=1~~~\text{and}~~~N_s=0\,,\\
      \text{$\mathcal{N}=2$ DBI:}&~~~~ N_f=2~~~\text{and}~~~N_s=1\,,\\
      \text{$\mathcal{N}=4$ DBI:}&~~~~ N_f=4~~~\text{and}~~~N_s=3\,.
  \end{array}
\ee
The scalars of the $\mathcal{N}=2$ and $\mathcal{N}=4$ supermultiplets are Dirac-Born-Infeld scalars, hence the switch of name from supersymmetric BI theory to the more commonly used supersymmetric DBI. 

As we have discussed, 4d pure non-supersymmetric BI has electromagnetic duality symmetry that acts as a $U(1)$ symmetry on the on-shell photon states. In supersymmetric BI, this becomes a $U(1)_R$ symmetry. Suppose the supercharge changes the charge by $r$, then if the highest weight state in the multiplet has helicity $h$ and charge $q$
\be 
   \begin{array}{lccccc}
   \text{state:}&~~|h\>~~ & ~~|h-\tfrac{1}{2}\>~~ & ~~|h-1\>~~ & \ldots
   \\
   U(1)_R: & q & q-r & q-2r & \ldots\;
   \end{array}
\ee
CPT conjugate states must have opposite charges. In particular, if the multiplet is CPT self-conjugate, as is the case for the $\mathcal{N}=4$ vector multiplet, then we must have $-q=q-4r$, i.e.~$r=q/2$. With $q=1$, as in pure BI, this fixes the $U(1)_R$ charges of the multiplet to be 
\be 
   \begin{array}{lccccccc}
   \text{state:}&~~|1\>~~ & ~~|\tfrac{1}{2}\>~~ & ~~|0\>~~ & ~~|-\tfrac{1}{2}\>~~&  ~~|-1\>~~
   \\[1mm]
   U(1)_R: & 1 & \tfrac{1}{2} & 0 & -\tfrac{1}{2} & -1
   \end{array}
\ee
which means that the $U(1)_R$-charges coincide with the helicity labels. In addition, the $\mathcal{N}=4$ theory admits an $SU(4)_R$ symmetry under which the vectors are singlets; so the non-abelian R-symmetry is not an electromagnetic duality symmetry.

When applied to $\mathcal{N}=4$ SYM, one can exclude the existence of a $U(1)_R$: the reason is that the cubic gluon interactions give rise to non-vanishing 3-particle amplitudes with helicities $++-$ and $--+$. This requires the vector charge, and hence all the other $U(1)_R$ charges, to vanish.\footnote{A parallel argument can be used to prove that there can be no $U(1)$ R-symmetry of $\mathcal{N}=8$ supergravity due to the graviton 3-particle self-interactions. See Section \ref{s:sg}.}

In $\mathcal{N}=4$ DBI, the $U(1)_R$ is allowed; there are for example no cubic interactions to forbid it. 
The existence of the $U(1)_R$ was noted in a CHY formulation of the $\mathcal{N}=4$ DBI amplitudes in \cite{Heydeman:2017yww}. 

In the supersymmetric theories \reef{SUSYDBI}, the Ward identities associated with the conservation of the $U(1)_R$ charge are, for the special case of amplitudes with only external photons, exactly the same as (\ref{EMamps}). Since self-dual and next-to-self dual amplitudes vanish in any supersymmetric theory, independent of the existence of a duality symmetry, the simplest class of potentially non-trivially duality-violating amplitudes are therefore the MHV sector starting at 6-point. Hence we now turn to study the MHV amplitudes. 

%%%%%%%%%%%%%%%%%%%%%%%%%%%%%
\subsection{All-Multiplicity 1-loop MHV Amplitudes in \texorpdfstring{$\mathcal{N}=4$}{N=4} DBI}
%%%%%%%%%%%%%%%%%%%%%%%%%%%%%

In this section we present a conjecture for the all-multiplicity 1-loop integrand of the MHV sector of $\mathcal{N}=4$ DBI in $d=4-2\epsilon$. As we argue, the expression we write down follows from combining two well-known conjectures, the dimension-shifting relation between self-dual and maximally supersymmetric Yang-Mills \cite{Bern:1996ja}, and the 1-loop version of the BCJ double-copy \cite{Bern:2010ue} applied to Born-Infeld models (\ref{gendcBI}). At $n=4$ and $n=6$, where alternative explicit results are available for comparison, we find exact agreement. 

It was conjectured in \cite{Bern:1996ja} that the 1-loop MHV integrand of $\mathcal{N}=4$ SYM is related to the 1-loop self-dual integrand of pure YM  theory as\footnote{Since finishing this paper, a proof of this conjecture has been presented by Britto, Jehu, and Orta \cite{Britto:2020crg}.}:
\begin{equation} \label{YMSD}
	\mathcal{I}_n^{\mathcal{N}=4\text{ SYM}}\big(1^+2^+\cdots i^-\cdots j^-\cdots n^+;l,\mu^2\big)
	=\frac{\ang{ij}^4}{2(\mu^2)^2}\mathcal{I}_n^{\text{YM}}\big(1^+2^+\cdots n^+;l,\mu^2\big).
\end{equation}
The relation was proven for $n\le 6$ and evidence was provided for its validity at any multiplicity \cite{Bern:1996ja}. We can write the 1-loop integrand for self-dual Yang-Mills in BCJ form
\begin{equation}
    \mathcal{I}_n^{\text{YM}}\big(1^+2^+\cdots n^+;l,\mu^2\big) = \sum_i \frac{c_i \,n_i^{\text{YM}}\big(1^+2^+\cdots n^+;l,\mu^2\big) }{d_i},
\end{equation}
where the sum over $i$ is taken over all trivalent 1-loop graphs with $c_i$ and $d_i$ the corresponding color factors and denominators respectively. If the 1-loop BCJ conjecture is correct, then we can always find a so-called \textit{generalized gauge} in which the numerators satisfy kinematic Jacobi relations \cite{Bern:2010ue}
\begin{equation}
\label{loopkinjac}
    n_i^{\text{YM}}+n_j^{\text{YM}}+n_k^{\text{YM}}=0 \hspace{5mm} \Longleftrightarrow \hspace{5mm} c_i+c_j+c_k=0.
\end{equation}
If we assume that such numerators exist then we can define 
\begin{equation}
\label{eq:dimshift_num}
n_i^{\mathcal{N}=4\text{ SYM}}\big(1^+2^+\cdots i^-\cdots j^-\cdots n^+;l,\mu^2\big)\,\equiv\,\frac{\ang{ij}^4}{2(\mu^2)^2}
\,n_i^{\text{YM}}\big(1^+2^+\cdots n^+;l,\mu^2\big).
\end{equation}
Then, further assuming the dimension-shifting relation (\ref{YMSD}), it follows that 
\begin{equation}
\label{N4integrandansatz}
     \mathcal{I}_n^{\mathcal{N}=4\text{ SYM}}\big(1^+\cdots i^-\cdots j^-\cdots n^+;l,\mu^2\big) = \sum_i \frac{c_i\, n_i^{\mathcal{N}=4\text{ SYM}}\big(1^+\cdots i^-\cdots j^-\cdots n^+;l,\mu^2\big) }{d_i}.
\end{equation}
The objects (\ref{eq:dimshift_num}) are BCJ numerators for $\mathcal{N}=4$ super Yang-Mills in some generalized gauge. Furthermore, since they are constructed by multiplying by an overall factor, these numerators must also satisfy the kinematic Jacobi relation (\ref{loopkinjac}). If the loop-level BCJ conjecture is correct then we can generate an expression for the MHV 1-loop integrand of $\mathcal{N}=4$ DBI by replacing the color factors $c_i$ in \reef{N4integrandansatz} with BCJ numerators of $\chi$PT (in any generalized gauge). This gives the following relation
\begin{equation} \label{N4DBIdc}
	\mathcal{I}_n^{\mathcal{N}=4\text{ DBI}}\big(1^+2^+\cdots i^-\cdots j^-\cdots n^+;l,\mu^2\big)
	\,=\,\frac{\ang{ij}^4}{2(\mu^2)^2}\,\mathcal{I}_n^{\text{BI}}\big(1^+2^+\cdots n^+;l,\mu^2\big).
\end{equation}
Using the explicit all-multiplicity expression for the self-dual integrand \reef{SDintegrand}, we then use \reef{N4DBIdc} to conjecture the following all-multiplicity expression for the 1-loop integrand in the MHV sector of $\mathcal{N}=4$ DBI
\begin{equation}
  \label{eq:dimshift_npt}
    \boxed{
\begin{aligned}
	&\mathcal{I}_{2n}^{\mathcal{N}=4\text{ DBI}}\left(1_\gamma^+2_\gamma^+\cdots i_\gamma^-\cdots j_\gamma^-\cdots 2n_\gamma^+;l,\mu^2\right)	\\
	&=\left(-\frac12\right)^{n} \ang{ij}^4\sq{12}^2\cdots \sq{2n-1,2n}^2\frac{(\mu^2)^{n-2}}{\prod_{i=1}^n\left[\left(l-\sum_{j =1}^{2i} p_j\right)^2+\mu^2\right]}+\mathcal{P}\left(2,3,\cdots,2n\right).
\end{aligned}
}
\end{equation}
The 1-loop double-copy construction was tested successfully at 4-point for the self-dual amplitude in Section  \ref{sec:DC4ptSD}. When the result \reef{BIallplusBCJ1} for $\mathcal{I}_4^{\text{BI}}\left(1_\gamma^+2_\gamma^+3_\gamma^+4_\gamma^+;l,\mu^2\right)$ is applied in \reef{N4DBIdc}, we obtain the 4-point MHV 1-loop integrand 
\begin{align}
\label{eq:4pt_dimshift}
	\mathcal{I}_4^{\mathcal{N}=4\text{ BI}}\left(1_\gamma^+2_\gamma^+3_\gamma^-4_\gamma^-;l,\mu^2\right)=\frac12[12]^2 \ang{34}^2\left(
	s^2 \mathcal{I}_2[p_{12}]+t^2 \mathcal{I}_2[p_{13}]+u^2 \mathcal{I}_2[p_{14}]\right),
\end{align}
where $\mathcal I_2$ is a scalar bubble integrand, whose integral $I_2$ in $4-2\epsilon$ dimensions is given in \eqref{eq:bubble_int}. Thus in the small-$\epsilon$ expansion we find

\be
 \label{N41L}
 \boxed{
  \mathcal{A}_4^\text{1-loop $\mathcal{N}=4$ DBI}(1^+_\gamma\,2^+_\gamma\,3^-_\gamma\,4^-_\gamma)
   = \frac{1}{2}[12]^2 \<34\>^2 
  \bigg[  s^2  I_2 (s) +t^2  I_2 (t) +u^2  I_2 (u) 
  \bigg]
  + \mathcal{O}(\epsilon) \,.
  }
\ee 
 The amplitude is UV divergent, and it is in fact the only MHV amplitude of (D)BI that has non-vanishing 4d cuts. Unitarity requires that these cuts factor into physical tree-amplitudes. Even though the complete integrand (\ref{eq:dimshift_npt}) is scheme-dependent, the values of these 4d cuts are not, and therefore give a non-trivial check on the proposal \reef{N4DBIdc}. In the following subsection we verify explicitly that the cut-constructible part of the 4-point MHV amplitude, constructed from the known tree-amplitudes, agrees exactly with (\ref{N41L}).

For $n>2$, i.e.~for 6-point and higher, the integrand vanishes as $\mu \to 0$, hence it has vanishing 4d cuts. Using the integral \reef{needthisint} derived in Appendix \ref{app:dimregintegrals}, we integrate \eqref{eq:dimshift_npt} for $n>2$ to find the rational local expression 
\be
  \label{N4all}
  \boxed{
  \begin{aligned}
	&\mathcal{A}_{2n}^{\text{1-loop $\mathcal{N}=4$ DBI}}\left(1_\gamma^+2_\gamma^+\cdots i_\gamma^-\cdots j_\gamma^-\cdots 2n_\gamma^+\right)\\[1mm]
	&=\frac{i}{16\pi^2}
	\frac{(-1)^{n+1} \<ij\>^4}{2^n (n-1)(n-2)}
	\left( \sq{12}^2\cdots \sq{2n-1,2n}^2+\mathcal{P}\left(2,3,\cdots,2n\right)\right) + \mathcal{O}(\epsilon).
	\end{aligned}}
\ee
In the following we compare the 4-point MHV 1-loop result (\ref{N41L}) with the prediction from unitarity and discuss the associated divergence and infinite local counterterm. We also compare our prediction for the 6-point MHV 1-loop amplitude (\ref{N4all}) with explicit results obtained in \cite{Wen:2020qrj} using the dimensional reduction of M5-brane tree-amplitudes. We find complete agreement in both cases.

%%%%%%%%%%%%%%%%%%%%%%%%%%%%%%%
\subsection{1-loop MHV in BI with \texorpdfstring{$\mathcal{N}$}{N}-Fold SUSY and Counterterms}
%%%%%%%%%%%%%%%%%%%%%%%%%%%%%%%

\paragraph{4-point.} Consider a Born-Infeld model with $N_v$ vectors coupled supersymmetrically to $N_f$ Weyl fermions and $N_s$ complex scalars. The 4-point MHV amplitude in this model has non-vanishing 4d cuts and it is therefore fairly straightforward to calculate from unitarity. We include the details for the calculation in Appendix \ref{s:MHV4BI}. The result is
\be
  \label{ppmm1-loop}
  \boxed{
  \begin{aligned}
  \mathcal{A}_4^\text{1-loop}(1^+_\gamma\,2^+_\gamma\,3^-_\gamma\,4^-_\gamma)
  &= [12]^2 \<34\>^2 
  \bigg[ \frac{N_v}{2} s^2  I_2 (s) +\Big( \frac{N_v}{5} +  \frac{N_f}{20} + \frac{N_s}{30}\Big)\big( t^2  I_2 (t)+ u^2  I_2 (u) \big)
  \bigg]
  \\
  & =\frac{1}{\epsilon}
  \frac{i}{16\pi^2} 
  [12]^2 \<34\>^2 
  \bigg[ \frac{N_v}{2} s^2 +\Big( \frac{N_v}{5} +  \frac{N_f}{20} + \frac{N_s}{30}\Big) \big(  t^2 + u^2 \big)
  \bigg]
  + \mathcal{O}(1)\,.
  \end{aligned}}
\ee 
For $N_v=1$ and $N_f = N_s = 0$ we obtain the pure Born-Infeld MHV amplitude. The $\mathcal{N}=1,2,4$ results are likewise obtained by setting $N_v=1$ and using \reef{SUSYDBI}. 
In particular, $N_v=1$, $N_f = 4$, and $N_s = 3$, reproduces the $\mathcal{N}=4$ DBI result \reef{N41L}, a non-trivial test of the conjectured relation \reef{N4DBIdc}. 
The 4-point MHV 1-loop amplitude of $\mathcal{N}=4$ DBI was  calculated previously by Shmakova \cite{Shmakova:1999ai} with the same result \reef{N41L}. 

In order to absorb the $1/\eps$ divergence in the 4-point MHV amplitudes, it follows from simple power-counting that we need a local counterterm of the form $\pa^4 F^4$. Little group scaling, Bose symmetry, and dimensional analysis show that there are two  independent local matrix elements, so there are two independent  $\pa^4 F^4$ operators on-shell. The general counterterm amplitude takes the form
\begin{equation}
  \label{ctab}
  \mathcal{A}_4^{\text{ct}}\big(1_\gamma^+ \,2_\gamma^+\, 3_\gamma^-\,4_\gamma^-\big) 
  = a\, [12]^2\langle 34\rangle^2 s^2+ b\, [12]^2\langle 34 \rangle^2 (t^2+u^2)\,,
\end{equation}
where $a$ and $b$ are constants. With particular choices of $a$ and $b$, we can cancel the UV divergence for all choices of  $N_f$ and $N_s$. 

Imposing $\mathcal{N}=4$ supersymmetry, the matrix element \reef{ctab} must satisfy the supersymmetry Ward identity
\begin{equation}
  \mathcal{A}_4^{\text{ct}}\big(1_\gamma^+\,2_\gamma^+\,3_\gamma^-\,4_\gamma^-\big) = \frac{[12]^4}{[13]^4} \,
  \mathcal{A}_4^{\text{ct}}\big(1_\gamma^+\,2_\gamma^-\,3_\gamma^+\,4_\gamma^-\big)\,,
\end{equation}
which requires
\begin{equation}
  a s^2+b t^2+b u^2 = b s^2+at^2+bu^2
  ~~\implies ~~
  ( a - b ) ( s^2 - t^2 ) = 0\,.
\end{equation}
So this is possible only if $a=b$. In other words, there is only one $\pa^4 F^4$ counterterm compatible with $\mathcal{N}=4$ supersymmetry. Thus, in $\mathcal{N}=4$ DBI, the UV divergence must be proportional to $s^2 + t^2 + u^2$, exactly as it is in \reef{N41L}. The counterterms associated with \reef{ctab} are easy to construct using spinorized fields \reef{spinorF} and external line Feynman rules \reef{FpFm}. We find
\be
   \begin{split}
      \Tr (\pa_\mu F_+ \pa^\mu F_+)\Tr (\pa_\nu F_- \pa^\nu F_-)
      &\longrightarrow
      4[12]^2\langle 34\rangle^2 \,s^2\,, \\[1mm]
      \Tr (\pa_\mu F_+ \pa_\nu F_+)\Tr (\pa^\mu F_- \pa^\nu F_-)
      &\longrightarrow
      2[12]^2\langle 34\rangle^2 \,(t^2+u^2)\,,
   \end{split}
\ee
where the trace refers to the spinor indices,
i.e.~$\Tr (\pa_\mu F_+ \pa^\mu F_+) = \pa_\mu (F_+)_a{}^b \pa^\mu (F_+)_b{}^a$. 
Linear combinations of these two operators  cancel the $1/\eps$ divergence in the 1-loop amplitude \reef{ppmm1-loop}. 

For the $\mathcal{N}=4$ supersymmetric case, the counterterm takes a particularly recognizable  form in terms of the 8-rank $t_8$-tensor known from the open string amplitude. Specifically,
\be 
  \label{t8ct}
  \begin{split}
  &(t_8)_{\mu_1\nu_1 \mu_2\nu_2 \mu_3\nu_3 \mu_4\nu_4} 
  \partial_\alpha F^{\mu_1\nu_1}  \partial_\beta F^{\mu_2\nu_2}  
  \partial^\alpha F^{\mu_3\nu_3}  \partial^\beta F^{\mu_4\nu_4}\\[1mm]
  &~~~= 
  2\,\Tr (\pa_\mu F_+ \pa^\mu F_+)\Tr (\pa_\nu F_- \pa^\nu F_-)
  + 4\Tr (\pa_\mu F_+ \pa_\nu F_+)\Tr (\pa^\mu F_- \pa^\nu F_-)
  \\[1mm]
  &~~~\longrightarrow 8[12]^2\langle 34\rangle^2 \,(s^2+ t^2+u^2)\,,
  \end{split}
\ee
giving the $\mathcal{N}=4$ supersymmetric matrix element \reef{N41L}.\footnote{The $(t_8) (\pa F)^4$ operator presented in \cite{Shmakova:1999ai} has a different index contraction that produces matrix elements with the wrong helicity structure.}

\vspace{2mm}
\paragraph{6-point.}
% The 6-point 1-loop MHV amplitude of $\mathcal{N}=4$ DBI was recently calculated from explicit CHY formulae for M5-brane tree-amplitudes using two methods \cite{Wen:2020qrj}. First, by dimensionally reducing the forward limit of 8-point M5-brane tree-amplitudes, and second, using generalized unitarity by imposing consistency with the M5-brane tree-amplitudes on 6d cuts. Their result agrees with \reef{N4all} for $n=3$. It was also noted in \cite{Wen:2020qrj} that the general form of the kinematic polynomial in \reef{N4all} is the only possible one compatible with the requirements of power counting, little group scaling, and supersymmetry. Our result \reef{N4all} is the exact result for the 1-loop MHV amplitude of $\mathcal{N}=4$ DBI, including its normalization, using the conjectured relation \reef{N4DBIdc}. 

The 6-point 1-loop MHV amplitude of $\mathcal{N}=4$ DBI was recently calculated from explicit CHY formulae for M5-brane tree-amplitudes using two methods \cite{Wen:2020qrj}. First, by dimensionally reducing the forward limit of 8-point M5-brane tree-amplitudes, and second, using generalized unitarity by imposing consistency with the M5-brane tree-amplitudes on 6d cuts. Their result agrees with \reef{N4all} for $n=3$ (6-particle scattering) up to a convention dependent phase factor. It was also noted in \cite{Wen:2020qrj} that the general form of the kinematic polynomial in \reef{N4all} is the only possible one compatible with the requirements of power counting, little group scaling, and supersymmetry. Our result \reef{N4all} is an exact expression for the 1-loop MHV amplitude of $\mathcal{N}=4$ DBI, including its normalization, using the conjectured relation \reef{N4DBIdc}. 

The expression (\ref{eq:dimshift_npt}) for the integrand has a curious structure if we try to interpret it in terms of Feynman diagrams. Each term in the sum over permutations corresponds to an $n$-gon loop diagram (where the total multiplicity is $2n$), with no contributions from $k$-gon loops with $k<n$. In our derivation this is a consequence of the fact that the non-supersymmetric SD integrands calculated in \cite{Elvang:2019twd} likewise have only $n$-gon contributions. Naively this appears to be in conflict with the results of \cite{Wen:2020qrj} where the existence of non-zero contact terms for the M5-brane tree amplitudes will give rise to diagrams with lower $k$-gon topology when glued together to form a loop. Whether or not an actual conflict exists is difficult to assess without an explicit calculation since such lower $k$-gon contributions can always be ``blown-up" to an $n$-gon by multiplying and dividing by appropriate propagator factors. We would therefore need the complete contribution to the integrand to assess if it can be algebraically manipulated into the form (\ref{eq:dimshift_npt}). Moreover, it is not completely clear that (\ref{eq:dimshift_npt}) and the results of \cite{Wen:2020qrj} need to agree at all orders in $\epsilon$ since they are calculated using different dimensional regularization schemes. The result (\ref{eq:dimshift_npt}) is constructed from the requirement that the $d$-dimensional cuts agree with the D5-brane tree amplitudes when $d=6$ since they are constructed as a double-copy from Yang-Mills. The expressions obtained in \cite{Wen:2020qrj} are constructed from the requirement that the $d$-dimensional cuts agree with the M5-brane tree amplitudes when $d=6$. These two approaches therefore make use of two different regularization schemes that both preserve $\mathcal{N}=4$ supersymmetry, and it is not clear that beyond $\mathcal{O}(\epsilon^0)$ they should agree. Nonetheless, since we do find explicit agreement for 6-particle scattering, it is possible that they do in fact coincide, and it would be interesting to extend the analysis of \cite{Wen:2020qrj} to 8-particle scattering and beyond to explore this.

Finally, the self-dual and next-to-self-dual sectors vanish in the presence of any amount of supersymmetry, hence the MHV amplitudes present the first potentially duality-violating sector in $\mathcal{N}=4$ DBI. 
For multiplicities beyond 4-point, the MHV result \eqref{N4all} has the important feature that it is completely local and therefore it \textit{can} be removed by the addition of a finite local counterterm. Thus the MHV duality-violating $\mathcal{N}=4$ DBI amplitudes can be set to zero, providing yet another piece of evidence that electromagnetic duality may be preserved at 1-loop.

%%%%%%%%%%%%%%%%%%%%%%%%%%%%%%%
\section{Rational Loop Amplitudes and Finite Counterterms}
%%%%%%%%%%%%%%%%%%%%%%%%%%%%%%%

\label{sec:rational}

Given the explicit results (\ref{SDallamp}) and (\ref{NSDallamp_grey}) for the SD and NSD duality-violating 1-loop amplitudes in pure BI theory and the MHV 1-loop amplitudes (\ref{N4all}) of $\mathcal{N}=4$ DBI, there is a clear motivation to attempt a general \textit{proof} that duality violation at 1-loop is always removable by adding an appropriate set of finite local counterterms. 

As we discuss in more detail below, at 1-loop, all duality-violating amplitudes are purely rational functions of spinor-helicity brackets. The problem of determining whether-or-not such rational functions are removable by adding higher-derivative (duality-violating) local operators to the classical action can be rephrased as the problem of whether their kinematic singularity structure resembles that of a tree-level scattering amplitude; we refer to such amplitudes as \textit{tree-like}. Explicitly, we define tree-like to mean that the rational 1-loop amplitudes $\mathcal{A}_n$ have two important properties:
\begin{itemize}
  \item \textit{Factorization}: If it is possible to find lower multiplicity on-shell amplitudes $\mathcal{A}_{n_L}$ and $\mathcal{A}_{n_R}$ with $n_L+n_R=n+2$ that can be glued together into an expression of the form 
\begin{equation}
  \sum_{X}\mathcal{A}_{n_L}(...,P_X)\mathcal{A}_{n_R}(-P_{\overline{X}},...),
\end{equation}
where we sum over all physical states $X$, $\overline{X}$ denotes a state with CP conjugate quantum numbers, and the remaining $n$ external states coincide with those of $\mathcal{A}_n$, then $\mathcal{A}_n$ must contain a (simple) pole at $P_X^2=0$ with this expression as its residue. 
\item \textit{Locality}: The rational amplitude $\mathcal{A}_n$ contains no additional \textit{spurious} singularities. 
\end{itemize}
These properties are guaranteed to hold for any expression constructed using Feynman rules derived from a local, Lorentz-invariant Lagrangian of a unitary model.
This includes both the genuine tree-level approximation to S-matrix
elements as well as the contributions from higher-derivative
counterterms equivalent in $\Lambda$ counting to 1-loop order in the EFT. As a consequence, if the purely rational 1-loop amplitudes are not tree-like, then no choice of finite local counterterms can cancel them. It is not \textit{a priori} obvious what the singularity structure of rational loop amplitudes should be. 

An illustrative example of a non-tree-like rational amplitude is provided by the 1-loop self-dual Yang-Mills amplitude at $n=4$ \cite{Bern:1993sx}
\begin{equation}
    \label{SDYM}
  \mathcal{A}_n^{\text{YM}}\left[1_g^+\,2_g^+\,3_g^+\,4_g^+\right] = -\frac{i}{96\pi^2}\frac{[12][34]}{\langle 12 \rangle \langle 34 \rangle}.
\end{equation}
Such an expression has complex multi-collinear singularity when $|1\rangle \propto |2\rangle \propto |3\rangle \propto |4\rangle$ that does not correspond to a physical factorization singularity, and hence this expression is \textit{non-tree-like}. The physical interpretation of this result is that such an amplitude is \textit{scheme-independent}, and cannot be removed by the addition of local counterterms to the Yang-Mills action. 

To make the problem explicit we briefly review the discussion of \cite{Ossola:2008xq}. Consider the representation of the $d=4-2\epsilon$ dimensional loop integrand constructed using Feynman rules. This is a sum over terms of the form
\begin{equation}
  \frac{N[l,q_i]}{(l+q_1)^2(l+q_2)^2...(l+q_{k})^2},
\end{equation}
where $q_i$ are region momenta. We consider only 1PI contributions since singularities arising from propagators which do not contain $l$ in the chopped off parts are not relevant for the discussion.
Hence the Feynman numerator $N[l]$ is a polynomial in both $l$ and $q_i$. We split the numerator into two pieces 
\begin{equation}
  N[l] = N^{(0)}[l] + N^{(1)}[l]\,,
\end{equation}
where $N^{(1)}[l]\rightarrow 0$ as $\epsilon\rightarrow 0$. The $N^{(1)}$ part may generate a contribution (which in \cite{Ossola:2008xq} is called $\mathcal{R}_2$) from $\epsilon/\epsilon$ cancellations against UV and IR divergences. Since this integrand is 1PI, the UV contributions from $N^{(1)}$ are always polynomial. The remaining contribution (called $\mathcal{R}_1$ in \cite{Ossola:2008xq}) arises from the rational parts of the triangle and box master scalar integrals generated by tensor reduction of the $N^{(0)}$ part of the numerator. One perspective on this is given by old-fashioned Passarino-Veltman reduction \cite{Passarino:1978jh}. Intermediate steps in the reduction of tensor integrals generates Gram determinant factors in the region momenta which in general contain spurious singularities. The general form of such rational contributions is quite complicated and not known explicitly in general. What is known is that they are not always tree-like, even in the case where the amplitudes are purely rational.

The pure BI results (\ref{SDallamp}) and (\ref{NSDallamp_grey})  and the MHV result (\ref{N4all}) in $\mathcal{N}=4$ DBI show that that these duality-violating 1-loop contributions are tree-like and can be set to zero by finite local counterterms. That expressions like (\ref{SDYM}) arise in the self-dual sector of Yang-Mills (and also Einstein gravity \cite{Bern:1998xc}) but not at 1-loop in Born-Infeld, as far as we know, could be taken as a hint that there is an essential difference between these models which is responsible for the absence of non-tree-like rational terms in the latter. A clue as to what this might be comes from the analysis in \cite{Bern:1995ix} of the factorization properties of QCD amplitudes. There, an argument was given (which assumed QCD-like power counting, though perhaps not in an essential way) that \textit{IR finite} 1PI integrals do not generate spurious singularities. If such an argument can be extended to integrals with Born-Infeld-like power counting, then it would imply that all duality-violating 1-loop amplitudes are tree-like rational functions. Here the essential property that distinguishes Born-Infeld from Yang-Mills or Einstein Gravity is the absence of 3-particle interactions that generate Feynman integrals with soft or collinear IR divergences at 1-loop. 

At present the statement: 
\be
\nonumber
\boxed{
\phantom{\bigg|}
\text{
\textit{no 3-point interactions + no 4d cuts} $\Rightarrow$ \textit{tree-like rational amplitudes}
}
\phantom{\bigg|}
}
\ee
remains a conjecture, but if proven it implies that all duality-violating 1-loop amplitudes in Born-Infeld can be completely cancelled by adding an appropriate set of finite local counterterms. To make the argument explicit, consider a 2-particle 4d cut of a 1-loop amplitude in Born-Infeld: 
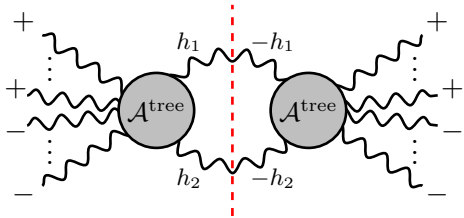
\begin{center}
   \begin{tikzpicture}[scale=1, line width=1 pt]
	\draw [vector] (-1.5,1)--(0,0);
	\node at (-1.4,0.65) {$\vdots$};
	\draw [vector] (-1.7,-0.2)--(-0,0);
	\node at (-1.4,-0.45) {$\vdots$};
	\draw [vector] (-1.5,-1)--(0,0);
	\draw [vector] (-1.7,0.2)--(-0,0);
	\draw [vector] (0,0) arc (160:20:1);
	\draw [vector] (0,0) arc (200:340:1);
	\draw [vector] (2,0)--(3.5,1);
	\draw [vector] (2,0)--(3.7,0.2);
	\node at (3.4,-0.45) {$\vdots$};
	\draw [vector] (2,0)--(3.5,-1);
	\node at (3.4,0.65) {$\vdots$};
	\draw [vector] (2,0)--(3.7,-0.2);
	\draw[black,fill=lightgray] (0,0) circle (3ex);
	\draw[black,fill=lightgray] (2,0) circle (3ex);
	\node at (0,0) {\small $\mathcal{A}^\text{tree}$};
	\node at (2,0) {\small $\mathcal{A}^\text{tree}$};
	\node [above left] at (-1.45,0.9) {$+$};
	\node [below left] at (-1.45,-0.85) {$-$};
	\node [above left] at (-1.55,0.015) {$+$};
	\node [below left] at (-1.55,-0.01) {$-$};
	\node [above right] at (3.4,0.9) {$+$};
	\node [below right] at (3.4,-0.85) {$-$};
	\node [above right] at (3.6,0.015) {$+$};
	\node [below right] at (3.6,-0.01) {$-$};
	\draw[red,dashed] (1,-1.4)--(1,1.4);
	\node at (0.43,0.91) {\footnotesize $h_1$};
	\node at (0.43,-0.9) {\footnotesize $h_2$};
	\node at (1.53,0.91) {\footnotesize $-h_1$};
	\node at (1.53,-0.9) {\footnotesize $-h_2$};
   \end{tikzpicture}
\end{center}
Here $h_1$ and $h_2$ are the helicities of the on-shell photons exchanged across the cut.  
Let the number of {\em external} positive and negative helicity photons on the LHS of the cut be $n_L^\pm$ and similarly on the RHS, $n_R^\pm$. Because each tree sub-amplitude satisfies the duality constraints (\ref{EMamps}), we have
\be
  n_L^+ = n_L^- - h_1 - h_2\,,~~~~
  \text{and}~~~~
  n_R^+ = n_R^- + h_1 + h_2\,.
\ee
Hence for the overall amplitude
\be
  n^+ = n_L^+ + n_R^+ = n_L^- + n_R^- = n^-\,.
\ee
This means that the 4d cut can only be non-zero when the 1-loop amplitude obeys the constraint $n^+ = n^-$ of duality. Any triple and quadruple 4d cuts must necessarily obey the same constraint, since they are further restrictions of the 2-particle 4d cuts. Thus any duality-violating 1-loop amplitude of Born-Infeld has vanishing 4d cuts and therefore must be purely rational. 

We now argue that any duality-violating 1-loop amplitude in BI theory can be set to zero by finite local counterterms {\em assuming} that this class of amplitudes have only standard factorizations into an on-shell tree amplitude and an on-shell 1-loop amplitude, i.e.~that they are tree-like. Consider first the 1-loop self-dual amplitude. Any factorization channel has vanishing residue since the  tree amplitude on the RHS is necessarily duality-violating and therefore vanishes:
\be\label{allplus1}
   \begin{tikzpicture}[scale=1, line width=1 pt]
	\draw [vector] (-1.5,1)--(0,0);
	\node at (-1.4,0.65) {$\vdots$};
	\draw [vector] (-1.7,-0.2)--(-0,0);
	\node at (-1.4,-0.45) {$\vdots$};
	\draw [vector] (-1.5,-1)--(0,0);
	\draw [vector] (-1.7,0.2)--(-0,0);
	\draw [vector] (0,0)--(3,0);
	\draw [vector] (3,0)--(4.5,1);
	\draw [vector] (3,0)--(4.7,0.2);
	\node at (4.4,-0.45) {$\vdots$};
	\draw [vector] (3,0)--(4.5,-1);
	\node at (4.4,0.65) {$\vdots$};
	\draw [vector] (3,0)--(4.7,-0.2);
	\draw[black,fill=white] (0,0) circle (3.8ex);
	\draw[black,fill=lightgray] (3,0) circle (3ex);
	\node at (0,0) {\small $\mathcal{A}^\text{1-loop}$};
	\node at (3,0) {\small $\mathcal{A}^\text{tree}$};
	\node [above left] at (-1.45,0.9) {$+$};
	\node [below left] at (-1.45,-0.85) {$+$};
	\node [above left] at (-1.55,0.015) {$+$};
	\node [below left] at (-1.55,-0.01) {$+$};
	\node [above right] at (4.4,0.9) {$+$};
	\node [below right] at (4.4,-0.85) {$+$};
	\node [above right] at (4.6,0.015) {$+$};
	\node [below right] at (4.6,-0.01) {$+$};
	\draw[red,dashed] (1.5,-1.1)--(1.5,1.1);
	\node at (0.98,0.35) {\footnotesize $h$};
	\node at (2.13,0.35) {\footnotesize $-h$};
   \end{tikzpicture}
   \raisebox{1.25cm}{~~~=~~~ 0\,.}
\ee
Because there are no factorization channels, the self-dual amplitude must be polynomial. 

Given tree-level duality, the 1-loop amplitudes with only a single negative helicity photon $\mathcal{A}_n(-+\dots+)$ factorize on simple poles into an $(n-2)$-point 1-loop self-dual amplitude $\mathcal{A}_{n-2}(+\dots+)$ and a 4-point tree amplitude $\mathcal{A}_4(++--)$:
\be\label{oneminus1}
   \begin{tikzpicture}[scale=1, line width=1 pt]
	\draw [vector] (-1.5,1)--(0,0);
	\node at (-1.4,0.65) {$\vdots$};
	\draw [vector] (-1.7,-0.2)--(-0,0);
	\node at (-1.4,-0.45) {$\vdots$};
	\draw [vector] (-1.5,-1)--(0,0);
	\draw [vector] (-1.7,0.2)--(-0,0);
	\draw [vector] (0,0)--(3,0);
	\draw [vector] (3,0)--(4.5,1);
	\draw [vector] (3,0)--(4.5,-1);
	\draw [vector] (3,0)--(4.7,0);
	\draw[black,fill=white] (0,0) circle (3.8ex);
	\draw[black,fill=lightgray] (3,0) circle (3ex);
	\node at (0,0) {\small $\mathcal{A}^\text{1-loop}_{n-2}$};
	\node at (3,0) {\small $\mathcal{A}_4^\text{tree}$};
	\node [above left] at (-1.45,0.9) {$+$};
	\node [below left] at (-1.45,-0.85) {$+$};
	\node [above left] at (-1.55,0.015) {$+$};
	\node [below left] at (-1.55,-0.01) {$+$};
	\node [above right] at (4.4,0.9) {$+$};
	\node [below right] at (4.4,-0.85) {$-$};
	\node at (5.,0.0) {$+$};
	\draw[red,dashed] (1.5,-1.1)--(1.5,1.1);
	\node at (0.98,0.35) {\footnotesize $+$};
	\node at (2.13,0.35) {\footnotesize $-$};
   \end{tikzpicture}.
\ee

The explicit expressions (\ref{SDallamp}) and (\ref{NSDallamp_grey}) for $\mathcal{A}^\text{1-loop}_n(++\dots+)$ and $\mathcal{A}^\text{1-loop}_n(-+\dots+)$ precisely have the above structure. In particular, the SD 1-loop amplitude (\ref{SDallamp}) is polynomial and the NSD 1-loop amplitude (\ref{NSDallamp_grey}) has precisely the factorization poles \reef{oneminus1} plus polynomial terms. 

This structure means that we can add local finite counterterms to the action so that the self-dual and next-to-self-dual 1-loop amplitudes are set to zero.  Here and below, this means that the amplitudes vanish up to order $\mathcal{O}(\eps)$ in dimensional regularization. Henceforth, let us suppose this has been done, i.e.
\be
  \label{1looptozeroA}
   \mathcal{A}^\text{1-loop}_n(++\dots+)=\mathcal{O}(\eps) 
   ~~~~\text{and}~~~~
   \mathcal{A}^\text{1-loop}_n(-+\dots+) =\mathcal{O}(\eps) \,.
\ee
Consider now the MHV amplitudes $\mathcal{A}^\text{1-loop}_n(--+\dots+)$. Electromagnetic duality of the BI tree amplitudes dictate that any factorization must involve either the self-dual or the next-to-self-dual 1-loop amplitudes; we write out the options explicitly
\be
\label{MHV1}
\begin{split}
&
   \begin{tikzpicture}[scale=1, line width=1 pt]
	\draw [vector] (-1.5,1)--(0,0);
	\node at (-1.4,0.65) {$\vdots$};
	\draw [vector] (-1.7,-0.2)--(-0,0);
	\draw [vector] (-1.5,-1)--(0,0);
	\draw [vector] (-1.7,0.2)--(-0,0);
	\draw [vector] (0,0)--(3,0);
	\draw [vector] (3,0)--(4.5,1);
	\draw [vector] (3,0)--(4.5,-1);
	\draw [vector] (3,0)--(4.7,0);
	\draw[black,fill=white] (0,0) circle (3.8ex);
	\draw[black,fill=lightgray] (3,0) circle (3ex);
	\node at (0,0) {\small $\mathcal{A}_{n-2}^\text{1-loop}$};
	\node at (3,0) {\small $\mathcal{A}_4^\text{tree}$};
	\node [above left] at (-1.45,0.9) {$+$};
	\node [below left] at (-1.45,-0.85) {$-$};
	\node [above left] at (-1.55,0.015) {$+$};
	\node [below left] at (-1.55,-0.01) {$+$};
	\node [above right] at (4.4,0.9) {$+$};
	\node [below right] at (4.4,-0.85) {$-$};
	\node at (5.,0.0) {$+$};
	\draw[red,dashed] (1.5,-1.1)--(1.5,1.1);
	\node at (0.98,0.35) {\footnotesize $+$};
	\node at (2.13,0.35) {\footnotesize $-$};
   \end{tikzpicture}
   ~~
%%%%
   \begin{tikzpicture}[scale=1, line width=1 pt]
	\draw [vector] (-1.5,1)--(0,0);
	\node at (-1.4,0.65) {$\vdots$};
	\draw [vector] (-1.7,-0.2)--(-0,0);
	\draw [vector] (-1.5,-1)--(0,0);
	\draw [vector] (-1.7,0.2)--(-0,0);
	\draw [vector] (0,0)--(3,0);
	\draw [vector] (3,0)--(4.5,1);
	\draw [vector] (3,0)--(4.5,-1);
	\draw [vector] (3,0)--(4.7,0);
	\draw[black,fill=white] (0,0) circle (3.8ex);
	\draw[black,fill=lightgray] (3,0) circle (3ex);
	\node at (0,0) {\small $\mathcal{A}_{n-2}^\text{1-loop}$};
	\node at (3,0) {\small $\mathcal{A}_4^\text{tree}$};
	\node [above left] at (-1.45,0.9) {$+$};
	\node [below left] at (-1.45,-0.85) {$+$};
	\node [above left] at (-1.55,0.015) {$+$};
	\node [below left] at (-1.55,-0.01) {$+$};
	\node [above right] at (4.4,0.9) {$+$};
	\node [below right] at (4.4,-0.85) {$-$};
	\node at (5.,0.0) {$-$};
	\draw[red,dashed] (1.5,-1.1)--(1.5,1.1);
	\node at (0.98,0.35) {\footnotesize $-$};
	\node at (2.13,0.35) {\footnotesize $+$};
   \end{tikzpicture}   
   \\
%%%%
  &\hspace{4cm}
   \begin{tikzpicture}[scale=1, line width=1 pt]
	\draw [vector] (-1.5,1)--(0,0);
	\node at (-1.4,0.65) {$\vdots$};
	\draw [vector] (-1.7,-0.2)--(-0,0);
	\draw [vector] (-1.5,-1)--(0,0);
	\draw [vector] (-1.7,0.2)--(-0,0);
	\draw [vector] (0,0)--(3,0);
	\draw [vector] (3,0)--(4.5,1);
	\draw [vector] (3,0)--(4.5,-1);
	\draw [vector] (3,0)--(4.7,0);
	\draw [vector] (3,0)--(4.6,-0.5);
	\draw [vector] (3,0)--(4.6,0.5);
	\draw[black,fill=white] (0,0) circle (3.8ex);
	\draw[black,fill=lightgray] (3,0) circle (3ex);
	\node at (0,0) {\small $\mathcal{A}_{n-4}^\text{1-loop}$};
	\node at (3,0) {\small $\mathcal{A}_6^\text{tree}$};
	\node [above left] at (-1.45,0.9) {$+$};
	\node [below left] at (-1.45,-0.85) {$+$};
	\node [above left] at (-1.55,0.015) {$+$};
	\node [below left] at (-1.55,-0.01) {$+$};
	\node [above right] at (4.4,0.9) {$+$};
	\node [above right] at (4.6,0.4) {$+$};
	\node at (5.,0.0) {$+$};
	\node [below right] at (4.6,-0.3) {$-$};
	\node [below right] at (4.45,-0.85) {$-$};
	\draw[red,dashed] (1.5,-1.1)--(1.5,1.1);
	\node at (0.98,0.35) {\footnotesize $+$};
	\node at (2.13,0.35) {\footnotesize $-$};
   \end{tikzpicture}   
\end{split}   
\ee
Since we have set the RHS 1-loop amplitudes to zero \reef{1looptozeroA}, there can be no contribution (at $\mathcal{O}(1)$) to the MHV 1-loop amplitude with $n>4$. It must therefore be polynomial and we can set it to zero with the help of local finite counterterms, i.e.~for $n> 4$
\be
  \label{1looptozeroB}
   \mathcal{A}^\text{1-loop}_n(--+\dots+) =\mathcal{O}(\eps) \,.
\ee
The 4-point MHV amplitude was calculated explicitly in Section \ref{sec:SD_MHV_relation} and as it has non-vanishing 4d cuts, it is UV divergent.  

It is clear that one can now proceed to check the factorization channels of the NMHV 1-loop amplitude and see that EM duality of the tree factor always requires the 1-loop sub-amplitude to be SD, NSD, or MHV. Since they vanish to order $\mathcal{O}(\eps)$, the NMHV 1-loop amplitude must be polynomial and we can proceed to set it to zero for $n>6$, 
\be
  \label{1looptozeroC}
   \mathcal{A}^\text{1-loop}_n(---+\dots+) =\mathcal{O}(\eps) \,.
\ee
For $n=6$, this argument fails because the 1-loop NMHV amplitude has non-vanishing 4d cuts and hence it is not a rational function. 

The argument extends in the obvious way to N$^k$MHV until the point where the duality-preserving amplitude with $n^+=n^-$ is reached. The duality-conserving amplitudes have non-vanishing 4d cuts and the factorization argument no longer applies. 

In the presence of any amount of supersymmetry, this argument continues to hold. In this case, \eqref{1looptozeroA} is modified to
\be
\label{SWI}
\mathcal{A}^\text{1-loop}_n(++\dots+)=0 
~~~~\text{and}~~~~
\mathcal{A}^\text{1-loop}_n(-+\dots+) =0 \,.
\ee 
For the factorization of MHV amplitudes in \eqref{MHV1}, the RHS 1-loop amplitudes vanish again, this time to any order in $\epsilon$ by \reef{SWI}. Thus the supersymmetric MHV amplitude can be removed by adding a finite local counterterm. This prediction is explicitly verified by the MHV 1-loop amplitudes \reef{N4all} with $n\ge 6$ in $\mathcal{N}=4$ DBI: they are indeed tree-like and the duality-violation can be removed by finite local counterterms. The argument extends as before to N$^k$MHV until $k=\frac{n}2-2$, which is the duality-preserving sector.

In conclusion, assuming that tree-like factorization of rational 1-loop amplitudes holds in BI theory, there exists a scheme (i.e.~a set of finite local counterterms) in which any duality-violating 1-loop amplitudes vanish. If true, this means that EM duality can be preserved at 1-loop order. As discussed in the Introduction, from one perspective this is surprising, since EM duality is an on-shell symmetry of the equations of motion rather than a traditional off-shell symmetry of the (covariant) action. 

Moreover, this analysis made no use of special properties of Born-Infeld beyond EM duality. It therefore applies to any 4d EM duality invariant model of nonlinear electrodynamics, such as the infinitely many models of the form (\ref{NLED}) which additionally satisfy the Gaillard-Zumino condition (\ref{GZ}).

%%%%%%%%%%%%%%%%%%%%%%%%%%%%%%

%%%%%%%%%%%%%%%%%%%%%%%%%%%%%%

%%%%%%%%%%%%%%%%%%%%%%%%%%%%%%
\section{Higher Derivative Corrections as a Double-Copy }
\label{sec:DChd}
%%%%%%%%%%%%%%%%%%%%%%%%%%%%%%
The KLT relations \cite{Kawai:1985xq} give closed-string tree amplitudes as sums of products of open-string tree amplitudes.
In the limit of infinite string tension ($\alpha' \to 0$), these relations reduce to field theory KLT relations that express (super)gravity tree amplitudes as sums of products of tree amplitudes of two (not necessarily the same) gauge theories.
In this section, we study the field theory KLT relations in the context of the double-copy \reef{BIequalschiPTYM} of Yang-Mills theory and $\chi$PT. In particular, we extend the double-copy relation to higher-derivative order with the purpose of examining the double-copy construction of the infinite and finite counterterms discussed in this paper. We also compare our results with the string effective action. 

\subsection{KLT Double-Copy}

The field theory KLT relation takes the form 
\begin{equation}
    \label{eq:KLT}
    \mathcal A_n^{A \otimes B} = \sum_{\alpha, \beta}  \mathcal A_n^A[ \alpha ] \, S [ \alpha | \beta ]\widetilde{\mathcal A_n^B} \,[ \beta ]\,,
\end{equation}
where $A$ and $B$ are theories with color-structure subject to constraints that we review below. The sum  over $\alpha$ and $\beta$ label sets of $( n - 3 ) !$ independent color orderings for the partial amplitudes.\footnote{We use square brackets for the arguments of a partial color-ordered amplitude and round brackets for the arguments of a full amplitude.} The KLT kernel $S [ \alpha | \beta ]$ is order $n-3$ in the Mandelstam variables in the field theory limit, but has an all-order in $\alpha'$ expression in string theory. 

The amplitudes in the theories $A$ and $B$ must satisfy a number of non-trivial conditions for \reef{eq:KLT} to hold; not all theories with color-structure can be double-copied. 
First of all, the amplitudes in the right-hand side of \eqref{eq:KLT} are color-ordered partial amplitudes. In the context of our discussion here, this  means that the full tree amplitudes admit an expansion of the form
\begin{equation}
    \label{eq:trace_decomposition}
    \mathcal A_n ( 1 2 \ldots n ) = \sum_{\sigma \in S_{n - 1}} \Tr \left ( t^{a_{\sigma_1}} t^{a_{\sigma_2}} \ldots t^{a_{\sigma_{n - 1}}} t^{a_n} \right ) \mathcal A_n [ {\sigma}, n ]\,,
\end{equation}
where $S_{n - 1}$ is the symmetric group of order $n - 1$.
In addition, the partial amplitudes must satisfy the following constraints, which reduces the number of independent partial amplitudes to $( n - 3 ) !$. These additional relations are
\begin{itemize}
    \item Cyclicity,
    \begin{equation}
        \label{eq:cyc}
        \mathcal A_n [ 1\, 2 \ldots n] = \mathcal A_n [ 2\, 3 \ldots n\, 1] = \mathcal A_n [ 3\, 4 \ldots n\, 1\, 2] = \ldots\,,
    \end{equation}
    as should be evident from the cyclicity of the trace of gauge group generators in \eqref{eq:trace_decomposition}.
    \item Kleiss-Kuijf (KK) relations \cite{Kleiss:1988ne},
    \begin{equation}
        \label{eq:kk}
        \mathcal A_n [ 1\, \beta\, 2\, \alpha] = ( - 1 )^{|\beta|} \sum_{\sigma \in \alpha  \shuffle \beta^T} \mathcal A_n [ 1\, 2\, \sigma]\,,
    \end{equation}
    where $| \beta |$ is the length of $\beta$ and $\alpha \shuffle \beta^T$ is the shuffle product of $\alpha$ and $\beta$ in reverse order.
    The special case of $\alpha$ being the empty list is the reflection relations.
    When $\beta$ has length 1, \reef{eq:kk} simply gives the $U(1)$ decoupling identity.
    \item 
    Fundamental Bern-Carrasco-Johansson (BCJ) identities \cite{Bern:2008qj},
    \begin{equation}
        \label{eq:bcj}
        \sum_{i = 2}^{n - 1} \left ( \sum_{j = 2}^i s_{jn} \right ) \mathcal A_n [ 1\, 2 \ldots i, n, i + 1 \ldots n - 1] = 0\,.
    \end{equation}
\end{itemize}

In the following, we restrict our study to 4-point amplitudes. For those the combined KK and BCJ relations give
\begin{equation}
    \label{eq:KK_BCJ_4pt}
    \mathcal A_4 [ 1 2 3 4 ] 
    = \frac t u \mathcal A_4 [ 1 2 4 3 ]
    = \frac t s \mathcal A_4 [ 1 3 2 4 ]
    = \frac t u \mathcal A_4 [ 1 3 4 2 ]
    = \frac t s \mathcal A_4 [ 1 4 2 3 ]
    = \mathcal A_4 [ 1 4 3 2 ]\,.
\end{equation}
All other orderings are cyclic permutations of the ones given here.
At 4-point, the explicit form of the field theory KLT relation  \eqref{eq:KLT} can be written
\begin{equation}
    \label{eq:KLTproduct_4pt}
    \mathcal A_4^{A \otimes B} ( 1 2 3 4 ) 
    \,=\, - \frac{1}{\Lambda^2} s \mathcal A_4^A [ 1 2 3 4 ] \mathcal A_4^B [ 1 2 4 3 ] 
    \,=\, - \frac{1}{\Lambda^2} \frac{s u}{t} \mathcal A_4^A [ 1 2 3 4 ] \mathcal A_4^B [ 1 2 3 4 ] \,,
\end{equation}
using in the second step from the identities \reef{eq:KK_BCJ_4pt}. Throughout this section we use the dimensionful scale $\Lambda$ in place of the physical couplings $g_{\text{YM}}$, $\lambda$ and $f_\pi$, these can restored straightforwardly at the end of the calculation.

We now turn to the study of higher-derivative corrections to the 4-point amplitudes of BI theory from the double-copy.

\subsection{Higher-Derivative Corrections to Born-Infeld}
\label{sec:higherD_KLT}

To extend the field theory KLT construction to include higher-derivative corrections, one must define what the double-copy means when higher-order terms are included. In a {\em top-down} approach, one uses the string theory prescription with $\alpha'$-corrections to both the KLT kernel and the BCJ relations. In a bottom-up approach, one parametrizes all possible higher-derivative corrections to the KLT kernel and BCJ relations and subject them to consistency conditions. In either approach, We find that for the 4-point calculations presented here, the absence of the first sub-leading correction means that we can work with the uncorrected field theory relations \reef{eq:KK_BCJ_4pt} and \reef{eq:KLTproduct_4pt} without any effect on the results presented here.\footnote{At higher-orders in the derivative expansion, one needs corrected versions of the BCJ relations and KLT kernel.}

With this setup, let us now consider the leading order higher-derivative corrections to the 4-point amplitudes of $\chi$PT and YM theory that satisfy the constraints of cyclicity, KK relations, and uncorrected BCJ relations \eqref{eq:KK_BCJ_4pt}. Additionally, we impose locality, unitarity, and the absence of higher-spin states in any  factorization channels. The details are presented in Appendix \ref{sec:YM_appendix}. We find:

\begin{itemize}
\item 
 $\chi$PT (also obtained in \cite{Elvang:2018dco}):
\begin{equation}
    \label{eq:chiPT_4pt}
    \mathcal A_4^{\chi\text{PT}} [ 1234 ] = \frac{1}{\Lambda^2} t \left ( c_0 + \frac{c_4}{\Lambda^4} \left ( s^2 + t^2 + u^2 \right ) + \frac{c_6}{\Lambda^6} s t u + \mathcal O ( \Lambda^{-8} ) \right )\,.
\end{equation}
\item 
  YM 
\bea
    \label{eq:YM_4pt_final}
    \mathcal A_4^\text{YM} [ 1^+ 2^+ 3^- 4^- ] &=& \frac{[ 12 ]^2 \langle 34 \rangle^2}{s u} \bigg ( \tilde a_0 + \frac{\tilde a_{4}}{\Lambda^4} t u + \frac{\tilde a_{6}}{\Lambda^6} s t u + \mathcal O ( \Lambda^{-8} ) \bigg )\,,
    \\[2mm]
    \label{eq:YM_pppm}
    \mathcal A_4^\text{YM} [ 1^+ 2^+ 3^+ 4^- ] &=& \frac{1}{\Lambda^2} \frac{[ 12 ]^2 [ 3 | p_1 | 4 \rangle^2}{s u} \bigg ( \tilde b_0 + \frac{\tilde b_6}{\Lambda^6} s t u + \mathcal O ( \Lambda^{-8} ) \bigg )\,,
    \\[2mm]
    \mathcal A_4^\text{YM} [ 1^+ 2^+ 3^+ 4^+ ] &=& \frac{\tilde c_2}{\Lambda^2} \frac{[ 12 ]^2 [ 34 ]^2 s + [ 13 ]^2 [ 24 ]^2 t + [ 14 ]^2 [ 23 ]^2 u}{s u} 
    \nonumber
    \\ 
    \label{eq:YM_4pt}
    &&+ \frac{\tilde c_6}{\Lambda^6} t \left ( [ 12 ]^2 [ 34 ]^2 + [ 13 ]^2 [ 24 ]^2 + [ 14 ]^2 [ 23 ]^2 \right ) + \mathcal O ( \Lambda^{-8} )\,.~~\phantom{3}
\eea
\end{itemize}
The leading contributions in $\mathcal A_4^\text{YM} [ 1^+ 2^+ 3^+ 4^+ ]$ and $\mathcal A_4^\text{YM} [ 1^+ 2^+ 3^+ 4^- ]$, as well as the sub-leading contribution of $\mathcal A_4^\text{YM} [ 1^+ 2^+ 3^- 4^- ]$ were also calculated in \cite{Dixon:1993xd}. An important feature of these results is that contributions of order $1/\Lambda^4$ are absent both in the $\chi$PT amplitude \reef{eq:chiPT_4pt} (where it is excluded by the BCJ constraints) and in the SD YM amplitude \reef{eq:YM_4pt} (where the one BCJ permissible term at this order has a pole corresponding to the exchange of a massless spin-3 particle, hence we exclude it). Substituting the above results in the double-copy formula \eqref{eq:KLTproduct_4pt} gives the following results for the amplitudes of Born-Infeld,
\begin{align}
    \label{eq:4pt_dc_ppmm}
    \mathcal A_4^\text{BI} ( 1^+ 2^+ 3^- 4^- ) 
    &= - \frac{[ 12 ]^2 \langle 34 \rangle^2}{\Lambda^4} \bigg ( \tilde a_0 + \frac{1}{\Lambda^4} \left ( 2 \tilde a_0 c_4 s^2 + \left ( \tilde a_4 - 2 \tilde a_0 c_4 \right ) t u \right ) \\ 
    &~~~~+ \frac{1}{\Lambda^6} \left ( \tilde a_6 + \tilde a_0 c_6 \right ) s t u + \mathcal O ( \Lambda^{-8} ) \bigg )\,,
    \nonumber
  \\[1mm]
\label{eq:4pt_dc_pppm}
    \mathcal A_4^\text{BI} ( 1^+ 2^+ 3^+ 4^- ) 
    &= - \frac{[ 12 ]^2 [ 3 | p_1 | 4 \rangle^2}{\Lambda^6} \bigg ( \tilde b_0 + \frac{\tilde b_0 c_4}{\Lambda^4} \left ( s^2 + t^2 + u^2 \right ) + \frac{\tilde b_0 c_6 + \tilde b_6}{\Lambda^6} s t u + \mathcal O ( \Lambda^{-8} ) \bigg )\,,
  \\[1mm]
    \label{eq:4pt_dc_pppp}
    \mathcal A_4^\text{BI} ( 1^+ 2^+ 3^+ 4^+ ) 
    &= - \frac{\tilde c_2}{\Lambda^6} \left ( [ 12 ]^2 [ 34 ]^2 s + [ 13 ]^2 [ 24 ]^2 t + [ 14 ]^2 [ 23 ]^2 u \right ) \\
    &~~~~
    - \frac{\tilde c_6 + 3 \tilde c_2 c_4}{\Lambda^{10}} s t u \left ( [ 12 ]^2 [ 34 ]^2 + [ 13 ]^2 [ 24 ]^2 + [ 14 ]^2 [ 23 ]^2 \right ) + \mathcal O ( \Lambda^{-12} )\,.\nonumber
\end{align}
At the leading order $\mathcal O ( \Lambda^{-4} )$, only the  duality-conserving MHV amplitude \eqref{eq:4pt_dc_ppmm} is non-zero and it matches the leading Born-Infeld amplitude if $\tilde a_0 = - 1$. This illustrates the idea of symmetry enhancement in the double-copy; we discuss this further in Section \ref{s:sg}.

The sub-leading contribution of the MHV amplitude \eqref{eq:4pt_dc_ppmm} is at the same order, $1/\Lambda^8$, as the 1-loop result of \eqref{ppmm1-loop}.
In fact, with the choice $c_4 = \frac{1}{32 \pi^2 \epsilon} \left ( \frac{7 N_v}{10} + \frac{N_f}{20} + \frac{N_s}{30} \right )$ and $\tilde a_4 = - \frac{1}{16 \pi^2 \epsilon} \left ( \frac{3 N_v}{10} - \frac{N_f}{20} - \frac{N_s}{30} \right )$ the two results match.
This means that the infinite counterterm necessary for the cancellation of the 1-loop UV divergence of $\mathcal A_4 ( 1^+ 2^+ 3^- 4^- )$ can be obtained from a double-copy construction.

The amplitudes of \eqref{eq:4pt_dc_pppm} and \eqref{eq:4pt_dc_pppp} have leading-order contributions $\mathcal O ( \Lambda^{-6} )$, which is higher than the leading tree-level BI amplitude but lower than any possible 1-loop contribution at $\mathcal O ( \Lambda^{-8} )$. The SD 4-point amplitude  \eqref{eq:4pt_dc_pppm} has no contribution at order $\mathcal O ( \Lambda^{-8} )$. This makes sense because there is no possible local finite operator at this order that can give rise to a NSD amplitude. This is also why the 1-loop NSD amplitude \eqref{genBIL} vanishes. 

It is very interesting that the $1/\Lambda^{8}$-term is missing from the SD amplitude \reef{eq:4pt_dc_pppp}. This stems from the lack of $1/\Lambda^4$ contributions in \reef{eq:chiPT_4pt} and \reef{eq:YM_4pt}, as well as the lack of $1/\Lambda^4$ terms in the corrected BCJ and KLT relations. However, such a $1/\Lambda^{8}$-term is needed to restore EM duality by cancelling the non-zero result of the SD 1-loop 4-point amplitude \reef{SD4pt}. 
Thus,  the finite quartic counterterm, required to restore electromagnetic duality at 1-loop level, {\em cannot} be obtained from a double-copy construction. We comment further on the potential implications of this result in Section \ref{open}.

Finally, let us briefly comment on 6-point. In the context of $\mathcal{N}=4$ DBI, the 6-point 1-loop MHV amplitude of the $n$-point result \reef{N4all} is polynomial and can be cancelled by a finite local counterterm that we have explicitly constructed using the KLT double-copy with higher-derivative corrections. This means that some counterterms needed to restore electromagnetic duality can be constructed via the double-copy with higher-derivatives while others cannot.

%%%%%%%%%%%%%%%%%%%%%%%%%%%%%%
\subsection{Comparison with the String Theory Effective Action}
\label{sec:stringEFT}
BI theory is the leading field-strength-dependent part of the open string effective action \cite{Fradkin:1985qd}. Higher-derivative corrections to this action have been obtained by considering the action at finite $\alpha'$, both in case of the bosonic open string \cite{Fradkin:1985qd} and the superstring \cite{Andreev:1988cb}. We now compare these results, with our construction of higher-derivative corrections to the BI model via the KLT product in Section \ref{sec:higherD_KLT}. This is done with the identification $\Lambda^{-2} = 2 \pi \alpha'$.

\paragraph{Bosonic open string.} The leading results at 4-point in the duality-violating sector, \eqref{eq:4pt_dc_pppm} and \eqref{eq:4pt_dc_pppp}, with choice of the Wilson coefficients, e.g.
\begin{equation}
\tilde b_0 = \frac 1 2
\qquad
\text{and}
\qquad
\tilde c_2 = - \frac 1 3\,
\end{equation}
agree with the bosonic open string action of \cite{Fradkin:1985qd}.

For the MHV amplitude, the KLT construction gives no $1/\Lambda^6$ term, so the leading order correction is order $1/\Lambda^8$. This is consistent with the open string effective action \cite{Fradkin:1985qd}.

\paragraph{Abelian Z-theory.}
 Bosonic open string amplitudes have been constructed via the KLT double-copy of abelian Z-theory and Yang-Mills with certain higher-derivative corrections \cite{Azevedo:2018dgo}. In order for this to be consistent with our construction in the previous section, the 4-point amplitude in $\chi$PT with higher derivative corrections \eqref{eq:chiPT_4pt} must reduce to the abelian Z-theory result \cite{Carrasco:2016ldy}. This is indeed the case upon choosing
\begin{equation}
c_0 = - \frac 1 2 \qquad c_4 = - \frac{1}{192} \qquad c_6 = \frac{3 \zeta_3}{16 \pi^3}\,.
\end{equation} 

\paragraph{Yang-Mills.}
Similarly, the 4-point Yang-Mills amplitudes \eqref{eq:YM_4pt_final}, \eqref{eq:YM_pppm} and \eqref{eq:YM_4pt} are found to agree with the corrected Yang-Mills amplitudes used in \cite{Azevedo:2018dgo} when
\begin{eqnarray}
\tilde a_0 = - 1 \qquad \tilde a_4 = - 1 \qquad \tilde b_0 = 1 \qquad \tilde b_6 = 0 \qquad \tilde c_2 = - \frac 2 3 \qquad \tilde c_6 = - 1\,.
\end{eqnarray}
Thus the results in Section \ref{sec:higherD_KLT} reduce to the KLT construction of bosonic string amplitudes in \cite{Azevedo:2018dgo} with a specific choice of free parameters.

\paragraph{Superstring.}
The SD and NSD sectors vanish in any supersymmetric context, so at 4-point we only have the MHV sector to compare with.
Furthermore, supersymmetry constrains the YM amplitude \eqref{eq:YM_4pt_final}, such that $\tilde a_4 = 0$.
This term in the amplitude comes from the $F^3$ correction of YM, which is only allowed in the absence of supersymmetry.
At the leading orders, the Born-Infeld MHV amplitude of \eqref{eq:4pt_dc_ppmm} agrees with the superstring MHV amplitude \cite{Andreev:1988cb} for the choice 
\begin{equation}
\tilde a_0 = - \frac 1 2 \qquad \qquad c_4 = \frac{1}{96}\,.
\end{equation}
There is no contribution at order $\alpha'^3 \sim \Lambda^{-6}$.

In \cite{Broedel:2013tta, Carrasco:2016ldy}, it was shown that the superstring amplitudes can be calculated as the KLT product of Yang-Mills theory and Z-theory. This too can be mapped onto the construction in Section \ref{sec:higherD_KLT} with a particular choice of Wilson coefficients.

%%%%%%%%%%%%%%%%%%%%%%%%%%%%%%
\section{Discussion}
\label{sec:disc}
%%%%%%%%%%%%%%%%%%%%%%%%%%%%%%

In this paper we have discussed various aspects the physics of the hidden electromagnetic duality symmetry of D3-brane worldvolume effective field theories. In particular, our focus has been on analyzing the consequences of such a symmetry for the physically observable S-matrix elements at sub-leading order in the EFT expansion. These sub-leading contributions are of two kinds: first, loop-level contributions from massless degrees-of-freedom present in the IR and, second, higher-derivative tree-level, or $\alpha'$ corrections, from integrating out massive states in the UV completion. Our work represents a first investigation of electromagnetic duality symmetry in this context and there are many avenues for further exploration. Here we first outline a number of open questions and then comment on similar duality symmetries in supergravity, both with respect to the double-copy and ideas of oxidation of symmetries from 3d to 4d. 

\subsection{Open Questions}
\label{open}

\noindent {\bf Loop-Level BCJ Double-Copy.}
In Section \ref{sec:DC4ptSD} we presented the first explicit example of a loop-level BCJ double-copy for a non-gravitational model. Using known color-kinematics duality satisfying numerators for self-dual Yang-Mills at 4-point, together with simple but non-color-kinematics duality manifesting numerators of $\chi$PT, we found that the result precisely matches the known self-dual Born-Infeld amplitude (\ref{SDallamp}) at all orders in the $\epsilon$-expansion. Since the loop-level double-copy \cite{Bern:2010ue} remains a conjecture, this successful matching can be taken as evidence that it extends to 1-loop in non-gravitational examples such as $\text{BI} = \text{YM}\otimes \chi\text{PT}$. It would be useful to have further examples, beyond 4-point and beyond 1-loop. Some of the relevant color-kinematics duality satisfying self-dual Yang-Mills numerators are known \cite{Bern:2013yya} but are quite complicated. A potentially simpler approach would be to construct color-kinematics duality satisfying numerators for $\chi$PT and form a double-copy with a simpler BCJ representation of the Yang-Mills amplitudes. 

\vspace{2mm}
\noindent {\bf 1-Loop Dimension-Shifting Relation.} 
Using the mysterious dimension-shifting relation between 1-loop integrands of self-dual Yang-Mills and the MHV sector of $\mathcal{N}=4$ super Yang-Mills together with the loop-level BCJ double-copy we have conjectured a representation of the integrand for the MHV sector of 1-loop $\mathcal{N}=4$ DBI (\ref{eq:dimshift_npt}) at all multiplicities. At $n=4$ the UV divergent result matches the physical amplitude we obtained from 4d cuts (and previously calculated in \cite{Shmakova:1999ai}) and for $n=6$ it agrees exactly with expressions recently derived from the dimensional reduction of M5-brane tree amplitudes in the forward limit \cite{Wen:2020qrj}. At higher-multiplicity our expression remains a conjecture, and it is important to verify its validity. With only even-point amplitudes and its lack of IR divergences at 1-loop, it is even possible that this simpler example could provide insight into the mechanism behind the dimension-shifting relations.  

\vspace{2mm}
\noindent {\bf IR Behavior and Tree-Like 1-loop Amplitudes.}  
 In Section \ref{sec:rational} we presented a conjecture, motivated by the analysis of 1-loop amplitudes in QCD \cite{Bern:1995ix}, that the absence of IR divergent Feynman integrals (as a consequence of the absence of 3-particle interactions) implies that purely rational 1-loop amplitudes in Born-Infeld are ``tree-like'' (in the sense defined in Section \ref{sec:rational}). As we showed in Section \ref{sec:rational}, if this conjecture is true, then in a 4d theory with classical electromagnetic duality, the 1-loop rational duality-violating amplitudes can \textit{always} be cancelled by adding finite local counterterms. Our explicit all-multiplicity results for the 1-loop SD and NSD amplitudes of pure BI theory and the MHV sector of $\mathcal{N}=4$ DBI are evidence of the conjecture. 
  If this limited conjecture is proven, an understanding of the structure of duality-violating amplitudes at 2-loops and beyond remains lacking. The status of electromagnetic duality symmetries of interacting quantum field theories at all-orders of perturbation theory is generally unknown.

\vspace{2mm}
\noindent {\bf Color-Kinematics vs.~Electromagnetic Duality?}
Finally, in Section \ref{sec:higherD_KLT} we have constructed the leading higher-derivative operators of Yang-Mills and $\chi$PT compatible with the tree-level KLT product. Contrary to the leading-order result, we find that generic double-copy constructible, higher-derivative operators do not conserve the duality charge. This is not surprising, duality invariance of the double-copy is not manifest at leading-order and is one of many examples of an unexpected \textit{symmetry enhancement} through the double-copy. It would be interesting if there was some better understanding of which Yang-Mills higher-derivative corrections lead to enhanced symmetries and which do not, and if there was some formulation of the double-copy that made this feature manifest. 

In some sense the result of the higher-derivative double-copy analysis is the worst of both worlds. The double-copy does not automatically generate duality satisfying amplitudes beyond leading-order, but neither does it generate all possible higher-derivative corrections to Born-Infeld. In particular, the matrix element corresponding to the finite local counterterm (\ref{4ptCT}) needed to restore duality invariance at 4-point in the self-dual sector of pure Born-Infeld cannot be generated. This suggests a potential tension between electromagnetic duality and color-kinematics duality at loop-level. 

The central problem is whether or not there exists a regularization scheme that is compatible with both notions of \textit{duality}. At present, the only known explicit examples of the loop-level BCJ double-copy make use of dimensional regularization and it is unclear if this is a strict requirement. Since electromagnetic duality is only a symmetry in exactly 4d, we would expect that it is broken in a generic dimensional scheme. If the duality symmetry is non-anomalous at loop-level then there are two logical possibilities: first, that there exists a \textit{special} dimensional regularization scheme which preserves the symmetry or second, that electromagnetic duality must be restored by including additional finite counterterm contributions, some of which may not be obtained via a double copy. Since there is presently no known duality preserving dimensional regularization scheme we will turn to an analysis of finite counterterms produced via the double-copy. 

For example, if we were to calculate a 2-loop amplitude, the 4-point self-dual amplitude must receive contributions from diagrams with an effective 1-loop topology and a single insertion of the counterterm (\ref{4ptCT}) of the form:
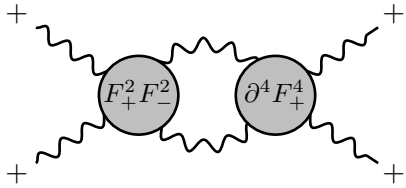
\begin{center}
   \begin{tikzpicture}[scale=0.9, line width=1 pt]
	\draw [vector] (-1.5,1)--(0,0);
	\draw [vector] (-1.5,-1)--(0,0);
	\draw [vector] (0,0) arc (160:20:1);
	\draw [vector] (0,0) arc (200:340:1);
	\draw [vector] (2,0)--(3.5,1);
	\draw [vector] (2,0)--(3.5,-1);
	\draw[black,fill=lightgray] (0,0) circle (3.5ex);
	\draw[black,fill=lightgray] (2,0) circle (3.5ex);
	\node at (0,0) {\small $F_+^2 F_-^2$};
	\node at (2,0) {\small $\partial^4 F_+^4$};
	\node [above left] at (-1.45,0.9) {$+$};
	\node [below left] at (-1.45,-0.85) {$+$};
	\node [above right] at (3.4,0.9) {$+$};
	\node [below right] at (3.4,-0.85) {$+$};
   \end{tikzpicture}
\end{center}
If the loop-level BCJ double-copy generates the complete loop amplitude it must include such contributions to the integrand. Here we find a problem. On any of the non-vanishing 4d cuts of this integrand, one half of the cut must be the local matrix element of the counterterm (\ref{4ptCT}) that we know cannot be generated as a tree-level BCJ double-copy. It is not clear that the existence of color-kinematics duality satisfying loop-level numerators strictly implies the existence of color-kinematics duality satisfying tree-level numerators on each cut. So this observation can at most be taken as indicative of a potential obstruction, rather than a firm argument. If such \textit{counterterm diagrams} cannot be generated by loop-level BCJ double-copy, and must be added to the amplitude by hand, then the advantages of the loop-level double-copy as a calculational tool are diminished. The calculation of the complete amplitude at $L$-loops requires a non-double-copy construction of an $L-1$-loop integrand with a counterterm insertion.

Conceptually this suggests two distinct definitions of Born-Infeld at the quantum level. One which is defined by the BCJ double-copy at all-loops but violates electromagnetic duality beginning at 1-loop, and another which is not a BCJ double-copy beyond tree-level but satisfies the selection rule (\ref{EMamps}) at all loop orders. At this stage such a dichotomy is just speculation with some supporting evidence. We do not even know if the double-copy construction is generally valid at higher-loops in Born-Infeld, or if we can continue to restore duality symmetry by adding further counterterms. As a further note, let us point out that the 6-point MHV amplitude $\mathcal{N}=4$ DBI can be set to zero by a finite local counterterm that we can in fact produce in a KLT-construction with higher-derivative corrections. It is not known if supersymmetry plays a role in this context. 
These questions clearly deserve further detailed study. 

%%%%%%%%%%%%%%%%%%%%%%%%%%%
\subsection{Supergravity: Double-Copy and Symmetry Oxidation}
\label{s:sg}

Supergravity with extended supersymmetry is an important class of models in which duality symmetries play a significant role. We review briefly how non-abelian R-symmetry in pure supergravity theories can be understood as duality symmetry, how it emerges as a symmetry enhancement in the double-copy of gauge theories, the partial breaking of duality-symmetry by higher-derivative terms, and we comment on the proposed oxidation of 3d symmetries to 4d supergravity.

\vspace{2mm}
\noindent {\bf Duality Symmetry in 4d Supergravity.} 
 The 4d super Poincar\'e algebra with $\mathcal{N}$ supercharges admits a maximal R-symmetry extension locally of the form $SU(\mathcal{N})_R\times U(1)_R$, which may or may not be realized in a given interacting model. In 4d extended supergravity, the graviton supermultiplet contains massless vector bosons, graviphotons $|\gamma^\pm\>$. The positive helicity states $|\gamma^+\rangle$ transform in an $\tbinom{\mathcal{N}}{2}$ dimensional representation of $SU(\mathcal{N})_R$ while the negative helicity states $|\gamma^-\rangle$ transform in the complex conjugate representation. Even if the full $SU(\mathcal{N})_R$ symmetry is realized in observables, it may not be manifest in the action. This depends on whether the $SU(\mathcal{N})_R$ representation is real or complex. In the former case it may lift to an off-shell symmetry acting on the field strength tensors. In the latter case, only the restriction to a real representation of a subgroup can possibly be realized off-shell and the remaining symmetries are seen only as duality invariance of the equations of motion. 
 Explicitly, for pure 4d supergravities, the graviphotons are in real representations for $\mathcal{N}=2,4$ and complex representations for $\mathcal{N}=5,6,8$. 
 
 Due to its self-interactions, the graviton cannot carry  $U(1)_R$ charge. Therefore if $q$ is the $U(1)_R$ charge of the supercharges $\mathcal{Q}$, a state schematically of the form $\mathcal{Q}^k|h^{+2}\rangle$ has $U(1)_R$  charge  $kq$. By CPT, the charges of the negative helicity multiplet must have the opposite signs of those for the positive helicity multiplet. If $q$ is non-zero, the graviphoton must be charged under the $U(1)_R$, but such a symmetry is not possible off-shell at the level of the action. Hence, if present, the $U(1)_R$ must act in pure supergravity as a duality symmetry. 
 
 $\mathcal{N}=8$ supergravity has a single CPT self-conjugate multiplet. The $U(1)_R$ charge of the state $|h^{-2}\rangle = \mathcal{Q}^8|h^{2}\rangle $ is $8q$, and since the graviton must be uncharged it must be that $q=0$. Hence, $\mathcal{N}=8$ supergravity cannot have a $U(1)_R$ duality symmetry and its  maximal R-symmetry group is $SU(8)_R$. As the graviphotons transform non-trivially under $SU(8)_R$, only an $SO(8)_R$ subgroup can be realized off-shell and the rest of the group acts as a non-abelian duality symmetry. 
 
\vspace{2mm}
\noindent {\bf Duality Symmetry in the Double-Copy.} 
The S-matrix of $\mathcal{N}=8$ supergravity can be obtained as the double-copy 
  \be
   \label{N8equalN4sq}
   \big ( \mathcal{N}=8 \text{ supergravity} \big ) = \big( \mathcal{N}=4 \text{ super Yang-Mills} \big) \otimes \big(\mathcal{N}=4 \text{ super Yang-Mills} \big)\,. 
\ee
In $\mathcal{N}=4$ super Yang-Mills, the vectors (gluons) are uncharged, so the $SU(4)_R$ is {\em not} an electromagnetic duality symmetry, but a regular global symmetry realized off-shell. In the double-copy \reef{N8equalN4sq}, the LHS directly inherits the $SU(4)_R\times SU(4)_R$ $R$-symmetry under which the graviphotons of the $\mathcal{N}=8$ supermultiplet are charged.  The KLT relations enhance $SU(4)_R\times SU(4)_R$ to the $SU(8)$ $R$-symmetry of the $\mathcal{N}=8$ supergravity tree amplitudes.

The 70 scalars of $\mathcal{N}=8$ supergravity are Goldstone modes of the spontaneous breaking of $E_{7(7)}$ to $SU(8)_R$.
 There are no states in the $\mathcal{N}=4$ SYM spectrum that have vanishing soft limits, and generically the individual terms in the KLT relations have $\mathcal{O}(1)$ soft limits. However, in the KLT sum, a cancellation takes place to ensure the vanishing soft limits of the Goldstone bosons. This is an example of a more general pattern: the double-copy of two states with soft behavior $\sigma_1$ and $\sigma_2$ respectively results in a state with soft behavior $\sigma \geq \sigma_1 + \sigma_2 +1$. This illustrates another type of symmetry enhancement in the double-copy.

Generic higher-derivative corrections to $\mathcal{N}=8$ supergravity need not respect the $SU(8)_R$. In tree-level string theory on $T^6$, the presence of the dilaton in the $\alpha'$-corrections breaks the $SU(8)_R$ to $SU(4)_R\times SU(4)_R$, for example through operators such as $\alpha'^3 e^{-6\phi}R^4$.
The $SU(4)_R\times SU(4)_R$ {\em is} a global symmetry of the tree-level closed string amplitudes with exclusively massless external states; open string amplitudes with 4d massless $\mathcal{N}=4$ states have $SU(4)_R$ symmetry and the closed string tree amplitudes inherit $SU(4)_R\times SU(4)_R$ via KLT \cite{Elvang:2010kc}.\footnote{The global symmetry  $SU(4)_R\times SU(4)_R$ is absent in the higher-genus superstring amplitudes (as expected in a theory of quantum gravity).} Note that this group with its 30 generators is larger than the 28-dimensional $SO(8)_R \subset SU(8)_R$ that can be realized off-shell. 

The existence of the $SU(4)_R\times SU(4)_R$ global symmetry in the 4d closed string tree amplitudes has an interesting origin. The 4d $\alpha'$-corrections explicitly break $E_{7(7)}$ to the $SO(6,6)$ of the $T^6$. The spontaneous breaking of 
$SO(6,6)$ to $SU(4)_R\times SU(4)_R$ produces 36 Goldstone bosons and these are exactly the scalars that in the double-copy are arise from the $6 \times 6$ SYM scalars in the 4d massless spectrum of the open string. For these 36 Goldstone scalars, the enhancement of the soft limits from  $\mathcal{O}(1)$ to vanishing takes place in KLT, but it does not happen for the dilaton and axion (obtained from the double-copy of opposite helicity gluons) or the remaining 32 scalars (from the double-copy of opposite helicity gluinos). This has been verified by explicit calculations \cite{Elvang:2010kc}.  

The two types of double-copy symmetry enhancements discussed here are in direct parallel with those studied in this paper for Born-Infeld theory. In the double-copy \reef{BIequalschiPTYM} of $\mathcal{N}=4$ DBI from $\mathcal{N}=4$ SYM times $\chi$PT, the $U(1)_R$ electromagnetic duality symmetry is emergent and so are the enhanced $\mathcal{O}(p^2)$  and $\mathcal{O}(|p\rangle)$ soft limits of the DBI scalars and Akulov-Volkov fermions respectively. That the double-copy provides these symmetry enhancements may appear like magic but given that the low-energy theories in these cases have these symmetries, it is also just a  necessary condition for the double-copy to work at all. 
With regard to higher-derivative corrections, we have seen that the double-copy generically gives duality-violating operators, however, the particular operator needed to restore duality invariance at 4-point 1-loop order in BI theory cannot be produced by the double-copy. 

There are other cases of emergent symmetries in double-copy constructions. In the double-copy 
\be
  \big(\text{gravity $\oplus$ dilaton $\oplus$ axion}\big) = 
  \text{YM} \otimes \text{YM}
\ee
the dilaton has an emergent $\mathbb{Z}_2$ \textit{dilaton parity}, which has no analogue in Yang-Mills. From one perspective this can be seen as an inherited property of the $SU(8)_R$ symmetry in the truncation of  $\mathcal{N}=8$ supergravity to gravity plus the dilaton-axion. 

A quite interesting case is that of $\mathcal{N}=4$ supergravity:
\be
  (\text{$\mathcal{N}=4$ supergravity}) = 
  \text{($\mathcal{N}=4$ SYM)} \otimes \text{YM}\,.
\ee
The two real scalars of $\mathcal{N}=4$ supergravity should be thought of as a dilaton-axion pair. They live in the $SU(1,1)/U(1)_R$ coset and as such they are the two Goldstone bosons of the breaking of $SU(1,1)$ to $U(1)_R$. The $U(1)_R$ emerges as a classical electromagnetic duality symmetry in the double-copy construction. At 
loop-level, the $U(1)_R$ was said to be anomalous \cite{Marcus:1985yy}, however, it was recently shown that the $U(1)_R$-violation can be removed by finite local counterterms \cite{Bern:2017rjw}, thus actually restoring the $U(1)_R$. This is very relevant for the 
study of the UV structure of $\mathcal{N}=4$ supergravity. There is a clear parallel to the possible removal of the BI and super-DBI electromagnetic duality violations at 1-loop level studied in this paper.  

For completeness, let us note that some recent discussions of S-duality in the context of the double-copy can be found in \cite{Huang:2019cja,
Alawadhi:2019urr}.

\vspace{2mm}
\noindent {\bf Dimensional Oxidation in Supergravity.}
In Section \ref{sec:EMduality} we proved that the 3d $U(1)$ symmetry of the 3d M2-brane theory oxidizes to electromagnetic $U(1)$ symmetry of the 4d D3-brane. It is interesting to consider the analogue of dimensional oxidation --- or absence thereof --- in extended supergravity. 

In this context the notion of dimensional oxidation has a longer history \cite{Marcus:1983hb, Cremmer:1999du, Keurentjes:2002xc}. As noted above, $\mathcal{N}=8$ supergravity in 4d has a sector of 70 scalars forming a sigma model on the coset space $E_{7(7)}/SU(8)_R$, with the additional 58 bosonic degrees-of-freedom transforming in linear representations of $SU(8)_R$. When dimensionally reduced to $d=3$, all 128 bosonic degrees-of-freedom become sigma model scalars parametrizing the coset space  $E_{8(8)}/SO(16)$ \cite{Marcus:1983hb}. There have recently appeared constructions, based on light-cone superspace, claiming that this enhanced symmetry oxidizes to $d=4$ at the level of the action \cite{Ananth:2017nbi}. Here we focus on a particular $U(1)$ subgroup and its manifestation in the physical S-matrix. The linearly realized $SO(16)$ in $d=3$ is rank 8, while the analogous linearly realized $SU(8)$ in $d=4$ is rank 7. There must therefore be an additional, conserved, additive charge generated by dimensional reduction. Indeed, this has been demonstrated explicitly using the CHY construction of the tree-level S-matrix of maximal supergravity; only amplitudes in the helicity conserving sector are non-vanishing when the external momenta are restricted to a 3d subspace \cite{Cachazo:2013iaa}. This would-be $U(1)$ symmetry would enhance the $SU(8)$ R-symmetry in $d=4$ to $U(8)$, analogously to the way the duality symmetry of the D3-brane enhances the $SU(4)$ to $U(4)$. But this symmetry is clearly broken (helicity non-conserving amplitudes in $d=4$ are generically non-vanishing), so why do we have oxidation in one case and not the other? In other words, why does the recursive argument given in Section \ref{sec:EMduality} fail for $\mathcal{N}=8$ supergravity? 

Certainly the recursive part of the argument remains valid, $n$-point tree-level graviton scattering amplitudes scale as $z^{2-2n}$ under a \textit{generic} holomorphic all-line shift. The failure is in the base case of the induction. The lowest multiplicity amplitudes in gravity are 3-point amplitudes of the form
\begin{equation}
\mathcal{A}^{\mathcal{N}=8\;\text{SUGRA}}_3\left(1_h^{+2}\,2_h^{+2}\,3_h^{-2}\right),
\end{equation}
which are non-zero despite being in the helicity non-conserving sector. Unlike the duality-violating 4-point amplitudes (\ref{4ptdv}), such an amplitude vanishes when all of the momenta are restricted to a 3d subspace \cite{Elvang:2015rqa}. If the would-be $U(1)$ symmetry (together with the rest of the $E_{8(8)}$ symmetry) does indeed oxidize to $d=4$ at the level of the action, then its implications for the physical S-matrix must be more subtle than the strong form of oxidation demonstrated for the D3-brane in Section \ref{sec:EMduality}. 

%%%%%%%%%%%%%%%%%%%%%%%%%%%%%%
\section*{Acknowledgements}
%%%%%%%%%%%%%%%%%%%%%%%%%%%%%%
We would like to thank Jacob Bourjaily, Dan Freedman, Simon Caron-Huot, Enrico Herrmann, Julio Parra-Martinez, Fei Teng, and Mukund Rangamani for useful discussions. 
This work was supported in part by the US Department of Energy under Grant No.~DE-SC0007859.
CRTJ was supported by a Leinweber Graduate Fellowship and a Rackham Predoctoral Fellowship from the University of Michigan. SP was supported by a Barbour Scholarship and Leinweber Summer Award from the University of Michigan. The work of MH is supported by the Knut and Alice Wallenberg Foundation under grant KAW 2018.0116.

%%%%%%%%%%%%%%%%%%%%%%%%%%%%%%
%%%%%%%%%%%%%%%%%%%%%%%%%%%%%%
\appendix

%%%%%%%%%%%%%%%%%%%%%%%%%%%%%%
\section{Momentum Shift for Recursion Relations}
\label{app:explicitshift}

In this appendix we provide an explicit $n\geq 6$ solution of the complex momentum shift presented in Section \ref{s:recrel4d3d}. In the spinor-helicity formalism, the shift takes the form \reef{eq:spinorshift} with $n$ parameters $a_i$ subject to the momentum conservation constraint 
\begin{equation}
     \sum_{i:\,h_i\ge 0}a_i |i][i|N + \sum_{j:\,h_j<0} a_j N|j\rangle \langle j|=0.
\end{equation}
At first glance this is 4 independent constraints, but since it actually involves a projection to a 3d subspace, momentum conservation in the $N^\mu$ direction is automatically satisfied. We therefore have to solve for 3 of the $a_i$ in terms of the remaining $n-3$. There are many ways to do this and each gives different, perfectly valid, momentum shifts. In this appendix, we present an explicit example of such a momentum shift. 

For every duality-violating amplitude with $n\geq 6$ there must be at least 4 particles of the same helicity.\footnote{For amplitudes with $n=4,5$ and fewer than 4 particles of the same helicity the steps to solve for $a_1,\; a_2, \; a_3$ are the same, but produce slightly different rational expressions.} Without loss of generality we  assume that the non-negative states have momentum labels $\{12 \cdots n_+\}$, with $n_+\geq 4$. We then project the above matrix equation onto spinor bases spanned by $[1|$ and $[2|$ on the left and $\bar N|3]$ and $\bar N|4]$ on the right, where $\bar N = N^\mu \bar\sigma_\mu$. For this, we need the following identity
\begin{equation}
    [ a|N \bar N|b] = -[ ab],
\end{equation}
which follows from the assumption that $N^\mu$ is a unit vector. The resulting system of equations is then
\begin{equation}
\renewcommand*{\arraystretch}{1.5}
    \begin{pmatrix}
    0 & [23][21] & 0 & [43][41] \\
    0 & [24][21] & [34][31] & 0 \\
    [13][12] & 0 & 0 & [43][42] \\
    [14][12] & 0 & [34][32] & 0 
    \end{pmatrix}
    \begin{pmatrix}
    a_1\\
    a_2\\
    a_3\\
    a_4
    \end{pmatrix}
    =
    \begin{pmatrix}
    -\sum\limits_{i=5}^{n_+}a_i[i3][i1]-\sum\limits_{j=n_++1}^{n}a_j[3|N|j\rangle[1|N|j\rangle\\
     -\sum\limits_{i=5}^{n_+}a_i[i4][i1]-\sum\limits_{j=n_++1}^{n}a_j[4|N|j\rangle[1|N|j\rangle\\
     -\sum\limits_{i=5}^{n_+}a_i[i3][i2]-\sum\limits_{j=n_++1}^{n}a_j[3|N|j\rangle[2|N|j\rangle\\
     -\sum\limits_{i=5}^{n_+}a_i[i4][i2]-\sum\limits_{j=n_++1}^{n}a_j[4|N|j\rangle[2|N|j\rangle
    \end{pmatrix}.
\end{equation}
This equation cannot be solved for $a_1,...,a_4$ since the matrix on the left-hand-side has vanishing determinant. This is a consequence of the fact that only three of the momentum conservation equations are linearly independent. So we can simply drop the last equation and solve the simpler non-degenerate system
\begin{equation}
\renewcommand*{\arraystretch}{1.5}
    \begin{pmatrix}
    a_1\\
    a_2\\
    a_3
    \end{pmatrix}
    =
    \left(
\begin{array}{ccc}
 0 & 0 & \frac{1}{[12][13]} \\
 \frac{1}{[21][23]} & 0 & 0 \\
 \frac{[24]}{[23][13][34]} & \frac{1}{[31][34]} & 0 \\
\end{array}
\right)
\begin{pmatrix}
    -\sum\limits_{i=4}^{n_+}a_i[i3][i1]-\sum\limits_{j=n_++1}^{n}a_j[3|N|j\rangle[1|N|j\rangle\\
     -\sum\limits_{i=4}^{n_+}a_i[i4][i1]-\sum\limits_{j=n_++1}^{n}a_j[4|N|j\rangle[1|N|j\rangle\\
     -\sum\limits_{i=4}^{n_+}a_i[i3][i2]-\sum\limits_{j=n_++1}^{n}a_j[3|N|j\rangle[2|N|j\rangle
    \end{pmatrix}.
\end{equation}
Since we want an all-line shift we choose $a_4,...,a_n$ to be any non-zero values. Importantly, the resulting shift \reef{eq:spinorshift} is a \textit{rational} function of the spinor brackets and therefore it is relatively simple to implement, both analytically and numerically. 

\section{Loop Integrals in Dimensional Regularization}
\label{app:dimregintegrals}
In this appendix, we present some useful 1-loop integrals in dimensional regularization.

\subsection{Scalar \texorpdfstring{$n$}{n}-gon in \texorpdfstring{$4+2p-2\epsilon$}{4 + 2p - 2 epsilon} Dimensions}

A well-known result is
\begin{equation} \label{standard}
   \int \frac{\text{d}^{4+2n-2\epsilon}l}{(2\pi)^{4+2n-2\epsilon}}\frac{1}{[l^2+\Delta]^n} = \frac{i}{(4\pi)^{n+2-\epsilon}}\frac{\Gamma\left(-2+\epsilon\right)}{(n-1)!}\Delta^{2-\epsilon}.
\end{equation}
Using this result with $\epsilon \to \epsilon' = \epsilon +n-p$ gives 
\begin{equation} \label{nonstandard}
\boxed{
   \int \frac{\text{d}^{4+2p-2\epsilon}l}{(2\pi)^{4+2p-2\epsilon}}\frac{1}{[l^2+\Delta]^n} = \frac{i}{(4\pi)^{p+2-\epsilon}}\frac{\Gamma\left(-2+\epsilon+n-p\right)}{(n-1)!}\Delta^{p+2-n-\epsilon}.}
\end{equation}
This result can also be obtained from Peskin \& Schroeder (A.44), but the form (\ref{nonstandard}) is directly useful for us.

%%%%%%%%%%%%%%%%%%%%%%%%
\subsection{\texorpdfstring{$n$}{n}-gon with \texorpdfstring{$p$}{p} Powers of \texorpdfstring{$l_{-2\epsilon}^2$}{l2(2 epsilon)}  in \texorpdfstring{$4-2\epsilon$}{4 - 2 epsilon} Dimensions}

For an $n$-gon diagram with massive momenta $P_i$ entering at each vertex, we now derive the following result:
\begin{equation}
\label{npinte}
\boxed{
\begin{aligned}
I_n^{(p)}&\equiv \int \frac{d^{4-2\epsilon}l}{(2\pi)^{4-2\epsilon}}\frac{( l_{-2\epsilon}^2)^{p}}{\prod_{i=1}^n\big(l-\sum_{j =1}^{2i} P_j\big)^2}
\\
&=
\frac{i}{(4\pi)^{2-\epsilon}}
        \frac{\Gamma(p-\epsilon)\Gamma(n-2-p+\epsilon)}{\Gamma(-\epsilon)}
        \int_0^1 dx_1\cdots dx_n \delta\left(\sum_{i=1}^nx_i-1\right)
        \Delta^{p+2-n-\epsilon}
\,,
\end{aligned}}
\end{equation}
where 
\begin{equation}
  \label{theDelta}
      \Delta = \sum_{i < j}^n x_i x_j \left ( \sum_{k = i + 1}^j P_k \right )^2\,.
\end{equation}
The dimension-shifting formula \cite{Bern:1995db} gives 
\begin{equation}
\begin{split}
	I_n^{(p)}
	&=
	(4\pi)^{p} \frac{\Gamma(p-\epsilon)}{\Gamma(-\epsilon)}\int \frac{d^{4+2p-2\epsilon}l}{(2\pi)^{4+2p-2\epsilon}}\frac{1}{\prod_{i=1}^n\big(l-\sum_{j =1}^{2i} P_j\big)^2}\,.
\end{split}
\end{equation}
Introducing Feynman parameters this can be rewritten as 
\begin{align}
\label{Inp1}
	I_n^{(p)}=(4\pi)^{p} \frac{\Gamma(p-\epsilon)}{\Gamma(-\epsilon)}(n-1)!\int_0^1 dx_1\cdots dx_n \delta\left(\sum_{i=1}^nx_i-1\right)\int \frac{d^{4+2p-2\epsilon}l}{(2\pi)^{4+2p-2\epsilon}}\frac{1}{\left(l^2+\Delta\right)^n}\,
\end{align}
where the compact from of $\Delta$ given in (\ref{theDelta}) was derived in Appendix C of \cite{Elvang:2019twd}. We evaluate the $l$-integral using (\ref{nonstandard}) to get
\begin{align}
	I_n^{(p)}
	&=
	\frac{i}{(4\pi)^{2-\epsilon}}
        \frac{\Gamma(p-\epsilon)\Gamma(n-2-p+\epsilon)}{\Gamma(-\epsilon)}
        \int_0^1 dx_1\cdots dx_n \delta\left(\sum_{i=1}^nx_i-1\right)
        \Delta^{p+2-n-\epsilon}	\,.
\end{align}
This completes the derivation of (\ref{npinte}).

When $p=n-2$, the integral (\ref{npinte}) is particularly simple and the Feynman parameter integral can be carried out trivially:
\begin{equation}
\begin{split}
I_n^{(n-2)}
&=
\frac{i}{(4\pi)^{2-\epsilon}}
        \frac{\Gamma(n-2-\epsilon)\Gamma(\epsilon)}{\Gamma(-\epsilon)}
        \int_0^1 dx_1\cdots dx_n \delta\left(\sum_{i=1}^nx_i-1\right)
        \Delta^{-\epsilon}
\,
\\
&\xrightarrow{\epsilon \to 0}
-\frac{i}{(4\pi)^{2}}
        \frac{\Gamma(n-2)}{(n-1)!}
        + \mathcal{O}(\epsilon)
\\
&=
-\frac{i}{(4\pi)^{2}}
        \frac{1}{(n-1)(n-2)}+ \mathcal{O}(\epsilon),
\end{split}
\end{equation}
i.e.~
\begin{equation}
\label{needthisint}
\boxed{
I_n^{(n-2)}=\int \frac{d^{4-2\epsilon}l}{(2\pi)^{4-2\epsilon}}\frac{( l_{-2\epsilon}^2)^{n-2}}{\prod_{i=1}^n\big(l-\sum_{j =1}^{2i} P_j\big)^2}
=
-\frac{i}{(4\pi)^{2}}
        \frac{1}{(n-1)(n-2)}+ \mathcal{O}(\epsilon)\,.
        }
\end{equation}
Another integral we repeatedly use in this paper is the bubble integral
\begin{equation}
\begin{aligned}
    I_2^{( p )} ( P^2 ) & = \int \frac{d^{4 - 2 \epsilon} l}{( 2 \pi )^{4 - 2 \epsilon}} \frac{( l_{-2 \epsilon}^2 )^p}{( l - \frac 1 2 P )^2 ( l + \frac 1 2 P )^2} \\
    & = \frac{i}{( 4 \pi )^{2 - \epsilon}} \frac{\Gamma ( p - \epsilon ) \Gamma ( - p + \epsilon )}{\Gamma ( - \epsilon )} \int d x \left ( P^2 x ( 1 - x ) \right )^{p - \epsilon}\,.
\end{aligned}
\end{equation}
In this simple case the Feynman integral can be evaluated exactly, giving
\begin{equation}
\label{eq:bubble_int}
    \boxed{
    I_2^{( p )} ( P^2 ) = \frac{i}{( 4 \pi )^{2 - \epsilon}} \frac{\Gamma ( p - \epsilon ) \Gamma ( - p + \epsilon ) \Gamma^2 ( 1 + p - \epsilon )}{\Gamma ( - \epsilon ) \Gamma ( 2 + 2 p - 2 \epsilon )} ( P^2 )^{p - \epsilon}.
    }
\end{equation}
%%

%%%%%%%%%%%%%%%%%%%%%%%%%%%%%%
\section{Born-Infeld Amplitudes}
%\label{sec:1loop}
\label{app:1loopcalc}

In this appendix, we calculate the 1-loop amplitudes with 4 external photons of pure Born-Infeld, using a combination of supersymmetric decomposition, traditional Feynman-diagrammatic methods, and generalized unitarity.

%%%%%%%%%%%%%%%%%%%%%%%%%%%%%%
\subsection{Tree-Level Amplitudes in BI and DBI}

We are primarily interested in amplitudes in pure BI theory, however, we  exploit computational tricks that use the supersymmetric completion of BI theory to supersymmetric DBI. The bosonic part of the $\mathcal{N}=4$ DBI action truncated to one complex scalar (i.e.~$\mathcal{N}=2$ DBI) takes the form
\begin{equation}
  S_{\text{D}3}[F_{\mu\nu},Z,\bar{Z}] = -\Lambda^4\int \text{d}^4x\sqrt{-\text{det}\left(\eta_{\mu\nu} +\Lambda^{-2} F_{\mu\nu} + \Lambda^{-4}\partial_{(\mu} Z\partial_{\nu)} \bar{Z}\right)}.
\end{equation}
The 4-scalar scattering processes are related by supersymmetry Ward identities 
\be
   \mathcal{A}_4\big(1_Z\, 2_Z\, 3_{\bar{Z}}\, 4_{\bar{Z}}\big)
   = \frac{\<12\>^2}{\<23\>^2}
       \mathcal{A}_4\big(1_\gamma^+\, 2_Z\, 3_\gamma^-\, 4_{\bar{Z}}\big)
   = \frac{\<12\>^2}{\<34\>^2}
   \mathcal{A}_4\left(1_\gamma^+\,2_\gamma^+\,3_\gamma^-\,4_\gamma^-\right),
\ee
to the 4-photon amplitude
\be
   \mathcal{A}_4\big(1_\gamma^+ \,2_\gamma^+\, 3_\gamma^-\, 4_\gamma^-\big) = [12]^2\langle 34\rangle^2\,.
\ee
All other 4-particle amplitudes vanish at leading order (tree-level), including those that violate the conservation of the duality charge (\ref{EMamps}):
\be
  \label{helviolA4}
  \mathcal{A}_4\big(1_\gamma^+\,2_\gamma^+\,3_\gamma^+\,4_\gamma^+\big)
  =
  \mathcal{A}_4\big(1_\gamma^-\,2_\gamma^+\,3_\gamma^+\,4_\gamma^+\big)
  =
  \mathcal{A}_4\big(1_\gamma^+\, 2_Z\, 3_\gamma^+\, 4_{\bar{Z}}\big) 
  = 0 \,. 
\ee
More generally, an $n$-particle process vanishes unless it involves an equal number of positive and  negative helicity photons (\ref{EMamps}).

\subsection{Feynman Rules}
\label{app:Feyn}

The vertex function of supersymmetric DBI with 2 external photons and 2 scalars is given by
\begin{equation}
    \label{eq:BIfeyn}
    \begin{aligned}
        V^{\mu\nu}(p_1,p_2, & p_3,p_4) \\
        = -2 i  \big [ & -p_3^\nu p_4^\mu (p_1\cdot p_2) - p_3^\mu p_4^\nu (p_1\cdot p_2) + p_2^\mu p_4^\nu (p_1\cdot p_3) + p_2^\mu p_3^\nu (p_1\cdot p_4) + p_1^\nu p_4^\mu (p_2\cdot p_3) \\
        & + p_1^\nu p_3^\mu (p_2\cdot p_4) - (p_1\cdot p_4)(p_2\cdot p_3) g^{\mu\nu} -(p_1\cdot p_3)(p_2\cdot p_4)g^{\mu\nu} \big ]\\
        - 2 i \big[ & - p_1^\nu p_2^\mu (p_3\cdot p_4) +  (p_1\cdot p_2)(p_3\cdot p_4)g^{\mu\nu} \big]\,.
    \end{aligned}
\end{equation}
Notice that the 2 expressions in square brackets are individually gauge invariant and correspond to the 2 independent (up to integration by parts and application of the equations of motion) Lagrangian operators with the desired mass dimension and external states.
The polarization vectors in spinor-helicity variables are 
\be
  \label{polvec}
  \eps_{i+}^\mu = -\frac{\<q_i | \bar{\sigma}^\mu | i ]}{\sqrt{2} \<q_i i\>}
  ~~~~\text{and}~~~~
  \eps_{i-}^\mu = -\frac{\< i | \bar{\sigma}^\mu | q_i ]}{\sqrt{2} [i q_i]},
\ee 
where $q_i$ are arbitrary reference spinors. 

%%%%%%%%%%%%%%%%%%%%%%%%%%%%%
\subsection{Duality-Violating 1-loop 4-Point Amplitudes}
\label{s:hviolating4pt}
%%%%%%%%%%%%%%%%%%%%%%%%%%%%%
For the computation of 1-loop 4-point amplitudes in the following, we are going to use a combination of on-shell techniques and Feynman rules. The  4-vector Feynman rule \eqref{eq:BIfeyn} is rather involved, but we can use a trick to calculate the duality-violating 1-loop amplitudes using the 
 (vector)$^2$-(scalar)$^2$ vertex instead.  The reason is that in any supersymmetrization of BI theory, the $++++$ and $+++-$ amplitudes \reef{helviolA4} must vanish at any loop-order. Thus, in $\mathcal{N}=1$ BI theory, the contributions from the vector in the loop must cancel against the contribution from the fermion, i.e. 
\be
  \mathcal{N}=1:~~~~\mathcal{A}_4^\text{V}(+++\pm)+\mathcal{A}_4^\text{F}(+++\pm) = 0 \,.
\ee  
Similarly, in $\mathcal{N}=2$ BI theory we have 
\be
  \mathcal{N}=2:~~~~\mathcal{A}_4^\text{V}(+++\pm)+2\mathcal{A}_4^\text{F}(+++\pm)
  +\mathcal{A}_4^\text{S}(+++\pm) = 0.
\ee  
Here the superscripts indicate contributions from a vector (V), Weyl fermion (F), or complex scalar (S) running in the loop. 
It follows that the pure BI result must equal the result from only scalars running in the loop:
\be
 \label{1loopSYSY}
 \mathcal{A}_4^\text{V}(+++\pm) = -\mathcal{A}_4^\text{F}(+++\pm)
 = \mathcal{A}_4^\text{S}(+++\pm) \,.
\ee
We exploit this in the following two subsections. 

Due to the form of the interaction, $\partial Z \partial \bar{Z} F F$, the general (vector)$^2$-(scalar)$^2$ vertex rule is proportional to two powers of the scalar-momenta, which we denote $l_1$ and $l_2$ in anticipation of the loop-calculations below.  The general Feynman rule is given in Appendix \ref{app:Feyn} and is gauge-invariant.

For later convenience, we write the vertex rules with on-shell photons and off-shell scalars compactly, using the spinor-helicity formalism, as
\be
   V(1_\gamma,2_\gamma, {l_1}_Z, {l_2}_{\bar{Z}}) 
   = V_{\mu\nu}(1_\gamma,2_\gamma)\, l_1^\mu l_2^\nu\,,
\ee
where
\begin{equation}
\label{Vppm}
\begin{aligned}
    V_\mn ( 1_\gamma^+, 2_\gamma^+ ) & = - i [ 12 ]^2 \eta_\mn - \frac 1 2 i \frac{\sq{12}}{\ang{12}} \operatorname{Tr} ( p_1 \bar \sigma_\mu p_2 \bar \sigma_\nu + p_1 \bar \sigma_\nu p_2 \bar \sigma_\mu ) + 2 i \frac{\sq{12}}{\ang{12}} (p_{1\mu} p_{2\nu} + p_{1\nu} p_{2\mu} ) \\
    V_\mn ( 1_\gamma^+, 2_\gamma^- ) & = i \langle 2 | \bar \sigma_\mu | 1 ] \langle 2 | \bar \sigma_\nu | 1 ].
\end{aligned}
\end{equation}
Since there are no 3-point interactions in BI theory, only bubble diagrams contribute to the 1-loop 4-point amplitudes. We use the following convenient parametrization of the loop momenta
\be
\label{scalarloopdiagram}
\raisebox{-2cm}{{\begin{tikzpicture}[scale=2, line width=1 pt]
\draw [vector] (-1,1/2)--(0,0);
\draw [vector] (-1,-1/2)--(0,0);
\draw [scalar] (0,0) arc (180:0:1/2);
\draw [scalar] (1,0) arc (0:-180:1/2);
\draw [vector] (1,0)--(2,1/2);
\draw [vector] (1,0)--(2,-1/2);
\draw [->] (0.35,0.7)--(0.65,0.7);
\draw [->] (0.35,-0.7)--(0.65,-0.7);
\draw [->] (-0.4,0.4)--(-0.7,0.55);
\draw [->] (-0.4,-0.4)--(-0.7,-0.55);
\draw [->] (1.4,0.4)--(1.7,0.55);
\draw [->] (1.4,-0.4)--(1.7,-0.55);
\node at (-0.45,0.65) {$p_1$};
\node at (-0.45,-0.65) {$p_2$};
\node at (1.5,0.65) {$p_3$};
\node at (1.5,-0.65) {$p_4$};
\node at (0.5,0.9) {$l_1 = l-\frac{1}{2}K$};
\node at (0.5,-0.9) {$l_2 = -l-\frac{1}{2}K$};
\end{tikzpicture}}}
\ee
where $K^\mu = p_1^\mu+p_2^\mu$ and similarly for other permutations of the external lines. 
The diagram in \reef{scalarloopdiagram} is given by the integral 
\be
  \begin{split}
  I_4^S
  &=\int \frac{\text{d}^dl}{(2\pi)^d}
  \frac{{V}\big(1_\gamma,2_\gamma,l_1,l_2\big){V}\big(3_\gamma,4_\gamma,-l_1,-l_2\big)}
  {\big(l+\frac{1}{2}K\big)^2\big(l-\frac{1}{2}K\big)^2}
  \\
  &= {V}_{\mu\nu}(1_\gamma,2_\gamma)\,
  {V}_{\rho\sigma}(3_\gamma,4_\gamma)
  \int \frac{\text{d}^dl}{(2\pi)^d} 
  \frac{l_1^\mu l_2^\nu l_1^\rho l_2^\sigma}{{\big(l+\frac{1}{2}K\big)^2\big(l-\frac{1}{2}K\big)^2}}
  \,,
  \end{split}
\ee
where the superscript $S$ is used to denote that this integral corresponds to the $s$-channel contribution. Since $l_1 = l-\frac{1}{2}K$ and $l_2 = -l-\frac{1}{2}K$, it is very useful to observe that 
\begin{itemize}
\item integrals with odd powers of loop momentum $l$ vanish, and
\item the photon-scalar vertices \reef{Vppm} have the property that 
\be
  \label{KdotV}
  K^\mu\, V_{\mu\nu}(1_\gamma,2_\gamma)  = K^\nu\, V_{\mu\nu}(1_\gamma,2_\gamma) = 0\,
  ~~~~\text{for}~~~K^\mu = p_1^\mu +p_2^\mu\,,
\ee
and similarly for the 34-vertex since $K^\mu = p_1^\mu +p_2^\mu = - p_3^\mu -p_4^\mu$. 
\end{itemize}
This means that in the integral numerator we can replace $l_1^\mu l_2^\nu l_1^\rho l_2^\sigma$  by $l^\mu l^\nu l^\rho l^\sigma$.

In Appendix \ref{app:intred} we use the Passarino-Veltman integral-reduction method to compute the two tensor integrals that we need in this paper, namely
\be 
\label{2tensorInt}
  \int \frac{\text{d}^dl}{(2\pi)^d} \frac{l^\mu l^\nu}{\big(l+\frac{1}{2}K\big)^2\big(l-\frac{1}{2}K\big)^2}
  = -\frac{1}{4(d-1)}\left[ K^2\eta^{\mu\nu} - K^{\mu}K^{\nu} \right]\, I_2(K^2)\,,
\ee
and
\bea \label{4tensorInt}
  \int \frac{\text{d}^dl}{(2\pi)^d} \frac{l^\mu l^\nu l^\rho l^\sigma}{\big(l+\frac{1}{2}K\big)^2\big(l-\frac{1}{2}K\big)^2}&&
  \\ 
  &&~~\hspace{-4.5cm} \nonumber
  = \frac{1}{16(d^2-1)}\left[(K^2)^2\eta^{(\mu\nu}\eta^{\rho\sigma)} 
  -K^2 K^{(\mu}K^\nu \eta^{\rho\sigma)} 
  +3 K^\mu K^\nu K^\rho K^\sigma\right]\, I_2(K^2)\,.
\eea
Here $I_2$ is the scalar-bubble integral defined as 
\be
  \label{masterbubble}
  I_2(K^2) \equiv 
  \int \frac{\text{d}^dl}{(2\pi)^d}\frac{1}{\big(l+\frac{1}{2}K\big)^2\big(l-\frac{1}{2}K\big)^2}
  =  \frac{i}{16\pi^2}\left[\frac{1}{\epsilon}-\log(K^2)\right] + \mathcal{O}(1)\,.
\ee
The property \reef{KdotV} therefore implies the very simple result of the scalar-loop integral:
\be
  \label{masterIS}
  \begin{split}
  I_4^S
  &= 
  \frac{(K^2)^2}{16(d^2-1)} 
    {V}_{\mu\nu}(1_\gamma,2_\gamma)\,{V}_{\rho\sigma}(3_\gamma,4_\gamma)
  \,\eta^{(\mu\nu}\eta^{\rho\sigma)}\, I_2(K^2)
  \\
  &=
  \frac{s^2}{240}
  \Big(
  V_\mu{}^{\mu}(1_\gamma,2_\gamma)\,
  V_\nu{}^{\nu}(3_\gamma,4_\gamma)
  +
  2 {V}_{\mu\nu}(1_\gamma,2_\gamma)\,
  {V}^{\mu\nu}(3_\gamma,4_\gamma)
  \Big)
  \, I_2(s)
  \,,
  \end{split}
\ee
where we have also used the symmetry of $V_\mn$ under the exchange $\mu \leftrightarrow \nu$.
Now it is a simple task to compute the desired duality-violating 1-loop amplitudes.

${\bullet }$ For the next-to-self-dual case, we have $V_\nu{}^{\nu}(3_\gamma^+,4_\gamma^-) = 0$, by the identity $\langle i | \bar \sigma^\mu | j ] \langle k | \bar \sigma_\mu | l ] = 2 \ang{ik} \sq{jl}$.
It is also easy to show that the second term in \reef{masterIS} vanishes since
\be
  \begin{split}
  & {V}_{\mu\nu}(1_\gamma^+,2_\gamma^+)\,
  {V}^{\mu\nu}(3_\gamma^+,4_\gamma^-) \\
  &~~
  = 
  - \sq{12}^2 \langle 4 | \bar \sigma^\mu | 3 ] \langle 4 | \bar \sigma_\mu | 3 ] - \frac{\sq{12}}{\ang{12}} \langle 1 | \bar \sigma^\mu | 2 ] \langle 2 | \bar \sigma^\nu | 1 ] \langle 4 | \bar \sigma_\mu | 3 ] \langle 4 | \bar \sigma_\nu | 3 ] + 4 \frac{\sq{12}}{\ang{12}} \langle 4 | p_1 | 3 ] \langle 4 | p_2 | 3 ] \\
  &~~ = 0\,.
  \end{split}
\ee

Hence the entire $s$-channel diagram vanishes: $I_4^S(+++-) = 0$. Since the $t$- and $u$-channel diagrams are simply permutations of the $+$-lines, we conclude that 
\be
  \label{eq:BI_1loop_pppm}
  \boxed{
  \mathcal{A}_4^\text{1-loop BI}\left(1_\gamma^+ 2_\gamma^+ 3_\gamma^+ 4_\gamma^-\right) = 0\,.
  }
\ee

${\bullet }$ For the self-dual case, let us define
\be
  Q_{\mu\nu\rho\sigma} \equiv {V}_{\mu\nu}(1_\gamma^+,2_\gamma^+)\,{V}_{\rho\sigma}(3_\gamma^+,4_\gamma^+)\,.
\ee

The external momenta and polarizations live in 4d, while the metric $\eta_{\mu\nu}$ arises from the loop-reduction and is therefore in $d= 4-2\eps$ dimensions, so $\eta_{\mu\nu}\eta^{\mu\nu} =  4-2\eps$. Using \eqref{Vppm}, we find
\be
   \label{Q1}
   Q_\mu^{~\,\mu}{}_\nu^{~\,\nu} = 
   \sq{12}^2 \sq{34}^2 
   \big[ 4 - \eta_{\mu\nu}\eta^{\mu\nu}\big]^2 
   =  4[12]^2 [34]^2\, \eps^2\,,
\ee
and
\be
   \label{Q2andQ3}
  {Q_{\mu\nu}}^{\mu\nu}= {Q_{\mu\nu}}^{\nu\mu}  
  = - 2 \eps\, [12]^2[34]^2 \,.
\ee
Using the results \reef{Q1} and \reef{Q2andQ3}, we arrive at 
\be
  \label{sumQs}
  Q_\mu^{~\,\mu}{}_\nu^{~\,\nu} + {Q_{\mu\nu}}^{\mu\nu} +{Q_{\mu\nu}}^{\nu\mu} 
  = 4[12]^2[34]^2 
  \Big[ 
    - \eps + \eps^2
  \Big]  \,.
  \ee
Note that there are no $\mathcal{O}(1)$ terms in \reef{sumQs}; this means that in the amplitude \reef{masterIS} the $1/\eps$ divergent terms and finite logarithms from \reef{masterbubble} vanish for the diagram, as expected. We are left with finite rational terms  generated from $\epsilon/\epsilon$ anomalies. The $s$-diagram \reef{masterIS}  simply evaluates to
\begin{align}
  I_4^S &=\frac{s^2}{240} \times 4[12]^2[34]^2 
  \Big[ 
    - \eps + \eps^2
  \Big]  \times I_2(s)
  \,\\
  & = -\frac{i}{960\pi^2}[12]^2[34]^2 s^2 + O(\eps)\,.
\end{align}
Summing over the three diagrams gives the full amplitude 
 \begin{equation} \label{BIloop}
\mathcal{A}_4^\text{1-loop BI}\big(1_\gamma^+ 2_\gamma^+ 3_\gamma^+ 4_\gamma^+\big) =  -\frac{i}{960\pi^2}\Big([12]^2[34]^2 s^2+[13]^2[24]^2 t^2 +[14]^2[23]^2 u^2\Big).
\end{equation}
Note that we automatically have a more general result too: using \reef{1loopSYSY} we can conclude that in a generalized BI theory with $N_v$ vectors, $N_f$ Weyl fermions, and $N_s$ complex scalars, we have 
 \begin{equation} \label{genBIloop}
 \begin{aligned}
 \mathcal{A}_4^\text{1-loop gen BI}\big(1_\gamma^+ 2_\gamma^+ 3_\gamma^+ 4_\gamma^+\big) 
 &=  
 \frac{-i}{960\pi^2} (N_v - N_f + N_s) \\
 & \hspace{2em} \times \Big([12]^2[34]^2 s^2+[13]^2[24]^2 t^2 +[14]^2[23]^2 u^2\Big).
 \\
 \mathcal{A}_4^\text{1-loop gen BI}\big(1_\gamma^+ 2_\gamma^+ 3_\gamma^+ 4_\gamma^-\big) 
 &=
 0\,.
 \end{aligned}
\end{equation}
Note that the self-dual amplitude vanishes when $(N_v - N_f + N_s)$ vanishes, as expected, for $\mathcal{N}=1,2,4$ supersymmetry.  

If we carry out the same calculation with external helicities  $++--$, we would compute the contribution of a complex scalar in the loop of a 1-loop MHV amplitude. But that does not help us compute the 1-loop MHV amplitude in pure BI theory, since the supersymmetry trick we used to calculate the duality-violating amplitudes does not apply. Instead, we use the fact that the MHV amplitude is cut-constructible and compute it using unitarity.

%%%%%%%%%%%%%%%%%%%%%%%%%%%%%%
\subsection{1-loop MHV Amplitude without Supersymmetry}
\label{s:MHV4BI}
%%%%%%%%%%%%%%%%%%%%%%%%%%%%%%

We compute the divergent part of the BI 1-loop MHV amplitude using unitarity. As a check, we also compute the MHV amplitude in a theory with $N_v$ vectors, $N_f$ fermions, and $N_s$ complex scalars. The loop-integrand is ambiguous by additive terms that integrate to zero. We construct a representative of the loop-integrand that factorizes correctly on all cuts. We then use integral-reduction to relate the result to the scalar-bubble integral. For the 1-loop amplitude $\mathcal{A}_4^\text{1-loop BI}(1^+_\gamma,2^+_\gamma,3^-_\gamma,4^-_\gamma)$ in pure BI theory, there are two distinct cuts to consider:
\be
\label{twocuts}
\raisebox{-2cm}{{\begin{tikzpicture}[scale=2, line width=1 pt]
\draw [vector] (-1,1/2)--(0,0);
\draw [vector] (-1,-1/2)--(0,0);
\draw [vector] (0,0) arc (180:0:1/2);
\draw [vector] (1,0) arc (0:-180:1/2);
\draw [vector] (1,0)--(2,1/2);
\draw [vector] (1,0)--(2,-1/2);
\draw [scalarnoarrow,color=red] (1/2,0.64)--(1/2,-0.64);
\draw [->] (0.35,0.7)--(0.65,0.7);
\draw [->] (0.35,-0.7)--(0.65,-0.7);
\draw [->] (-0.4,0.4)--(-0.7,0.55);
\draw [->] (-0.4,-0.4)--(-0.7,-0.55);
\draw [->] (1.4,0.4)--(1.7,0.55);
\draw [->] (1.4,-0.4)--(1.7,-0.55);
\node at (-0.45,0.65) {$p_1$};
\node at (-0.45,-0.65) {$p_2$};
\node at (1.5,0.65) {$p_3$};
\node at (1.5,-0.65) {$p_4$};
\node at (0.5,0.9) {$l_1 = l-\frac{1}{2}K$};
\node at (0.5,-0.9) {$l_2 = -l-\frac{1}{2}K$};
\node at (-1.1,0.55) {$+$};
\node at (-1.1,-0.55) {$+$};
\node at (2.1,0.55) {$-$};
\node at (2.1,-0.55) {$-$};
\end{tikzpicture}}}
~\hspace{1.0cm}~
\raisebox{-2cm}{{\begin{tikzpicture}[scale=2, line width=1 pt]
\draw [vector] (-1,1/2)--(0,0);
\draw [vector] (-1,-1/2)--(0,0);
\draw [vector] (0,0) arc (180:0:1/2);
\draw [vector] (1,0) arc (0:-180:1/2);
\draw [vector] (1,0)--(2,1/2);
\draw [vector] (1,0)--(2,-1/2);
\draw [scalarnoarrow,color=red] (1/2,0.64)--(1/2,-0.64);
\draw [->] (0.35,0.7)--(0.65,0.7);
\draw [->] (0.35,-0.7)--(0.65,-0.7);
\draw [->] (-0.4,0.4)--(-0.7,0.55);
\draw [->] (-0.4,-0.4)--(-0.7,-0.55);
\draw [->] (1.4,0.4)--(1.7,0.55);
\draw [->] (1.4,-0.4)--(1.7,-0.55);
\node at (-0.45,0.65) {$p_1$};
\node at (-0.45,-0.65) {$p_3$};
\node at (1.5,0.65) {$p_2$};
\node at (1.5,-0.65) {$p_4$};
\node at (0.5,0.9) {$l_1' = l-\frac{1}{2}K'$};
\node at (0.5,-0.9) {$l_2' = -l-\frac{1}{2}K'$};
\node at (-1.1,0.55) {$+$};
\node at (-1.1,-0.55) {$-$};
\node at (2.1,0.55) {$+$};
\node at (2.1,-0.55) {$-$};
\end{tikzpicture}}}
\ee
In the first diagram $K=p_1 + p_2$ and in the second $K'=p_1 + p_3$. In addition, one needs the $3 \lra 4$ permutation of the second diagram. The cut-constructible part of the amplitude is then the sum of the $s$-, $t$-, and $u$-channel cuts $C_s$, $C_t$, and $C_u$. Note that each diagram comes with a symmetry factor of $1/2$.

The cut puts the internal lines $l_1$ and $l_2$ on-shell and each diagram factorizes into a product of tree-amplitudes for each possible on-shell state that can appear in the loop. For the $s$-channel diagram we have
\be
  \begin{split}
  C_s &= 
  \mathcal{A}_4^\text{tree}\big(1^+_\gamma\,2^+_\gamma\,{l_1}^-_\gamma\,{l_2}^-_\gamma\big)
  \mathcal{A}_4^\text{tree}\big(-{l_1}^+_\gamma\,-{l_2}^+_\gamma\,3^-_\gamma\,4^-_\gamma\big)
  \\
  &= [12]^2 \< l_1 l_2 \>^2 [l_1 l_2]^2 \<34\>^2
  \\
  &= [12]^2  \<34\>^2 s^2, 
  \end{split}
\ee
where we used that\footnote{Note the ambiguity in this rewriting: we could also have written $\< l_1 l_2 \>^2 [l_1 l_2]^2 = \big((l_1-l_2)^2\big)^2 = 16(l^2)^2$. The resulting two forms of the integrand differ by terms that integrate to zero. 
} 
$\< l_1 l_2 \>^2 [l_1 l_2]^2 = \big((l_1+l_2)^2\big)^2 = (K^2)^2 = s^2$. Thus the contribution from the $s$-channel diagram is (including the symmetry factor of $1/2$)
\be
  \label{MHVsChan}
  I_s = \frac{1}{2}  
  \int \frac{\text{d}^dl}{(2\pi)^d}\frac{C_s}{\big(l+\frac{1}{2}K\big)^2\big(l-\frac{1}{2}K\big)^2}
  = \frac{1}{2}  [12]^2  \<34\>^2 s^2 \, I_2(s),
\ee
where $I_2(s)$ is the scalar bubble integral \reef{masterbubble}. The $t$-channel cut is a little more involved. Here we have
\be
  \begin{split}
  C_t &= 
  2 \mathcal{A}_4^\text{tree}\big(1^+_\gamma\,3^-_\gamma\,{l'_1}^+_\gamma\,{l'_2}^-_\gamma\big)
  \mathcal{A}_4^\text{tree}\big(-{l'_1}^-_\gamma\,-{l'_2}^+_\gamma\,2^+_\gamma\,4^-_\gamma\big)
  \\
  &= 2[1 l'_1]^2 \< 3 l'_2 \>^2 [2 l'_2]^2 \<l'_1 4\>^2
  \\
  &=2 \<4 | \bar\sigma_\mu | 1] \<4 | \bar\sigma_\nu | 1] \<3 | \bar\sigma_\rho | 2] \<3 | \bar\sigma_\sigma | 2]
  \,
  {l'_1}^\mu {l'_1}^\nu {l'_2}^\rho {l'_2}^\sigma.
  \end{split}
\ee
The factor of 2 takes into account that one also has to include the opposite helicity assignments for the internal lines. Odd powers of the loop-momentum $l$ vanish in the integral
\be
  I_t = \frac{1}{2}  
  \int \frac{\text{d}^dl}{(2\pi)^d}\frac{C_t}{\big(l+\frac{1}{2}K'\big)^2\big(l-\frac{1}{2}K'\big)^2}\,.
\ee 
Again, we have included a symmetry factor of $1/2$.  
Using $l'_1 = l + \frac{1}{2} K'$ and $l'_2 = -l + \frac{1}{2} K'$, the integral can be simplified to the scalar bubble integral \reef{masterbubble} with the help of the tensor integrals  \reef{2tensorInt} and \reef{4tensorInt}. After some manipulations of the resulting expressions, one arrives at the simple result
\be  
  \label{MHVtChan}
 I_t = \frac{1}{5} [12]^2 \<34\>^2   t^2\, I_2 (t)\,.
\ee
The result of the $u$-channel diagram is obtained by taking $3 \lra 4$ of $I_t$. We can now write the divergent part of the 1-loop MHV 4-point amplitude in pure BI theory as
\be
  \mathcal{A}_4^\text{1-loop BI}(1^+_\gamma\,2^+_\gamma\,3^-_\gamma\,4^-_\gamma)
  = [12]^2 \<34\>^2 
  \Big[ \frac{1}{2} s^2 I_2 (s) + \frac{1}{5} t^2 I_2 (t)+ \frac{1}{5} u^2 I_2 (u)
  \Big]\,.
\ee 
Thus the divergent part is simply
\be
  \mathcal{A}_4^\text{1-loop BI}(1^+_\gamma\,2^+_\gamma\,3^-_\gamma\,4^-_\gamma)
  =
  \frac{1}{\epsilon}
  \frac{i}{16\pi^2} 
  [12]^2 \<34\>^2 
  \Big[ \frac{1}{2} s^2 + \frac{1}{5} t^2 + \frac{1}{5} u^2
  \Big]
  + \mathcal{O}(1)\,,
\ee
where the $\mathcal{O}(1)$-terms are regulator dependent. We have also calculated this amplitude using unitarity and the results agree. 

%%%%%%%%%%%%%%%%%%%%%%%%%%%%%%%%%%%%%%
\subsection{1-loop MHV Amplitude with Supersymmetry}
\label{s:MHV4susy}
%%%%%%%%%%%%%%%%%%%%%%%%%%%%%%%%%%%%%%

We now perform a check on our calculation by computing the MHV amplitude in a theory with fermions and scalars coupled to the vector supersymmetrically, i.e.~such that the supersymmetry Ward identities hold:
\be
  \mathcal{A}_4\big(1_\gamma^+\,2_Z\,3_\gamma^-\,4_{\overline{Z}}\big)
  = \frac{\<23\>}{\<34\>}
  \mathcal{A}_4\big(1_\gamma^+\,2_\psi^+\,3_\gamma^-\,4_\psi^-\big)
  = \frac{\<23\>^2}{\<34\>^2}
  \mathcal{A}_4\big(1^+_\gamma\,2^+_\gamma\,3^-_\gamma\,4^-_\gamma\big)\,.
\ee
Amplitudes with two same-helicity vectors and two scalars or two fermions vanish; this  means that there are no $s$-channel cuts when a fermion or scalar runs in the loop. The calculation of the $t$ and $u$-channel cuts proceed as for the vector. With $N_v$ vectors, $N_f$ Weyl fermions, and $N_s$ complex scalars, the result for the divergent part of the MHV 1-loop amplitude is then
\be
  \label{eq:ppmm1-loop}
  \begin{aligned}
  \mathcal{A}_4^\text{1-loop gen} & (1^+_\gamma\,2^+_\gamma\,3^-_\gamma\,4^-_\gamma) \\
  &= [12]^2 \<34\>^2 
  \bigg[ \frac{N_v}{2} s^2 I_2 (s) +\Big( \frac{N_v}{5} +  \frac{N_f}{20} + \frac{N_s}{30}\Big)\big( t^2 I_2 (t)+ u^2 I_2 (u) \big)
  \bigg]
  \\
  & =\frac{1}{\epsilon}
  \frac{i}{16\pi^2} 
  [12]^2 \<34\>^2 
  \bigg[ \frac{N_v}{2} s^2 +\Big( \frac{N_v}{5} +  \frac{N_f}{20} + \frac{N_s}{30}\Big) \big(  t^2 + u^2 \big)
  \bigg]
  + \mathcal{O}(1)\,.
  \end{aligned}
\ee 
As an independent check, we also calculated this amplitude using unitarity. It is also reproduced by the expression for the bubble coefficient given in \cite{ArkaniHamed:2008gz} that connects the behavior of the tree amplitudes under BCFW shifts to the UV divergences at 1-loop.  

With $\mathcal{N}=4$ supersymmetry, i.e. in ~$\mathcal{N}=4$ DBI, we have $N_v=1$, $N_f=4$, and $N_s = 3$, which gives
\be
 \label{N41loop}
  \begin{aligned}
  \mathcal{A}_4^\text{1-loop $\mathcal{N}=4$}(1^+_\gamma\,2^+_\gamma\,3^-_\gamma\,4^-_\gamma)
  &=
 [12]^2 \<34\>^2 
  \frac{1}{2} \bigg[ s^2 I_2 (s) + t^2 I_2 (t)+ u^2 I_2 (u)
  \bigg]
  \\
  & =\frac{1}{\epsilon}
  \frac{i}{32\pi^2} 
  [12]^2 \<34\>^2 
  \bigg[ s^2 +t^2 + u^2 
    \bigg]
  + \mathcal{O}(1)\,.
  \end{aligned}
\ee 
This result for the 1-loop amplitude in $\mathcal{N}=4$ DBI reproduces that found in \cite{Shmakova:1999ai}.

%%%%%%%%%%%%%%%%%%%%%%%%

\section{Passarino-Veltman Integral Reduction}
\label{app:intred}
We use the Passarino-Veltman integral reduction method to evaluate the 1-loop integrals in section \ref{app:1loopcalc}. We outline the method here and use it to reduce the needed tensor integrals to scalar-bubble integrals. 

The loop-integrands encountered in section \ref{app:1loopcalc} are all of the form
\be
    \mathcal{I} =  \frac{N[l;p_i;\epsilon_i]}{\big(l-\frac{1}{2}K\big)^2\big(l+\frac{1}{2}K\big)^2}\,,
\ee
where $l$ is the loop-momentum, $\eps_i$ are polarization vectors, $p_i$ are on-shell external momenta, and $K$ is a sum of external momenta. 

The numerator $N[l;p_i;\epsilon_i]$ is ambiguous by terms that integrate to zero: there are two types
\begin{description}
\item[Property 1.] Any term with odd powers of the loop-momentum vanishes. 
\item[Property 2.] Any term proportional to 
$\big(l-\frac{1}{2}K\big)^2$ or $\big(l+\frac{1}{2}K\big)^2$ vanishes. The reason is that an integral with such a factor can be put in the form of scaleless integrals 
\begin{equation}
  \int \frac{\text{d}^dl}{(2\pi)^4} \frac{1}{l^2},\;\;\; \int \frac{\text{d}^dl}{(2\pi)^4} \frac{l^\mu l^\nu}{l^2}, \;\;\; \int \frac{\text{d}^dl}{(2\pi)^4}(1), \;\;\;\text{...}
\end{equation}
that are zero in dimensional regularization. 
\end{description}
We now use these two properties to derive the integrals \reef{2tensorInt} and \reef{4tensorInt} we need for the computations in section \ref{app:1loopcalc}:

\begin{itemize}
\item Integral 1
\be 
\label{Int1}
  \int \frac{\text{d}^dl}{(2\pi)^d} \frac{(K\cdot l) l^{\mu_1} \cdots l^{\mu_n}}
  {\big(l+\frac{1}{2}K\big)^2\big(l-\frac{1}{2}K\big)^2}
  = 0\,.
\ee
{\em Proof:} if $n$ is even, the integral vanishes by Property 1 above.  If $n$ is odd, write 
$(K\cdot l) = \big(l + \frac{1}{2} K \big)^2 - l^2 - \frac{1}{4}K^2$. Then the first term makes the integral vanish by Property 2, while the last two terms integrate to zero by Property 1.
\item Integral 2
\be 
\label{Int2}
  \int \frac{\text{d}^dl}{(2\pi)^d} \frac{l^2}
  {\big(l+\frac{1}{2}K\big)^2\big(l-\frac{1}{2}K\big)^2}
  =  - \frac{1}{4}K^2\,  I_2(K^2)\,.
\ee
{\em Proof:} write $l^2 = \big(l + \frac{1}{2} K \big)^2 - (K \cdot l) - \frac{1}{4}K^2$. The first two term integrate to zero. So we are left with $- \frac{1}{4}K^2$ times the scalar bubble-integral $I_2(K^2)$.
\item Integral 3 (the integral in \reef{2tensorInt})
\be 
\label{Int3}
  \int \frac{\text{d}^dl}{(2\pi)^d} \frac{l^\mu l^\nu}{\big(l+\frac{1}{2}K\big)^2\big(l-\frac{1}{2}K\big)^2}
  = -\frac{1}{4(d-1)}\left[ K^2\eta^{\mu\nu} - K^{\mu}K^{\nu} \right]\, I_2(K^2)\,.
\ee
{\em Proof:} there are only two possible tensor structures for the integral, so we can write an ansatz as $A_1 \eta^{\mu\nu} K^2 + A_2 K^\mu K^\nu$, where $A_1$ and $A_2$ are constant numbers. Dot in $K^\mu$ and use Integral 1 \reef{Int1} to conclude that $A_1 + A_2 = 0$. Then contract with  $\eta_{\mu\nu}$ and use Integral 2 \reef{Int2} to show that $A_1 d + A_2 = - \frac{1}{4} I_2(K^2)$. Solving for the unknown constants gives 
$A_1 = -A_2 = - \frac{1}{4(d-1)} I_2(K^2)$ and the result \reef{Int3} follows.
\item Integral 4 
\be \label{Int4}
  \int \frac{\text{d}^dl}{(2\pi)^d} \frac{(l^2)^2}{\big(l+\frac{1}{2}K\big)^2\big(l-\frac{1}{2}K\big)^2} 
   = \frac{1}{16}(K^2)^2\, I_2(K^2)\,.
\ee
{\em Proof:} rewrite 
$(l^2)^2 = \big( \big(l + \frac{1}{2} K \big)^2 - (K \cdot l) - \frac{1}{4}K^2\big) l^2$.
The first two term integrates to zero by Properties 1 and 2, so this leaves $- \frac{1}{4}K^2$ times Integral 2. Then \reef{Int4} simply follows from \reef{Int2}.
\item Integral 5 (the integral in \reef{4tensorInt})
\bea \label{Int5}
  \int \frac{\text{d}^dl}{(2\pi)^d} \frac{l^\mu l^\nu l^\rho l^\sigma}{\big(l+\frac{1}{2}K\big)^2\big(l-\frac{1}{2}K\big)^2}&&
  \\ 
  &&~~\hspace{-4.5cm} \nonumber
  = \frac{1}{16(d^2-1)}\left[(K^2)^2\eta^{(\mu\nu}\eta^{\rho\sigma)} 
  -K^2 K^{(\mu}K^\nu \eta^{\rho\sigma)} 
  +3 K^\mu K^\nu K^\rho K^\sigma\right]\, I_2(K^2)\,,
\eea
where we have normalized the symmetrizations such that
\be
  \eta^{(\mu\nu}\eta^{\rho\sigma)} =\eta^{\mu\nu}\eta^{\rho\sigma}
  + \eta^{\mu\rho}\eta^{\nu\sigma} + \eta^{\mu\sigma}\eta^{\nu\rho},
\ee  
and $K^{(\mu}K^\nu \eta^{\rho\sigma)}  = K^{\mu}K^\nu \eta^{\rho\sigma} + \text{5 nontrivial perms}$.

{\em Proof:} the tensor structure allows only three possible terms, so we write an ansatz
\be
 B_1\, (K^2)^2\eta^{(\mu\nu}\eta^{\rho\sigma)} 
 + B_2 \,K^2 K^{(\mu}K^\nu \eta^{\rho\sigma)} 
 +B_3\, K^\mu K^\nu K^\rho K^\sigma,
\ee
for the value of \reef{Int4}. We know from Integral 1 that the integral must vanish whenever $K$ is dotted in, so from dotting in $K_\mu K_\nu K_\rho K_\sigma$ and 
$K^{\mu}K^\nu \eta^{\rho\sigma}$ we learn that 
\be
  3B_1 + 6 B_2 + B_3 = 0 
  ~~~~\text{and}~~~~
  (d+2) B_1 + (d+5) B_2 + B_3 = 0\,.
\ee
Finally we contract with $\eta^{\mu\nu}\eta^{\rho\sigma}$ and use Integral 4 to conclude that
\be
  d(d+2) B_1 + 2 (d+2) B_2 + B_3 =  \frac{1}{16}\, I_2(K^2)\,.
\ee
Solving for the coefficients $B_1$, $B_2$, and $B_3$ gives \reef{Int5}.

\end{itemize}

%%%%%%%%%%%%%%%%%%%%%%%%%%%%%%
\section{Higher-Derivative Corrections of Yang-Mills}
\label{sec:YM_appendix}

In this appendix, we calculate possible higher-derivative corrections to the 4-point amplitudes of YM that are used in Section \ref{sec:higherD_KLT} to calculate corrections to 4-point BI amplitudes using the KLT double-copy. 

We construct BCJ-compatible ansatze for three helicity configurations that have the following properties:
\begin{itemize}
	\item \textbf{Relabeling:} Different color-orderings with the same helicity structure have to be related by momentum relabelling.
	\item \textbf{Locality:} The amplitude has at most simple poles at $s, t, u = 0$. Compatibility with BCJ relations \eqref{eq:KK_BCJ_4pt} and locality dictate that $\mathcal{A}_4^{\text{YM}}\left[1234\right]$ cannot have a pole at $t = 0$.
	\item \textbf{Unitarity:} 4-point amplitudes factorize on simple poles into products of 3-point amplitudes. Since 3-point amplitudes of massless particles are fixed by little group scaling, residues on the $s$- and $u$-channel poles are constrained by the spins of particles in our spectrum. Since no fermions can be exchanged and gravitons cannot carry a color-charge, any particles exchanged must have spin 0 and spin 1.
\end{itemize}

%%%%%%%%%%%%%%%%%%%%%%%%%%%%%%
\subsection{The \texorpdfstring{$++++$}{++++} Amplitude}
\label{app:pppp}
We begin with the 4-point self-dual YM amplitude. While it is well known that at leading order the self-dual YM amplitude is vanishing, there can exist non-zero higher-derivative terms, which we evaluate in this section.

Since all external states have the same helicity, all color orderings of the self-dual amplitude are related by momentum relabeling. With this in mind we write the following BCJ and locality-compatible ansatz,
\begin{align}
     \mathcal A_4^\text{YM} [ 1^+ 2^+ 3^+ 4^+ ]= & \frac1{su}\bigg(\tilde c_0 \left ( [ 12 ]^2 [ 34 ]^2 + [ 13 ]^2 [ 24 ]^2 + [ 14 ]^2 [ 23 ]^2 \right ) \nonumber \\
    & + \frac{\tilde c_2}{\Lambda^2} \left ( [ 12 ]^2 [ 34 ]^2 s + [ 13 ]^2 [ 24 ]^2 t + [ 14 ]^2 [ 23 ]^2 u \right ) \nonumber \\
    & + \frac{\tilde c_4}{\Lambda^4} \left ( s^2 + t^2 + u^2 \right ) \left ( [ 12 ]^2 [ 34 ]^2 + [ 13 ]^2 [ 24 ]^2 + [ 14 ]^2 [ 23 ]^2 \right ) \nonumber \\
    &\left. + \frac{\tilde c_6}{\Lambda^6} s t u \left ( [ 12 ]^2 [ 34 ]^2 + [ 13 ]^2 [ 24 ]^2 + [ 14 ]^2 [ 23 ]^2 \right ) + \mathcal O ( \Lambda^{-8} )\right)\,.
\end{align}

One can show that $\tilde{c_0}$ produces a non-zero residue that does not correspond to a particle exchange and $\tilde{c_4}$ corresponds to a higher spin exchange, so they must both be zero by unitarity. Thus we finally conclude that the 4-point YM amplitude takes the form
\begin{multline}
    \label{eq:YM_4pt_app}
    \mathcal A_4^\text{YM} [ 1^+ 2^+ 3^+ 4^+ ] = \frac{\tilde c_2}{\Lambda^2} \frac{[ 12 ]^2 [ 34 ]^2 s + [ 13 ]^2 [ 24 ]^2 t + [ 14 ]^2 [ 23 ]^2 u}{s u} \\ + \frac{\tilde c_6}{\Lambda^6} t \left ( [ 12 ]^2 [ 34 ]^2 + [ 13 ]^2 [ 24 ]^2 + [ 14 ]^2 [ 23 ]^2 \right ) + \mathcal O ( \Lambda^{-8} )\,.
\end{multline}
The $\mathcal O ( \Lambda^{-2} )$ contribution of this amplitude was also calculated in \cite{Dixon:1993xd} and our result is in perfect agreement.
To our knowledge, the $\mathcal O ( \Lambda^{-6} )$ contribution is new.

\subsection{The \texorpdfstring{$+++-$}{+++-} Amplitude}
\label{app:pppm}
We now move on to the next-to-self-dual amplitude. We know this to be vanishing at leading order and here we calculate sub-leading higher-derivative corrections.

Since not all external particles have the same helicity, different color orderings can be related using cyclicity, in addition to simple relabeling. For example, 
\begin{equation}
    \mathcal A_4^\text{YM} [ 1^+ 2^+ 4^- 3^+ ] = \mathcal A_4^\text{YM} [ 3^+ 1^+ 2^+ 4^- ] = \mathcal A_4^\text{YM} [ 1^+ 2^+ 3^+ 4^- ] \big |_{1 \to 3 \to 2 \to 1}\,.
\end{equation}
Compatibility with the BCJ relations and locality fixes our ansatz to be,
\begin{equation}
    \mathcal A_4^\text{YM} [ 1^+ 2^+ 3^+ 4^- ] = \frac{1}{\Lambda^2} \frac{[ 12 ]^2 [ 3 | p_1 | 4 \rangle^2}{s u} \left(\tilde b_0 + \frac{\tilde b_4}{\Lambda^4} \left ( s^2 + t^2 + u^2 \right ) + \frac{\tilde b_6}{\Lambda^6} s t u + \mathcal O ( \Lambda^{-8} )\right)\,.
\end{equation}
The $\tilde{b}_4$ corresponds to the exchange of a spin-3 particle and is thus excluded by unitarity. This leaves us with
\begin{equation}
    \label{eq:YM_pppm_app}
    \mathcal A_4^\text{YM} [ 1^+ 2^+ 3^+ 4^- ] = \frac{1}{\Lambda^2} \frac{[ 12 ]^2 [ 3 | p_1 | 4 \rangle^2}{s u} \left ( \tilde b_0 + \frac{\tilde b_6}{\Lambda^6} s t u + \mathcal O ( \Lambda^{-8} ) \right )\,.
\end{equation}
It is interesting to note that the first term factorizes into a 3-point YM amplitude and a 3-point amplitude consistent with an $F^3$-type interaction. As in the self-dual case, we check the leading contribution to our result against the results of \cite{Dixon:1993xd} and we find perfect agreement.

\subsection{The \texorpdfstring{$++--$}{++--} Amplitude}
\label{app:ppmm}
Finally, let us examine the MHV 4-point amplitude. Using momentum relabeling and cyclicity, we can rewrite the BCJ relations as
\begin{equation}
    \begin{split}
        & \frac u t \mathcal A_4^\text{YM} [ 1^+ 2^+ 3^- 4^- ] = \mathcal A_4^\text{YM} [ 2^+ 1^+ 3^- 4^- ] = \mathcal A_4^\text{YM} [ 1^+ 2^+ 3^- 4^- ] \Big |_{1 \leftrightarrow 2} \ \text{and} \\
        & \frac u t \mathcal A_4^\text{YM} [ 1^+ 2^+ 3^- 4^- ] = \mathcal A_4^\text{YM} [ 1^+ 2^+ 4^- 3^- ] = \mathcal A_4^\text{YM} [ 1^+ 2^+ 3^- 4^- ] \Big |_{3 \leftrightarrow 4}\,.
    \end{split}
\end{equation}
With the above constraints, a local ansatz for the amplitude is
\begin{multline}
    \label{eq:YM_4pt_intermediate}
    \mathcal A_4^\text{YM} [ 1^+ 2^+ 3^- 4^- ] \\ = \frac{[ 12 ]^2 \langle 34 \rangle^2}{s u} \bigg ( \tilde a_0 + \frac{\tilde a_{21}}{\Lambda^2} s + \left ( \frac{\tilde a_{41}}{\Lambda^4} s^2 - \frac{\tilde a_{42}}{\Lambda^4} t u \right ) + \left ( \frac{\tilde a_{61}}{\Lambda^6} s^3 - \frac{\tilde a_{62}}{\Lambda^6} s t u \right ) + \mathcal O ( \Lambda^{-8} ) \bigg ).
\end{multline}
$\tilde{a}_{21}$, $\tilde{a}_{41}$ and $\tilde{a}_{61}$ correspond to exchanges of spin-2, spin-3 and spin-4 particles respectively and are hence all set to zero by unitarity. If we continue at higher orders in the derivative expansion, we see that this pattern continues.
We either find local contributions to the amplitude or pole-terms with residues that correspond to the exchange of  higher-spin particles.

To summarize, the YM 4-point amplitude with higher-derivative corrections has the form
\begin{equation}
\label{eq:YM_4pt_final_app}
\mathcal A_4^\text{YM} [ 1^+ 2^+ 3^- 4^- ] = \frac{[ 12 ]^2 \langle 34 \rangle^2}{s u} \left ( \tilde a_0 + \frac{\tilde a_{4}}{\Lambda^4} t u + \frac{\tilde a_{6}}{\Lambda^6} s t u + \mathcal O ( \Lambda^{-8} ) \right )\,.
\end{equation}
In the above, the leading term matches the well-known Parke-Taylor formula for Yang-Mills if we choose $\tilde a_0 = - g_\text{YM}^2$. Also note that we have redefined $\tilde a_{42} = - \tilde a_4$ and $\tilde a_{62} = - \tilde a_6$ to simplify the notation.

%%%%%%%%%%%%%%%%%%%%%%%%%%%%%%
%%%%%%%%%%%%%%%%%%%%%%%%%%%%%%
\bibliographystyle{JHEP}
\bibliography{BIBib.bib}

%%%%%%%%%%%%%%%%%%%%%%%%%%%%%%
%%%%%%%%%%%%%%%%%%%%%%%%%%%%%%

\end{document}